\documentclass{llncs}
\usepackage{makeidx}  
\usepackage{url}      
\usepackage{graphicx}
\usepackage{subfigure}
\usepackage{amssymb,amsmath}
\usepackage{amsfonts}
\graphicspath{{Pics/}} 

\begin{document}
\frontmatter

\pagestyle{headings}

\mainmatter

\title{A Fast and Robust Patient Specific\\Finite Element Mesh Registration Technique:\\Application to 60 Clinical Cases}

\titlerunning{Patient Specific Finite Element Mesh Registration Technique}

\author{Marek Bucki\inst{1} \and Claudio Lobos\inst{2} \and Yohan Payan\inst{1,3}}

\authorrunning{M. Bucki, C. Lobos, Y. Payan}   

\tocauthor{Marek Bucki (University Joseph Fourrier), Claudio Lobos (University Joseph Fourrier), Yohan Payan (University Joseph Fourrier)}

\institute{
TIMC-IMAG Laboratory, UMR CNRS 5525, University Joseph Fourier,\\38706 La Tronche, France, \email{Marek.Bucki@imag.fr} 
\and
Departamento de Inform\'atica, Universidad T\'ecnica Federico Santa Mar\'{\i}a, Av. Vicu\~na Mackenna 3939, Zip: 8940897, San Joaqu\'{\i}n, Santiago, Chile
\and
PIMS-Europe, UMI CNRS 3069, 200-1933 West Mall,\\University of British Columbia, Vancouver BC, V6T 1Z2, Canada
}

\maketitle

\begin{abstract}
Finite Element mesh generation remains an important issue for patient specific biomechanical modeling. While some techniques make automatic mesh generation possible, in most cases, manual mesh generation is preferred for better control over the sub-domain representation, element type, layout and refinement that it provides. Yet, this option is time consuming and not suited for intraoperative situations where model generation and computation time is critical. To overcome this problem we propose a fast and automatic mesh generation technique based on the elastic registration of a generic mesh to the specific target organ in conjunction with element regularity and quality correction. This Mesh-Match-and-Repair (MMRep) approach combines control over the mesh structure along with fast and robust meshing capabilities, even in situations where only partial organ geometry is available. The technique was successfully tested on a database of 5 pre-operatively acquired complete femora CT scans, 5 femoral heads partially digitized at intraoperative stage, and 50 CT volumes of patients' heads. In the latter case, both skin and bone surfaces were taken into account by the mesh registration process in order to model the face muscles and fat layers. The MMRep algorithm succeeded in all 60 cases, yielding for each patient a hex-dominant, Atlas based, Finite Element mesh with submillimetric surface representation accuracy, directly exploitable within a commercial FE software.
\end{abstract}


\hyphenation{re-gis-tra-tion ana-to-mi-cal pro-blem re-pre-sents ca-lib-ra-tion re-gu-la-ri-ty li-mi-ted re-tro-gna-thic ap-pro-xi-ma-tion po-si-ti-ve-ly mo-de-ling as-so-cia-ted di-gi-ti-zed}

\newcommand{\R}{{\mathbb{R}}}

\section{Introduction}

Physically based models are now widely used in the field of biomedical engineering to represent human organs' geometrical and mechanical behaviors. Among them, numerical models based on the Finite Element Method \cite{HughesFEM87} became very popular because of their ability to address the complex geometries, the anisotropic material properties and the specific boundary conditions associated with living tissues. 

In the field of medical imaging, the first Finite Element (FE) models were mostly used to better understand and validate a given surgical treatment, to model physiological behaviours or to provide virtual simulators for clinicians. In these frameworks, models were limited to a single generic model for each study (the terminology ``Atlas'' was often used).

More recently, applications in the domain of Computer Assisted Planning and Computer Aided Surgery sparked the need for patient-specific FE models representing the modeled organ geometry reconstructed from patient medical image data, such as computed tomography (CT) or magnetic resonance imaging (MRI). In most cases, organs are identified in these data sets by means of manual, semi-automatic or automatic segmentation tools that extract shape information (3D points, contours and/or surfaces) necessary for the generation of the Finite Element mesh representing the volume of the organ.

For both segmentation and FE mesh generation phases, manual intervention is often required which can make this procedure long and tedious. This is especially true in situations where the pre- or intraoperative time window or clinician availability to perform these delicate tasks is limited.

This paper addresses the second phase, namely the FE mesh generation, with the introduction of a procedure - the Mesh-Match-and-Repair (MMRep) algorithm - that allows a fast and fully-automatic patient-specific FE mesh generation.

Although FE models were already used in the field of biomedical engineering at the beginning of the 1980s \cite{Jaspers80,Huiskes80}, researchers only began focusing on automatic mesh generation in the early 2005 \cite{Couteau00,Gibson03,Viceconti04,Taddei04,Luboz05,Liao05,Shim07,Sigal08,Grosland09}. The primary motivation for automatic mesh generation algorithms was that patient specific FE models could be routinely used by the clinicians. In that domain, the vast majority of automatic mesh generators for living tissues produce tetrahedral meshes \cite{Molino2003,George02,Alliez2005,Si05}. Based on a 3D surface representing the external geometry of the organ, these algorithms produce a volume made of high quality 3D tetrahedra and sometimes allow for adaptive mesh refinement (\cite{Si06}, Tetgen\footnote{\url{http://tetgen.berlios.de}}).

Some researchers in biomechanics continue to propose hand-made FE meshes \cite{Chabanas00,Luboz04,Wittek07}, mainly for two reasons. Firstly, they argue that it is important to be able to identify sub-regions associated to anatomical sub-structures inside the 3D FE mesh (e.g. the ventricles, the tumor and the hemispheres of the brain FE mesh proposed by Wittek et al. in \cite{Wittek07}). These sub-regions, corresponding to sets of elements, are labeled and associated with specific constitutive behaviors and boundary conditions. Secondly, they tend to prefer hexahedra over tetrahedra, based on numerical considerations \cite{Benzley95} as well as the fact that for incompressible and/or nearly incompressible materials, 4-noded tetrahedra with linear shape functions tend to lock and become overly stiff \cite{HughesFEM87}. Despite some improvements to this method of creating FE meshes this work still requires an excessive amount of manual effort to achieve satisfactory results.

In order to still benefit from this manual design while providing automatic FE mesh generation, techniques termed ``registration methods'' or ``morphing methods'' were recently proposed \cite{Couteau00,Castellano-Smith01,Sigal08,Grosland09}. The main premise is to start with a predefined ``generic'' (or ``template'') FE mesh that represents the organ. This mesh is manually designed to include any necessary sub-regions and hexahedra, and to preserve element quality, orientation and density in regions that require it for numerical simulation. This template is then automatically morphed onto the patient data (3D landmark points, contours and/or surfaces) that was extracted from the segmentation of the medical images. This process generates a patient-specific mesh adapted to the geometries of the anatomical structures extracted from patient data, with a mesh topology that is similar to that of the template (same nodes and elements organization).

Our group was the first to initiate this principle of mesh morphing with the introduction of the Mesh-Matching algorithm \cite{Couteau00}. Since then, we encountered two strong limitations with our mesh morphing principle. Both are due to the fact that the template mesh quality after registration can be strongly decreased when the morphing algorithm induces excessive spatial distortions. Therefore, the first consequence is the inability to maintain the regularity of some elements, which disables FE analysis from being carried out. This concern was partially discussed in \cite{Luboz05}. The second consequence of the morphing method is that the elements shape qualities are decreased in some regions of the template mesh, which leads to lower accuracy in the numerical simulation \cite{Field00,Kwok2000,Shewchuk2002a}.

This paper aims at introducing our latest algorithms concerning (1) the elastic registration method that guarantees a $C^1$-diffeomorphic transform; and (2) the mesh repair technique that ensures that the produced mesh complies with both regularity and quality criteria. The MMRep approach was applied and evaluated on 60 clinical cases, which is, to our knowledge, the largest database ever tested in the literature for a patient-specific FE mesh registration method.

The MMRep algorithm is a two-step sequential procedure. Firstly, the patient data and the Atlas mesh surface nodes are registered using the \textbf{elastic deformation procedure} described in \S \ref{SecElasticregistration}. This deformation is then applied to the inner Atlas nodes yielding a FE mesh that represents the modeled domain with sufficient accuracy. As a consequence of this deformation, the Atlas elements may suffer distortions and become either ``irregular elements'' which make FE analysis impossible, or ``poor quality elements'' in which case the computation, although feasible, can exhibit numerical instabilities. To recover mesh regularity and reach an acceptable quality level, the \textbf{mesh repair procedure} described in \S \ref{SecMeshreparation} is carried out on the deformed mesh.

\section{Elastic registration}
\label{SecElasticregistration}

\subsection{Registration overview}

We define an elastic registration function as a mapping $\emph{R}: \R^3 \rightarrow \R^3$ that superimposes a source point cloud \emph{S} onto a target, or ``destination'', data set \emph{D}, which can either be a point cloud or a surface mesh. The computed elastic registration procedure complies with continuum mechanics conditions on motion \cite{Belytschko06} as $\emph{R}$ defines a $C^1$\textbf{-diffeomorphic}, \textbf{non-folding} and \textbf{one-to-one} correspondence between geometric data sets, as demonstrated respectively in \S \ref{SecC1Diff}, \S \ref{SecControlOverSpaceDistortion} and \S \ref{SecRegistrationInversion}. 

The input source points set is initially embedded in a deformable ``virtual elastic grid''. We arbitrarily set the shape of the grid to be the bounding box of the points, extended by a 10\% margin. The considered deformation \emph{R} is formed by successive elementary grid deformations noted $r$, all having the desired regularity properties, much in the same way as proposed in \cite{Rueckert06}.

The regular grid is progressively refined in order to increase registration accuracy and the expression of the compound registration function is:

\begin{equation*}
\emph{R} =
\underbrace{r_J^{N_J} \circ\ldots \circ r_J^1}_{G_J}
\circ \ldots \circ
\underbrace{r_1^{N_1} \circ \ldots \circ r_1^1}_{G_1}
\end{equation*}

where $G_j, j=1,\ldots,J$ are successive grid refinements, and $r_j^i, i=1,\ldots,N_j$, elementary deformations performed at grid level $j$. As level indications are irrelevant in the following demonstrations, the deformation \emph{R} is simply written as:

\begin{equation}
\emph{R} = r_N \circ \ldots \circ r_1
\label{EqTotalRegistration}
\end{equation}

where $N = \sum_{j=1}^J N_j$.

If required, the inverse registration, $R^{-1}$, can be computed by combining the elementary inverses in the reverse order of the direct registration, thus: $\emph{R}^{-1} = r_1^{-1} \circ \ldots \circ r_N^{-1}$.

At each step $n$, the choice of the elementary deformation $r_n$ to be applied to the source data is driven by the minimization of a ``registration energy'' $E$ which measures the similarity between the deformed source points set and the destination data set \emph{D}. As geometrical shape similarity is sought, $E$ is defined as the sum of Euclidean distances between \emph{S} and \emph{D}:

\begin{equation}
E(r_n \circ \emph{R}_{n-1} \,) = \sum_{\textbf{s} \in \emph{S}} d(r_n(\emph{R}_{n-1}(\textbf{s})),\emph{D} \,)
\label{EqRegistrationEnergy}
\end{equation}

where $\emph{R}_{n-1} = r_{n-1} \circ \ldots \circ r_1$ represents the registration function assembled at step $n-1$, and the operator $d(\cdot,\cdot)$ can either be a point-to-point or a point-to-surface distance measure, depending on the nature of \emph{D}.

In order to speed up the computations of energy $E$, a distance map is generated from the destination data set \emph{D} prior to registration \cite{Saito94}. Distance map voxel dimensions are $1\times1\times1$mm in order to achieve submillimetric surface representation. The sampled space region covers the bounding box of the considered destination data, extended by a 5\% margin. 

Point-to-surface registrations are performed in all three use cases discussed in \S \ref{SecExperimentalevaluation}. For each destination Atlas mesh, signed point-to-surface distances are measured using surface orientation information. A positive distance is recorded in the distance map for points lying outside the closed destination surface, and a negative distance is recorded for points lying inside. The distance map computations can take several minutes, which is not a limitation to our approach, even in the intraoperative context illustrated in \S \ref{SecPartialFemora}, as the Atlas meshes are processed pre-operatively. At registration time, source-to-destination distance evaluations are done by trilinear interpolation within the signed distance map and the absolute value of the result is retained.

The $r_n$ functions are successively chosen so that $E$ decreases optimally at each step. To this end the virtual grid is subdivided into a number of regular hexahedrons called ``cells''. Each node of the grid is considered separately and the gradient of the registration energy is computed as function of the node's position. The opposite of this vector defines the node's preferred displacement, that is, the one that will lead to the greatest registration energy decrease achievable by moving the considered node.

Let $\emph{S}_0 := \emph{S}$ be the initial source points set, and let $\emph{S}_i$ be the source points set at iteration $i$. Of all nodes, the one leading to the highest energy decrease is chosen and its preferred displacement is applied while all the other nodes remain fixed. This nodal displacement is propagated throughout the neighboring grid cells and the affected source points are moved accordingly to generate $\emph{S}_{i+1}$, the new set of deformed source points.

Once this basis deformation step applied, the virtual grid returns to its initial regular configuration and the source points set $\emph{S}_{i+1}$ is embedded within it. A new iteration can be computed by taking the new configuration $\emph{S}_{i+1}$ as input. This Eulerian strategy allows large deformations of \emph{S} to be achieved without having to maintain large-strain consistency of the virtual mesh, as would be the case if the virtual grid strictly followed the deformation with each iteration.

The regularity of the grid before each iteration also saves computational time by allowing the interpolation of a node displacement throughout its neighboring cells to be computed using a template unitary cell. The iterations stop when no significant energy decrease can be achieved by moving any grid nodes at the current grid refinement level. In our implementation, we have set this stop threshold $T_E$ to 1\%, and the registration iterations continue as long as energy $E$ can be reduced by more than 1\%, i.e. $(E_i-E_{i+1})/E_i > T_E$. Thus, the number of iterations performed at each refinement step is variable.

The above iterative loop describes the procedure at a given grid refinement level. In order to maintain the spatial consistency of \emph{S} during the assembly of the registration transformation, a top-down hierarchical approach has been implemented. The iterative assembly of the registration function \emph{R} starts at the coarsest grid level. Once the deformation search at the current level has been exhausted, the grid is refined by subdividing each cell into 8 smaller ones in an octree method.

\begin{figure}[tb]
\begin{center}
	\subfigure[]{\includegraphics[width=0.20\linewidth]{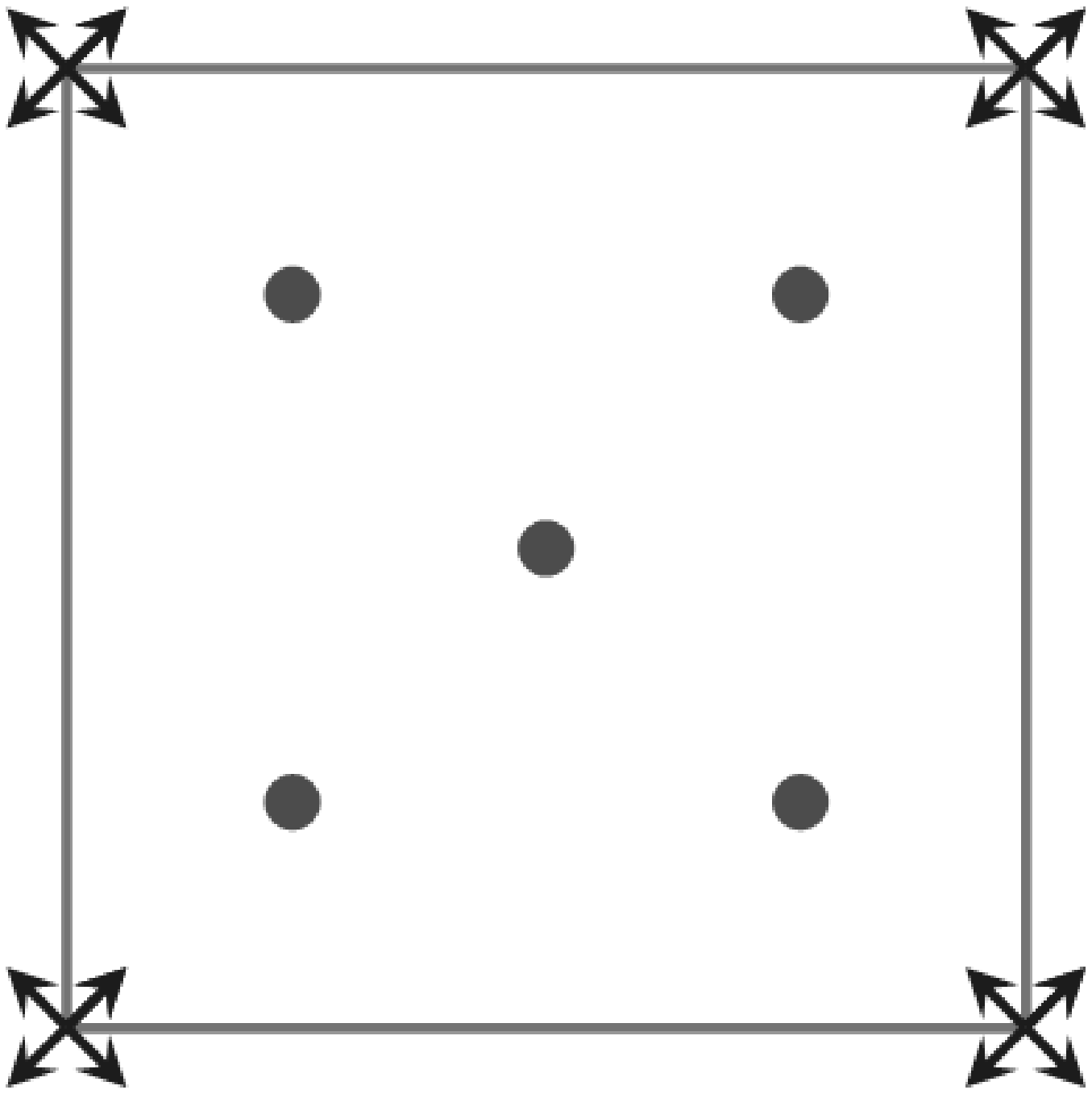}}
	\hspace{0.5cm}
  \subfigure[]{\includegraphics[width=0.20\linewidth]{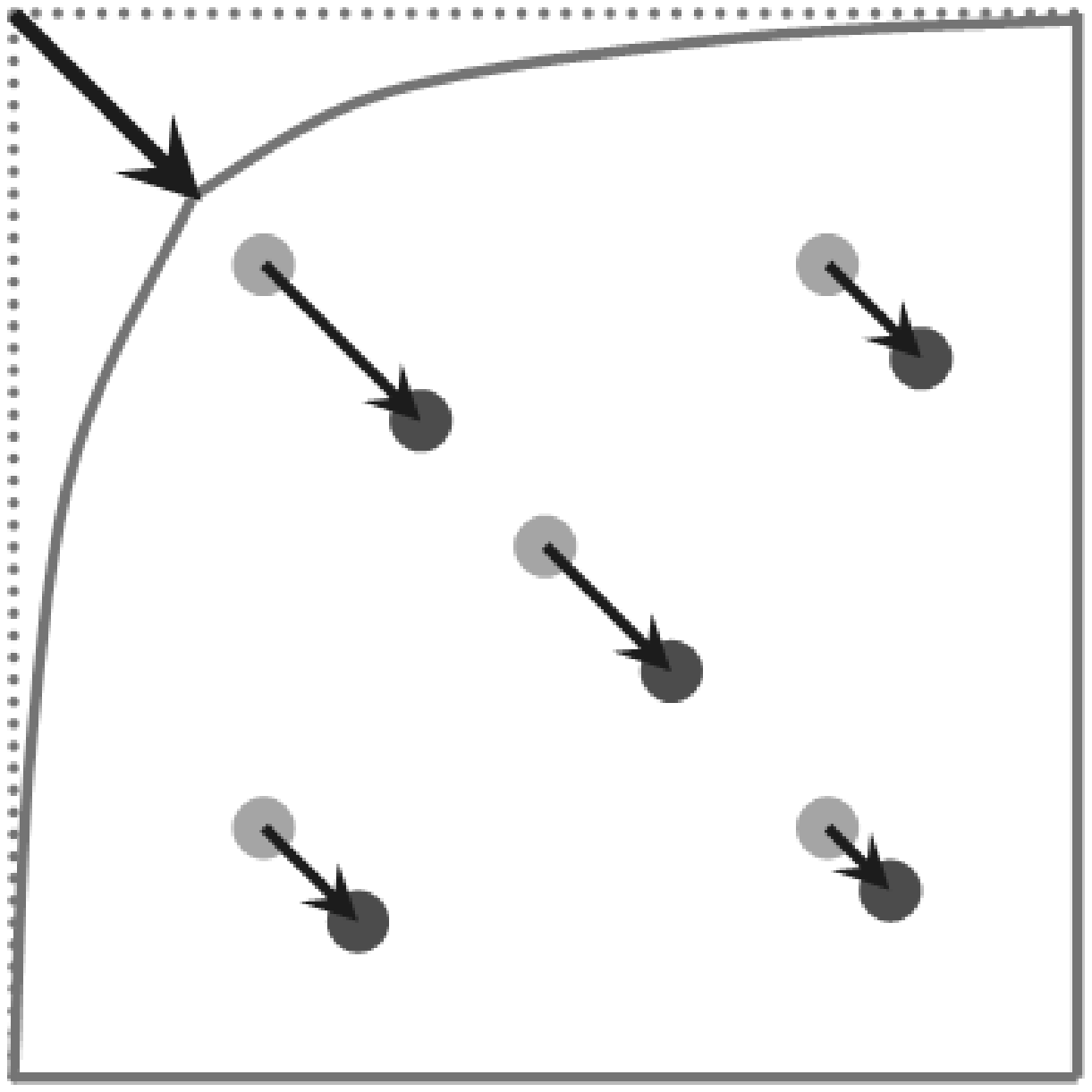}}
	\hspace{0.5cm}
	\subfigure[]{\includegraphics[width=0.20\linewidth]{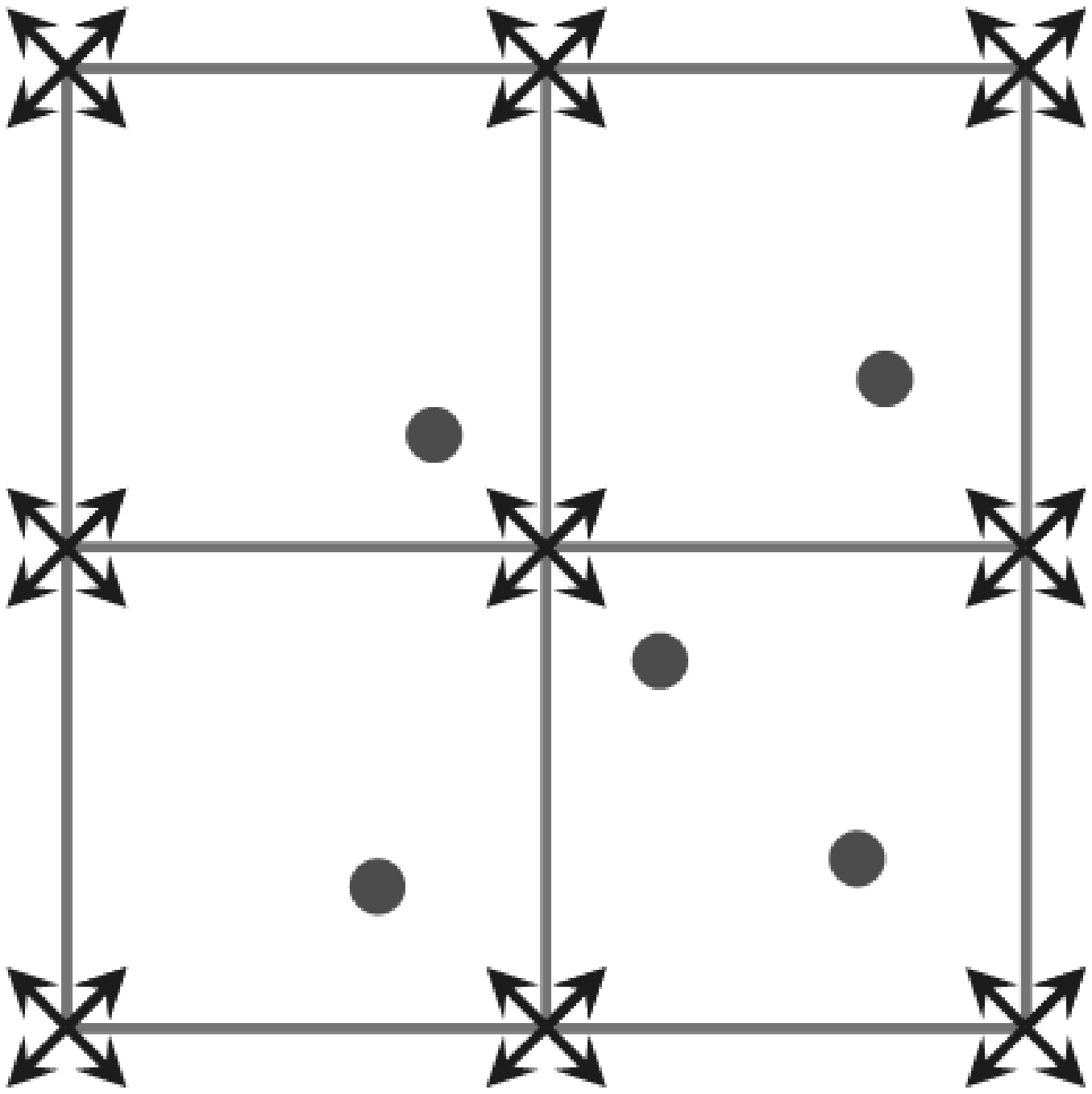}}
	\hspace{0.5cm}
  \subfigure[]{\includegraphics[width=0.20\linewidth]{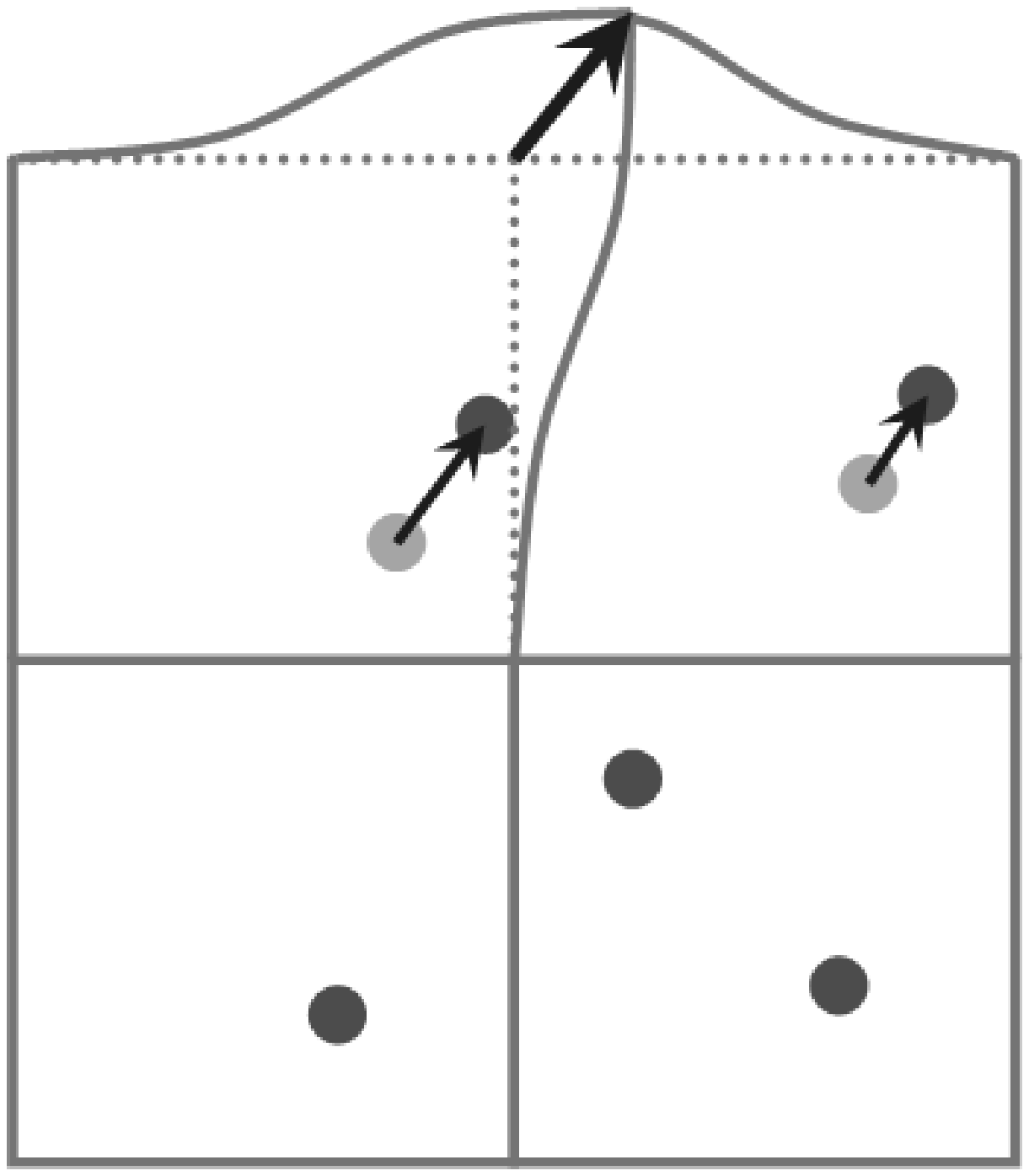}}
	\caption{Elastic registration overview. (a) $\emph{S}_0$ at refinement level 1. (b) $\emph{S}_1$ after deformation at level 1. (c) $\emph{S}_1$ at refinement level 2. (d) $\emph{S}_2$ after deformation at level 2.}
	\label{fig:EMatchOverview}
\end{center}
\end{figure}

Figure \ref{fig:EMatchOverview} illustrates in 2D the multi-grid iterative registration technique. The source points \emph{S} are represented by the grey dots; for clarity \emph{D} is not shown. In (a) the initial set $\emph{S}_0:=\emph{S}$ is embedded within the virtual square grid discretized at the coarsest level 1, and the energy gradients are computed at the 4 mesh nodes. In (b) the optimal node displacement is applied to the grid and the source points move producing a new source set $\emph{S}_1$. If no significant energy decrease can be generated at level 1, the grid is refined at level 2, (c), $\emph{S}_1$ is embedded within it, and the energy gradients are computed at the 9 nodes of the mesh. In (d) the best nodal displacement is applied and the resulting source points set $\emph{S}_2$ is computed.

At each grid refinement level, the size of the registered source features is approximately that of the current grid cell. The virtual grid nodal displacement is only applied if the resulting grid deformation leads to a registration energy decrease. As all the source points located in the grid cells surrounding the displaced node are altered, the source features significantly smaller than the current cell size have limited impact on the registration sequence. The dimensions of the smallest cell reached during the top-down refinement descent thus roughly define the size of the specific features present in \emph{S} that may not be registered if the corresponding feature is absent from \emph{D}. 

By primarily focusing on larger source features, this hierarchical approach also reduces the influence of noise usually found in patient data gathered from pre-operative medical image segmentation (see \S \ref{SecCompleteFemora}) or from intraoperative acquisitions (see \S \ref{SecPartialFemora}). During the evaluation of the MMRep technique the smallest cell size was set to 1mm, which required about 8 or 9 grid subdivision levels as the typical model size was about 25cm and $250\text{mm} / 2^8 \approx 1\text{mm}$.

Furthermore, the mechanical regularization strategy described in \S \ref{SecMechanicalregularization} limits excessive space distortions due to the presence of noise by monitoring the magnitude of potential elastic energy in the source space throughout the deformation process. Accurate registration of outliers has indeed a prohibitive mechanical cost which, in our procedure, discards elementary deformations attempting to register them.

The optimization procedure is done by a gradient descent technique \cite{Press92}. The energy gradients are evaluated at each node by the Finite Differences method. The line search in the direction of the gradient descent is performed by approximating the energy curve by a parabola $P(t)=at^2+bt+c$. The parameters $a$, $b$ and $c$ are deduced from: the current value of $E$ at the considered node, corresponding to $P(0)$; the slope of $E$ in the descent direction, defining $P'(0)$; and the value of $E$ at the maximally displaced node position (see \S \ref{SecControlOverSpaceDistortion}), yielding $P(1)$. Finally, the optimal descent step in the current direction is given by $t_\text{min} = -b/(2a)$.

\subsection{$C^1$-differentiability of the deformation}
\label{SecC1Diff}

At each registration step, the displacement applied to a given grid node $\textbf{n}$ is propagated to the source points located in cells surrounding $\textbf{n}$ by means of a ``weight'' function $w_\textbf{n}: \R^3 \rightarrow [0,1]$.

Let $\overrightarrow{u}$ be the displacement applied to node $\textbf{n}$, and $\textbf{s}$ be a source point found in a cell affected by the movement of $\textbf{n}$. The displacement propagated to $\textbf{s}$ is $w_\textbf{n}(\textbf{s})\overrightarrow{u}$. Similarly to the shape functions in the Finite Element Method, the support of the weight function $w_\textbf{n}$ is the union of the cells neighboring $\textbf{n}$. The deformation consistency is further ensured by the two following conditions: $w_\textbf{n}(\textbf{n})=1$ and $w_\textbf{n}=0$ at the boundary of its support.

From the above, it follows that the elementary elastic registration function $r$ created by the displacement $\overrightarrow{u}$ of node $\textbf{n}$ has the following expression:

\begin{equation}
r: \R^3 \rightarrow \R^3, \textbf{s} \mapsto \textbf{s} + w_\textbf{n}(\textbf{s})\overrightarrow{u}
\label{eq:BasisRegDefinition}
\end{equation}

$C^1$\textbf{-differentiability} of the elementary registration function $r$ stems from the $C^1$-differentiability of the weight function $w_\textbf{n}$ and the Jacobian matrix of $r$ is given by:

\begin{equation}
J_r :=
\frac{\partial r}{\partial \textbf{s}} =
\textit{Id}_{3 \times 3}
+
\left( \begin{array}{c} u_x \\ u_y \\ u_z \end{array} \right) \left( \begin{array}{ccc} \frac{\partial w_\textbf{n}}{\partial x} & \frac{\partial w_\textbf{n}}{\partial y} & \frac{\partial w_\textbf{n}}{\partial z} \end{array} \right)
\label{eq:BasisRegJacobian}
\end{equation}

Now, let $\textbf{s}_0 \in \R^3$ be an arbitrary source point and $\textbf{s}_n$ this point transformed after applying $n$ elementary registration steps. Then $\{ \textbf{s}_n, n \in [ 0, N ] \}$ is the set of all successive positions of $\textbf{s}_0$ and, according to Eq. \ref{EqTotalRegistration}, $\emph{R}(\textbf{s}_0) = \textbf{s}_N$. The Jacobian matrix of \emph{R} computed at point $\textbf{s}_0$ is given by the following chain rule: 

\begin{equation}
J_\emph{R}(\textbf{s}_0) :=
\frac{\partial \emph{R}}{\partial \textbf{s}}(\textbf{s}_0) = 
\frac{\partial r_N}{\partial \textbf{s}}(\textbf{s}_{N-1}) \;
... \;
\frac{\partial r_2}{\partial \textbf{s}}(\textbf{s}_1) \;
\frac{\partial r_1}{\partial \textbf{s}}(\textbf{s}_0)
\label{eq:RegJacChainRule}
\end{equation}

All weight functions $w_\textbf{n}$ stem from a ``template'' weight function ``$w$'' defined on the $[0,1]^3$ grid cell and associated to node $(1,1,1)$. Let $\pi$ be a third degree polynom defined by $\pi(t) = t^2(3-2t)$, such as $\pi(0)=0$, $\pi(1)=1$, $\pi'(0)=0$ and $\pi'(1)=0$. The template weight function $w$ is defined on $[0,1]^3$ as:

\begin{equation}
w(\textbf{s}) = w(s_1,s_2,s_3) := \pi(s_1)\pi(s_2)\pi(s_3)
\label{eq:WeightFunction}
\end{equation}

\begin{figure}[tb]
\begin{center}
	\subfigure[]{\includegraphics[width=0.30\linewidth]{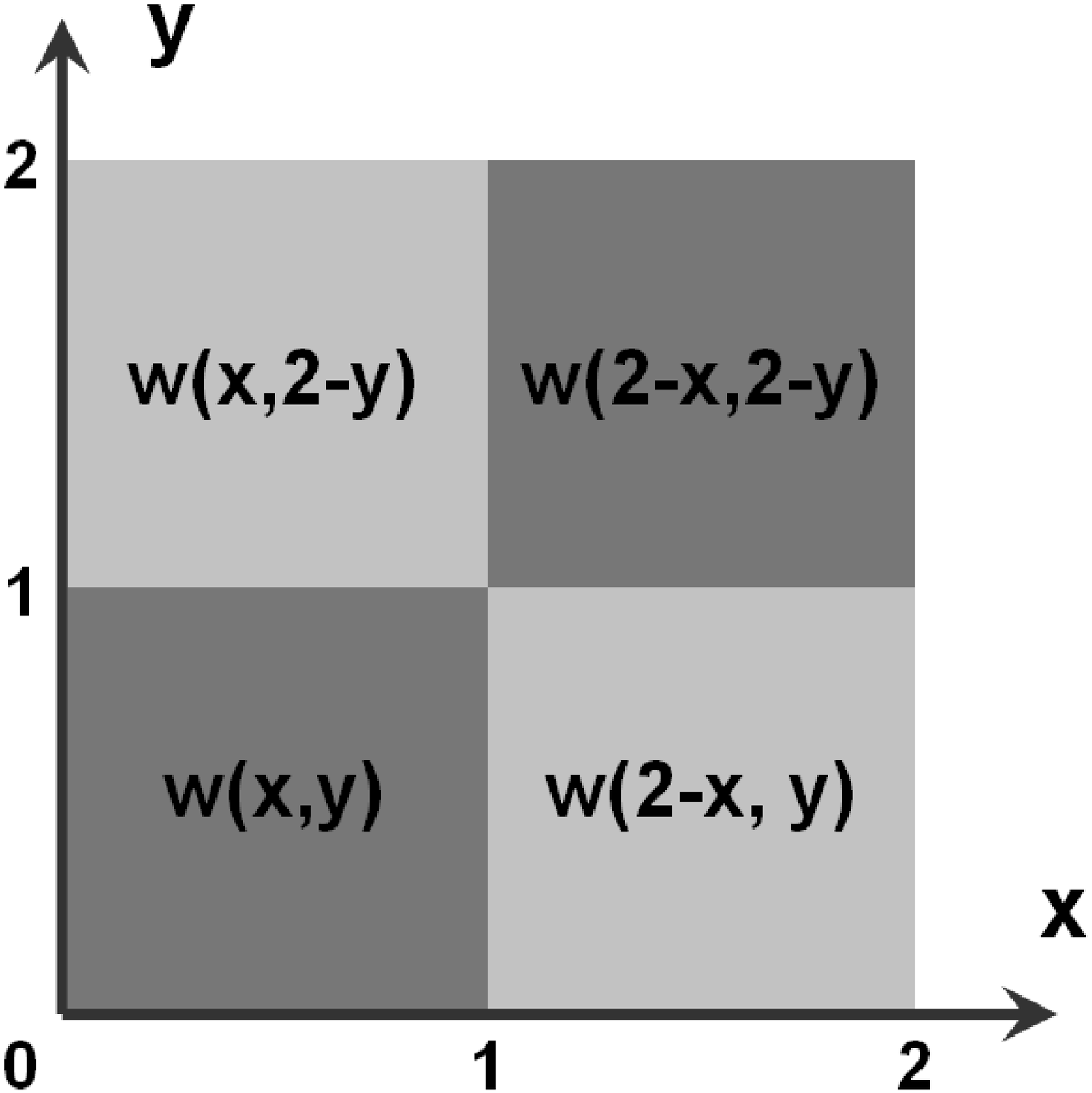}}
	\hspace{0.4cm}
  \subfigure[]{\includegraphics[width=0.30\linewidth]{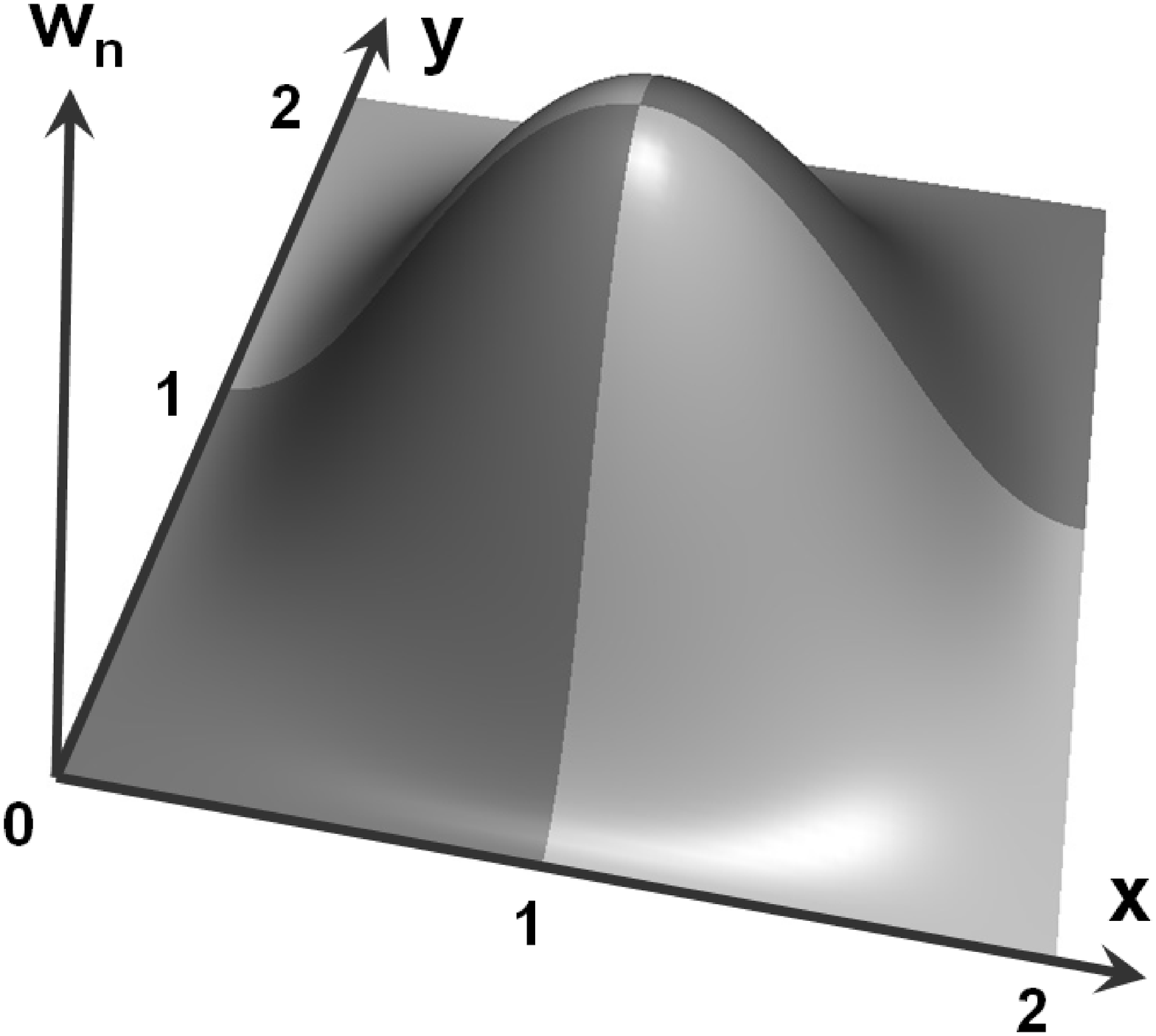}}
	\hspace{0.4cm}
  \subfigure[]{\includegraphics[width=0.30\linewidth]{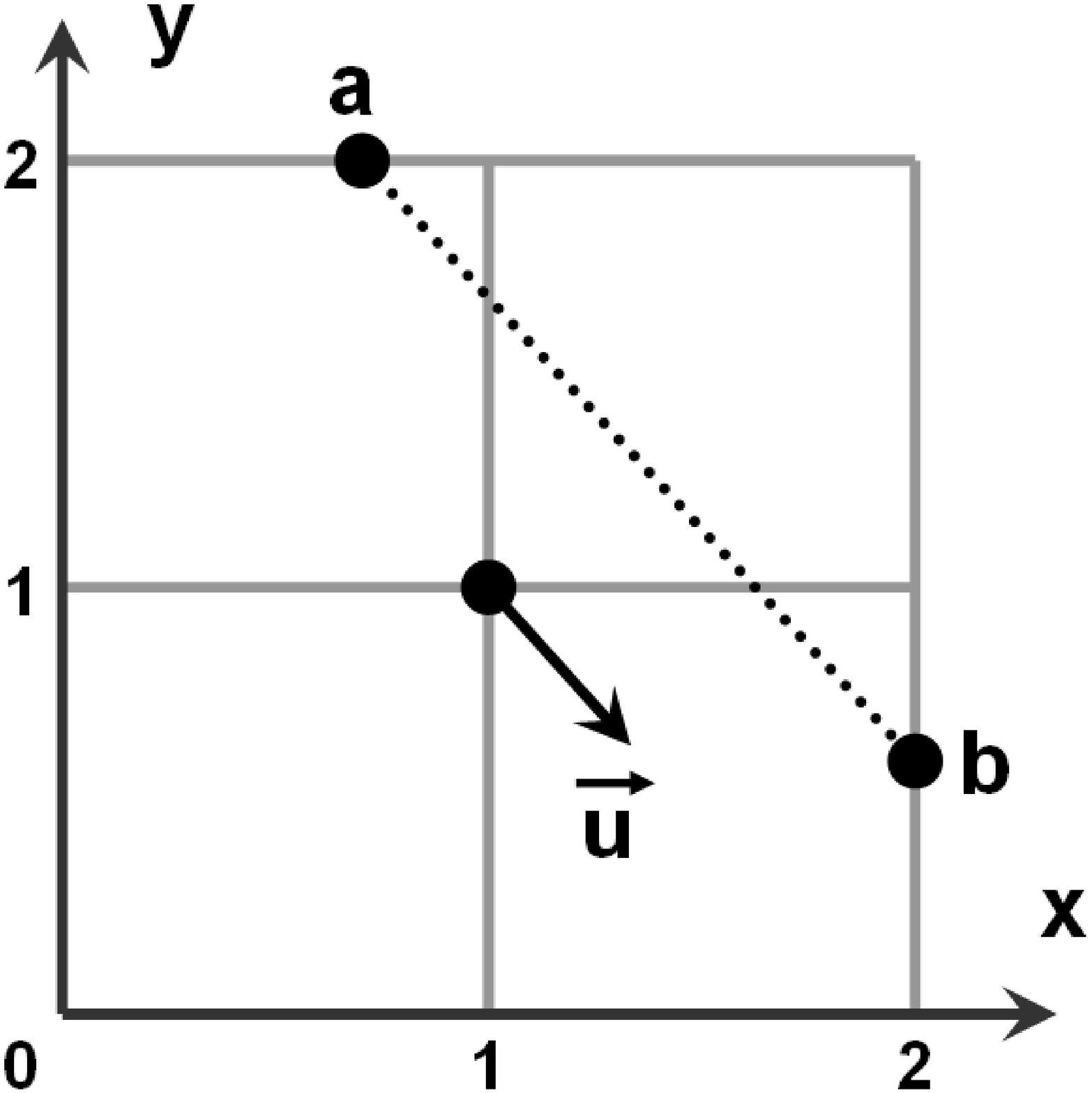}}
	\caption{(a) Variable changes required in 2D for the assembly of $w_\textbf{n}$ on a neighborhood of 4 cells centered on node $\textbf{n}=(1,1)$. (b) Value of $w_\textbf{n}$ plotted over the four 2D cells. (c) An elementary deformation leaves all segments $[\textbf{a},\textbf{b}]$ parallel to the applied nodal displacement unchanged.}
	\label{fig:Sigmoides1D2D}
\end{center}
\end{figure}

A specific weight function $w_\textbf{n}$ is defined on the union of neighboring cells around the displaced node $\textbf{n}$ by variable change and scaling in order to adapt the canonic $[0,1]^3$ domain to the cell size within the actual grid. Fig. \ref{fig:Sigmoides1D2D}-a illustrates on a 2D grid the variable changes needed to construct from $w$ the weight function $w_\textbf{n}$, with $\textbf{n}=(1,1)$, defined on a 2$\times$2 cells neighborhood of $\textbf{n}$. Fig. \ref{fig:Sigmoides1D2D}-b shows the 3D plot of the resulting weight function $w_\textbf{n}: \R^2\rightarrow[0,1]$.

The two other regularity properties of the elementary registration functions $r$, namely \textbf{bijection} and \textbf{non-folding}, are enforced by limiting the amplitude of nodal displacements, as described in the following sections.

\subsection{Control over space distortion}
\label{SecControlOverSpaceDistortion}

Space \textbf{folding} can be mathematically expressed as follows. Consider a positively oriented set of three infinitesimal vectors, $d\textbf{X}_1$, $d\textbf{X}_2$ and $d\textbf{X}_3$, placed in undeformed space at point $\textbf{s}$ and defining the volume $dV$. The signed value of $dV$ can be computed as the determinant of the matrix formed by the three vectors.

\begin{displaymath}
dV = (d\textbf{X}_1 \times d\textbf{X}_2) \cdot d\textbf{X}_3 =
\left \vert
\begin{array}{ccc}
d\textbf{X}_1 & d\textbf{X}_2 & d\textbf{X}_3 \\
\end{array}
\right \vert
\end{displaymath}

$\emph{R}$ is said to be \textbf{locally non-folding} if the deformed infinitesimal volume $dv=\emph{R}(dV)$ defined by the three deformed vectors $d\textbf{x}_1$, $d\textbf{x}_2$ and $d\textbf{x}_3$, remains positive, or, in other words, if \emph{R} does not change the orientation of space in the neighborhood of $\textbf{s}$. 

The differential form of \emph{R} gives us the expression of the deformed vectors:

\begin{displaymath}
\forall i=1,2,3, \; d\textbf{x}_i = \frac{\partial \emph{R}}{\partial \textbf{s}}(\textbf{s}) \; d\textbf{X}_i
\end{displaymath}

and the relation between $dv$ and $dV$ is:

\begin{displaymath}
dv =
\left \vert
\begin{array}{ccc}
d\textbf{x}_1 & d\textbf{x}_2 & d\textbf{x}_3 \\
\end{array}
\right \vert =
\left \vert
\frac{\partial \emph{R}}{\partial \textbf{s}}(\textbf{s})
\right \vert
\left \vert
\begin{array}{ccc}
d\textbf{X}_1 & d\textbf{X}_2 & d\textbf{X}_3 \\
\end{array}
\right \vert =
\vert J_\emph{R}(\textbf{s}) \vert \; dV
\end{displaymath}

where $J_\emph{R}(\textbf{s}) := (\partial \emph{R} /\partial \textbf{s})(\textbf{s})$ is the Jacobian of \emph{R} given by the chain rule in Eq. \ref{eq:RegJacChainRule}. It follows that an application \emph{R} is \textbf{non-folding} if its Jacobian is strictly positive, i.e.:

\begin{equation}
\forall \textbf{s} \in \R^3, \; \vert J_\emph{R}(\textbf{s}) \vert > 0
\label{eq:NonFolding}
\end{equation}

If the value of the Jacobian at a given point is greater than 1, the deformation locally \textbf{stretches} space and if its value is smaller than 1, space is locally \textbf{compressed}.

From Eq. \ref{eq:RegJacChainRule} it follows that \emph{R} is non-folding if and only if all $r_i, i=1,\ldots,N$ are non-folding, since:

\begin{displaymath}
\left \vert J_\emph{R} \right \vert = 
\left \vert \frac{\partial \emph{R}}{\partial \textbf{s}} \right \vert =
\prod_{i=1}^N \left \vert \frac{\partial r_i}{\partial \textbf{s}} \right \vert =
\prod_{i=1}^N \left \vert J_{r_i} \right \vert
\end{displaymath}

Let's now see how to ensure that each individual elementary registration function is non-folding. Using Eq. \ref{eq:BasisRegJacobian}, the non-folding condition on $r$ becomes: 

\begin{equation}
\forall \textbf{s} \in \R^3, \; \vert J_r(\textbf{s}) \vert = 1 + \overrightarrow{\triangledown}w_\textbf{n}(\textbf{s}) \cdot \overrightarrow{u} > 0
\label{eq:NonFolding2}
\end{equation}

Given the above expression of the template weight function $w$, the magnitude of the gradient of a specific weight function $w_\textbf{n}$ defined on a $[0,L]^3$ grid cell has an upper bound of $3\sqrt{3}/(2L)$. To ensure non-folding it is thus sufficient to limit the amplitude of nodal displacements: 

\begin{displaymath}
\| \overrightarrow{u} \| \le 38\% \; L < \frac{2}{3\sqrt{3}} \; L
\end{displaymath}

In our implementation of the algorithm this value has been reduced to \textbf{10\% of the current grid cell size}, so as to not only achieve non-folding but also limit space distortion at each elementary registration step. As $\|\overrightarrow{u}\| \le L/10$ and $\|\overrightarrow{\triangledown}w_\textbf{n}\| \le 3\sqrt{3}/(2L)$, it follows that:

\begin{equation}
1-\frac{3\sqrt{3}}{20} \le 1+\overrightarrow{\triangledown}w_\textbf{n}(\textbf{s}) \cdot \overrightarrow{u} \le 1+\frac{3\sqrt{3}}{20}
\label{eq:DistortionBounds}
\end{equation}

The above equation gives us, for all cell sizes, approximate\footnote{$3\sqrt{3}/20\simeq0.259$} Jacobian lower and upper bounds: $0.74 < \vert J_r \vert < 1.26$, and hence $0.74^N < \vert J_\emph{R} \vert < 1.26^N$.

\subsection{Registration inversion}
\label{SecRegistrationInversion}

In this section we will show that under the non-folding constraint, the deformation $\emph{R}$ is a bijection and discuss how its inverse, $R^{-1}$, can be computed with a pre-defined level of accuracy for any point $\textbf{q} \in \R^3$.

The registration function $\emph{R}$ is one-to-one if the same is true of each elementary registration $r$. To prove that a non-folding elementary registration function is also a bijection, consider a 2D deformation $r$, as defined in Eq. \ref{eq:BasisRegDefinition}, created by applying the displacement $\overrightarrow{u}$ to the central node of the 4 cells group depicted in Fig. \ref{fig:Sigmoides1D2D}-c.

Given that all the displacements applied to the source points are collinear with $\overrightarrow{u}$, parallel segments such as $[\textbf{a},\textbf{b}]$ in Fig. \ref{fig:Sigmoides1D2D}-c, map onto themselves. Furthermore, points lying outside or on the boundary of the 4-cell group are left unchanged. To prove that $r$ is a bijection it is thus sufficient to show that it defines a one-to-one application $[\textbf{a},\textbf{b}] \rightarrow [\textbf{a},\textbf{b}]$, for any segment $[\textbf{a},\textbf{b}]$ parallel to $\overrightarrow{u}$.

As the considered segment is parallel to $\overrightarrow{u}$, there exists a scalar $\beta$ such as $\textbf{b} = \textbf{a} + \beta \overrightarrow{u}$. Moreover, all segment points $\textbf{p} \in [\textbf{a},\textbf{b}]$ can be written as $\textbf{p} = \textbf{a} + \rho \overrightarrow{u}$, with $\rho \in [0,\beta]$, leading to $r(\textbf{p})$ being rewritten as follows:

\begin{displaymath}
r(\textbf{p}) =
\textbf{p} +  w_\textbf{n}(\textbf{p}) \overrightarrow{u} =
\textbf{a} + (\rho + w_\textbf{n}(\textbf{a} + \rho \overrightarrow{u})) \overrightarrow{u}
\end{displaymath} 

The deformation $r$ is thus a bijection $[\textbf{a},\textbf{b}] \rightarrow [\textbf{a},\textbf{b}]$ if and only if the mapping $f: \rho \mapsto \rho + w_\textbf{n}(\textbf{a} + \rho \overrightarrow{u})$ is also a bijection $[0,\beta] \rightarrow [0,\beta]$, which we shall prove is true under the non-folding hypothesis.

Deformation $r$ leaves the segment extremities unchanged, $r(\textbf{a})=\textbf{a}$ and $r(\textbf{b})=\textbf{b}$, which, using the above expressions, leads to $f(0)=0$ and $f(\beta)=\beta$, respectively. Furthermore, the derivative of $f$ is $f' = 1 + \overrightarrow{\triangledown}w_\textbf{n} \cdot \overrightarrow{u}$. As a consequence, if $r$ is non-folding then, according to Eq. \ref{eq:NonFolding2}, we have $f' > 0$ which implies that the strictly monotonic function $f$ is a bijection $[0,\beta] \rightarrow [0,\beta]$ and so is $r: [\textbf{a},\textbf{b}] \rightarrow [\textbf{a},\textbf{b}]$, which concludes our proof.

Another issue is computing the inverse registration $\emph{R}^{-1} = r_1^{-1} \circ \ldots \circ r_N^{-1}$ with a user-defined accuracy $\epsilon$, measured in undeformed space. The task consists in finding, for a given deformed point $\textbf{q} \in \R^3$, an undeformed point $\textbf{p} \in \R^3$, such as $\| \textbf{p} - \emph{R}^{-1}(\textbf{q}) \| < \epsilon$. We will first show how to accurately compute the inverse of an elementary registration function $r$, and then use these results to accurately inverse the compound registration $\emph{R}$. 

Given $\textbf{q} \in [\textbf{a},\textbf{b}]$, $\textbf{q} = \textbf{a} + \mu \overrightarrow{u}$, finding $\textbf{p} = \textbf{a} + \rho \overrightarrow{u}$ such as $\textbf{p} = r^{-1}(\textbf{q})$ reduces to solving $\rho=f^{-1}(\mu)$ on $[0,\beta]$. The mapping $f$ is a 9$^\text{th}$ degree polynomial (see Eq. \ref{eq:WeightFunction}) and as no analytical expression of the solution is available, it must be approximated iteratively using, for example, a Newton-Raphson procedure.

Let $\overline{\textbf{p}} = \textbf{a} + \overline{\rho} \overrightarrow{u} = r^{-1}(\textbf{q})$ be the searched point, and $\textbf{p} = \textbf{a} + \rho \overrightarrow{u} \approx r^{-1}(\textbf{q})$ its iteratively computed approximation. The approximation error in undeformed space can be rewritten as:

\begin{equation}
\| \textbf{p} - r^{-1}(\textbf{q}) \| = \| \textbf{p} - \overline{\textbf{p}} \| = \vert \rho - \overline{\rho} \vert \| \overrightarrow{u} \|
\label{eq:ApproxInUndefSpace}
\end{equation} 

In order to compute the inverse registration with the desired accuracy level $\epsilon$, the exact solution $\overline{\rho} = f^{-1}(\mu)$ must be approximated by $\rho$ so that:

\begin{displaymath}
\vert \rho - \overline{\rho} \vert < \frac{\epsilon}{\| \overrightarrow{u} \|}
\end{displaymath} 

As the value of $\overline{\rho}$ is unknown, the approximation error must be computed in deformed space. To this end the above expression is rewritten and, using the finite increments theorem, yields:

\begin{equation}
\vert \rho - \overline{\rho} \vert = \vert f^{-1}(f(\rho)) - f^{-1}(\mu) \vert < M \, \vert f(\rho) - \mu \vert
\label{eq:ApproxInDefSpace}
\end{equation} 

The above constant $M$ is computed using the relation $(f^{-1})' = (f')^{-1}$, which in conjunction with Eq. \ref{eq:DistortionBounds}, gives $1/1.26 < (f^{-1})' < 1/0.74 = M$. This leads us to the conclusion that in order to compute an $\epsilon\,$-accurate inverse of $\textbf{q}$ in undeformed space, Newton-Raphson iterations must be carried out until the approximate parameter $\rho$ satisfies:

\begin{displaymath}
\vert f(\rho) - \mu \vert < 0.74 \, \frac{\epsilon}{\| \overrightarrow{u} \|}
\end{displaymath} 

Now let's compute the inverse of a compound registration function $\emph{R}$. To control the accumulated error overhead at each elementary inversion step we will use the fact, demonstrated by combining Eq. \ref{eq:ApproxInUndefSpace} and \ref{eq:ApproxInDefSpace}, that the inverse of any elementary registration function $r$ is \textbf{$\boldsymbol M$-Lipschitz} continuous, i.e.:

\begin{displaymath}
\forall \textbf{q}, \textbf{p} \in \R^3, \| r^{-1}(\textbf{q}) - r^{-1}(\textbf{p}) \| < M \, \| \textbf{q} - \textbf{p} \|
\end{displaymath} 

For the sake of clarity, let $\emph{R} = r_3 \circ r_2 \circ r_1$ be the considered registration function and $\textbf{s}_3 \in \R^3$ a point in deformed space which inverse, $\emph{R}^{-1}(\textbf{s}_3)$, is sought with accuracy $\epsilon$. Adopting the notation convention used to derive Eq. \ref{eq:RegJacChainRule} above, we have:

\begin{displaymath}
\emph{R}^{-1}(\textbf{s}_3) =
(r_1^{-1} \circ r_2^{-1} \circ r_3^{-1})(\textbf{s}_3) =
(r_1^{-1} \circ r_2^{-1})(\textbf{s}_2) =
r_1^{-1}(\textbf{s}_1) =
\textbf{s}_0
\end{displaymath} 

\begin{figure}[tb]
\begin{center}
	\includegraphics[width=0.55\linewidth]{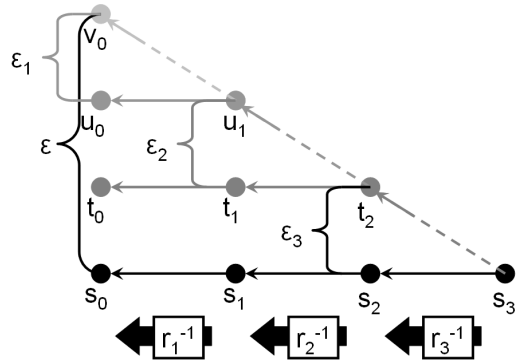}
	\caption{Approximation of $\textbf{s}_0=\emph{R}^{-1}(\textbf{s}_3)$ by $\textbf{v}_0$. Horizontal arrows represent exact inversions and oblique dashed arrows iterative inverse approximations. Left: undeformed space, right: deformed space.}
	\label{fig:CompoundInversion}
\end{center}
\end{figure}

Fig. \ref{fig:CompoundInversion} illustrates the 3-step computation of $\emph{R}^{-1}(\textbf{s}_3)$ along with error accumulation due to approximations performed at each elementary inversion step. Undeformed space is represented on the left, and deformed space on the right of the figure. Approximation steps are shown as dashed oblique arrows, and exact inversions as horizontal arrows.

We shall now establish the relation between $\epsilon$, the selected tolerance in undeformed space, and successive elementary approximation errors $\epsilon_3$, $\epsilon_2$ and $\epsilon_1$.

\textbf{Inversion of $r_3$.} $\textbf{s}_2 = r_3^{-1}(\textbf{s}_3)$ is approximated by $\textbf{t}_2$ with accuracy $\epsilon_3$ in $r_3$-undeformed space: $\epsilon_3 > \| \textbf{t}_2 - \textbf{s}_2 \|$. Using the Lipschitz constant $M$, this error propagates to the left as: $\epsilon_3 \, M > \| \textbf{t}_1 - \textbf{s}_1 \|$, and $\epsilon_3 \, M^2 > \| \textbf{t}_0 - \textbf{s}_0 \|$.

\textbf{Inversion of $r_2$.} $\textbf{t}_1 = r_2^{-1}(\textbf{t}_2)$ is approximated by $\textbf{u}_1$ with accuracy $\epsilon_2$ in $r_2$-undeformed space: $\epsilon_2 > \| \textbf{u}_1 - \textbf{t}_1 \|$. As in the previous step, this error propagates to the left as: $\epsilon_2 \, M > \| \textbf{u}_0 - \textbf{t}_0 \|$.

\textbf{Inversion of $r_1$.} Finally, $\textbf{u}_0 = r_1^{-1}(\textbf{u}_1)$ is approximated by $\textbf{v}_0$ with accuracy $\epsilon_1$ in initial undeformed space: $\epsilon_1 > \| \textbf{v}_0 - \textbf{u}_0 \|$.

The final error in undeformed space $\| \textbf{v}_0 - \textbf{s}_0 \|$ can be upper bound as:

\begin{displaymath}
\| \textbf{v}_0 - \textbf{s}_0 \| \leq
\| \textbf{v}_0 - \textbf{u}_0 \| + \| \textbf{u}_0 - \textbf{t}_0 \| + \| \textbf{t}_0 - \textbf{s}_0 \| <
\epsilon_1 + \epsilon_2 \, M + \epsilon_3 \, M^2
\end{displaymath}

The above expression can be generalized to any compound registration function $\emph{R} = r_N \circ \ldots \circ r_1$. In order to meet the desired accuracy standard $\epsilon$, the computational effort can be spread among all $N$ elementary inversions by setting the Newton-Raphson approximation threshold for each $r_n^{-1}, n=1,\ldots,N$ to:

\begin{displaymath}
\epsilon_n = \frac{\epsilon}{N \, M^{n-1}}
\end{displaymath}

This is a reasonable heuristic as $\epsilon_n$ is smaller for higher values of $n$, corresponding to finer grid levels and hence, smaller nodal displacements (10\% of grid cell size - see \S \ref{SecControlOverSpaceDistortion}). Solution search intervals are thus narrower, and higher inversion accuracy is thus easier to achieve than at coarser grid levels.

There is a final consideration on registration inversion: if the cubic polynomials $\pi$ used to define the shape function $w$ in Eq. \ref{eq:WeightFunction} are replaced with the identity $i: [0,1]\rightarrow[0,1], t \mapsto t$, then $f$ becomes a 3$^\text{rd}$ degree polynomial and the exact inverse of the registration function \emph{R} can be computed in a single iteration by using the well-known Tartaglia-Cardan formula. Nevertheless, this enhancement comes at a price as the smoothness of the registration function \emph{R} is degraded from $C^1$ to $C^0$.

\subsection{Asymmetric registration}
\label{SecAsymmetricregistration}

The ``asymmetric'' energy $E$ defined in Eq. \ref{EqRegistrationEnergy} handles situations where the features present in \emph{S} are present in \emph{D} but the reciprocal is not necessarily true. A situation can occur in which the Atlas mesh only represents a fraction of the organ features available within the patient data (see \S \ref{SecFaces}), requiring that the elastic registration \emph{R} is computed as transforming the Atlas to fit the patient data. The patient specific mesh is then obtained by application of this transform to the generic mesh i.e. \textbf{Patient mesh = \emph{R}(Atlas)}. 

Conversely, if only partial information about the patient's organ is available, its overall shape has to be inferred from the a priori knowledge carried by the generic Atlas mesh (see \S \ref{SecCompleteFemora} and \ref{SecPartialFemora}). This is done by taking the patient data as source points, and computing the registration function \emph{R} that fits the source points onto the Atlas. The patient specific mesh is obtained by inverting the resulting elastic deformation, i.e. \textbf{Patient mesh = \emph{R}$^{\textbf{-1}}$(Atlas)}.

\subsection{Mechanical regularization}
\label{SecMechanicalregularization}

In order to avoid excessive space distortions, a mechanical regularization term is monitored during the registration process, thereby upgrading the virtual elastic ``grid'' concept to virtual elastic ``solid''. As the elastic registration compensates for inter-individual morphological variations and does not model a physical deformation, the underlying ad-hoc mechanical properties are not related to the actual rheology of the organ under study.

Using the notation from Eq. \ref{EqRegistrationEnergy}, we now define the Jacobian matrix of the overall registration function considered at iteration $n$, and taken at material point $X$, as:

\begin{displaymath}
J_n(X) := \frac{\partial \emph{R}_n}{\partial X}(X) = \frac{\partial (r_n \circ \ldots \circ r_1)}{\partial X}(X)
\end{displaymath}

Equation \ref{eq:RegJacChainRule} shows that this matrix can be updated after each elementary registration step by first multiplying the previous matrix with the Jacobian matrix of the new elementary deformation $r_{n+1}$ taken at the current position of the considered material point, $(r_n \circ \ldots \circ r_1)(X)$.

During the registration assembly, the elastic energy stored in the virtual solid is measured at a set of material points $\{ X^i \}_i$, evenly distributed among the initial source data, and set to ``probe'' the space distortions induced by the accumulated elementary deformations. To do so, at each iteration $n$ the Green-Lagrange strain tensor $\epsilon_n$ is derived from the above mentioned Jacobian matrix, as $\epsilon_n = (J_n^TJ_n - I)/2$, and the stress tensor $\sigma_n$ is related to the strain tensor $\epsilon_n$ by a linear constitutive equation $\sigma_{n\,ij} = D_{ijkl} \, \epsilon_{n\,kl}$.

The potential elastic energy generated by the deformation $\emph{R}_n$ at control point $X^i$ can be computed using the Total Lagrangian formulation, as:

\begin{displaymath}
W^i_n = \int_{V^i} \epsilon_n : \sigma_n \, dX
\end{displaymath}

where $V^i$ is the volume in the initial source configuration associated to, or ``monitored'' by the material point $X^i$. In order to preserve fast registration computation, the above integral is approximated as $W^i_n \approx \vert V^i \vert \, \epsilon_n (X^i) : \sigma_n (X^i)$, where $\vert V^i \vert$ is the measure in the initial configuration of the volume $V^i$ associated to the material control point $X^i$.

The sum of the contributions of all control points $\{ X^i \}_i$ gives an approximation of the total potential elastic energy stored in the deformed source space at iteration $n$, as $W_n = \sum_i W^i_n$. This measure is taken into account at each deformation step to select, among all possible elementary deformations, the registration function $r_n$ which offers the best ratio between registration energy decrease and elastic energy increase.

To this end, at each iteration $n$ and for each candidate deformations $r_n$ the associated registration energy decrease $\Delta E_n > 0$ is computed as $\Delta E_n = E(\emph{R}_{n-1})-E(r_n \circ \emph{R}_{n-1})$. The change in potential elastic energy $\Delta W_n$ associated to each $r_n$ is also computed as $\Delta W_n = W_n - W_{n-1}$ and the following selection algorithm is applied.

\begin{enumerate}
	\item Among all candidate elementary deformations, only the deformations $r_n$ leading to a relative registration gain greater than the stop threshold $T_E$ are considered, i.e. the ones for which $\Delta E_n / E_{n-1} > T_E$.
	\item Among those, if deformations such as $\Delta W_n < 0$ can be found, the one yielding the highest registration energy decrease is chosen.
	\item Otherwise, the deformation $r_n$ having the highest $\Delta E_n / \Delta W_n$ ratio is applied.
\end{enumerate}

Rule 1 merely implements the stop criterion mentioned above. If no elementary registration function satisfying this condition can be found, the iterations stop. Rule 2 favors space ``decompression'' if it goes with satisfactory registration enhancement. Finally Rule 3 is applied in the most general case to select the deformation offering the best trade-off between space distortion and registration gain.

It is important to emphasize here that although the elastic grid initial state is restored prior to each elementary registration step (Eulerian approach), the mechanical energy terms $W_n$ computed above keep track of all the deformations accumulated in $\emph{R}_n$ at step $n$ (Lagrangian formulation). They are therefore appropriate for measuring space distortion as the registration advances.

As a relationship between inter-individual shape variations and mechanical behavior could not be determined, the simple St. Venant-Kirchoff mechanical framework was selected. An isotropic soft compressible material using an empirically set Poisson's ratio $\nu=0.2$ and a Young's modulus $E=1\text{Pa}$ was implemented. The value of $E$ has no effect on the registration sequence as it is not the sum but the ratio between registration and elastic energy that conditions each elementary deformation.

Finally, in our implementation, 1000 control points are used. The $\{ X^i \}_{i=1,\ldots,1000}$ are distributed in the bounding box of the source data, on a regular $10 \times 10 \times 10$ grid and all $\vert V^i \vert$ are set to 1/1000 of the bounding box volume.

\subsection{Multiple structures registration}
\label{SecMultiplestructuresregistration}

Prior to mesh registration, a subset of the Atlas mesh nodes needs to be labeled as anatomical features that can be identified within the patient data.

In the case of the femur models (see \S \ref{SecCompleteFemora} and \ref{SecPartialFemora}) the surface mesh nodes are labeled as ``bone surface'' as they lie on the cortical surface of the bone, which can easily be extracted from a Computed Tomography (CT) volume.

As for the face model, two families of nodes are defined, labeled ``skin'' for the exterior surface nodes registered onto the patients' faces skin, and ``bone'' for the interior surface nodes that correspond to the skull features and need to be registered onto segmented patients' skulls (see \S \ref{SecFaces}). This application example illustrates the capability of our procedure to capture the shapes of multiple anatomical structures modeled within a unique generic FE mesh.

When multiple anatomical structures need to be recovered, the registration of the FE mesh onto the patient data is driven by the minimization of an energy term $E^\text{lab}$ which measures the fit between the labeled nodes and their corresponding anatomical structures:

\begin{displaymath}
E^\text{lab}(\emph{R} \,) = \sum_{l=1}^{L} \sum_{\textbf{s} \in \emph{S}_l} d(\emph{R}(\textbf{s}),\emph{D}_l \,)
\end{displaymath}

where $l=1, \ldots, L$ are the predefined labels and $\emph{S}_l$ and $\emph{D}_l$ are the corresponding source and destination regions respectively. When multiple labels are defined, a distance map is computed for each $\emph{D}_l$ subset of the destination data.

The elastic grid deformation is driven by the $E^\text{lab}$ energy computed over the sets of labeled source nodes and destination surfaces. The unlabeled source points, on the other hand, are embedded within the elastic grid and passively follow the deformation without contributing to the energy term.

\section{Mesh repair}
\label{SecMeshreparation}

Although the elastic registration algorithm described above strongly limits space distortions, the registered mesh may exhibit irregular or low quality elements that need to be untangled before proceeding to FE analysis. Indeed, the non-folding nature of the registration function is a local property ensuring that space orientation is preserved. While this is locally true at every point in space, the property does not hold when considering finite structures such as element edges and, as a consequence, irregular elements may appear after the application of the smooth and non-folding elastic deformation to the initially regular Atlas mesh.

Element \textbf{regularity} is assessed by considering the mapping $F: \xi \mapsto x$, between the element parent (or reference) coordinates system $(\xi_1, \xi_2, \xi_3)$ and the actual element coordinates $(x_1, x_2, x_3)$. The Jacobian $J(\xi)$ is the determinant of the matrix $\partial F / \partial \xi$. Its value represents, at a given reference point $\xi$, the local volume transformation between the parent and actual element configuration.

An element is said to be regular if $J(\xi)>0$ for all $\xi$ in the parent configuration. Element regularity is usually assessed by considering the value of $J$ at specific points such as the integration points or the element nodes \cite{HughesFEM87}. We call a node $\textbf{n}$ ``irregular'' if it has a negative Jacobian $J_\textbf{n}^e$, computed in element $e$. A mesh is said ``regular'' if all the elements, and hence all nodes, are regular. An irregular mesh is not suitable for Finite Element analysis as the singularity of the mapping $F$ leads to modeling inconsistency.

Element \textbf{quality}, on the other hand, is a measure of the conformity of its shape, which reflects the evenness of the discretization of the modeled domain. There is a great variety of quality measures and their relevance is dependent on the considered element type and computations to be carried out \cite{Field00,Shewchuk2002a}.

Given the fact that fast and robust tetrahedral discretization solutions are already available, such as the widely used TetGen software, or have been proposed in the literature \cite{Molino2003,Frey2004,Alliez2005}, we focus our work on hexhedral-dominant meshes and demonstrate our mesh generation technique using a popular quality measure, well suited for hexahedrons and wedges: the Jacobian ratio (JR) \cite{Knupp2000b}. 

The JR is defined for a given node $\textbf{n}$ considered within element $e$. Its value is the ratio between the Jacobian at node $\textbf{n}$ in $e$, and the maximal nodal Jacobian value in element $e$, $J_{max}^e = \text{Max}_{\textbf{m} \in e} \{ J_\textbf{m}^e \}$, thus:

\begin{displaymath}
\text{JR}_\textbf{n}^e = \frac{J_\textbf{n}^e}{J_{max}^e}
\end{displaymath}

By comparing, the Jacobian value of each node with the maximal value within the considered element, the JR gives an indication of the contribution of a node to the overall element distortion, as opposed to local nodal distortion between the parent and actual configuration measured by the Jacobian alone. Element quality with respect to parent configuration is proportional to the JR value, ranging from 0 to 1.

Mesh repair is a two-fold process: first all the elements in the mesh are inspected and, if necessary, their nodes' positions are adapted so as to recover \textbf{regularity}; then a second relaxation procedure is carried out on the mesh if the achieved \textbf{quality} levels are unacceptable. Both steps can affect the nodes positions and alter the organ surface representation achieved after the elastic registration step.

In the absence of a formal proof that an acceptable mesh configuration exists and can be found by the relaxation procedure in all situations, the aim of this study is to evaluate the performance of MMRep on a database of diverse organ shapes and clinical use cases. The results described in \S \ref{SecCompleteFemora}, \ref{SecPartialFemora} and \ref{SecFaces} suggest that the smoothness of the elastic registration strongly limits spatial distortion making it possible to recover both mesh regularity and quality without interfering with the prior surface registration, by applying small displacements to a limited subset of the surface nodes. 

A large number of nodes in a mesh can make the repair procedure computationally prohibitive. This complexity can be greatly reduced using a local relaxation strategy i.e. grouping all irregular or poor quality nodes into ``regions'' defined in such a way that nodal corrections applied inside a region leave the outside mesh configuration unchanged. This local repair strategy makes it possible to perform all relaxation procedures independently on each repair region identified within the mesh. It also decreases significantly the computational complexity as the number of degrees of freedom to be considered in a region is usually small.

The untangling of an irregular region $R$ consists in finding a configuration where all the Jacobians in $R$, $\{ J_j \}_{j \in R}$, are positive. This nodal relaxation can be formulated as a maximization procedure driven by a ``regularity energy'' $E_R$. $E_R$ is expressed as the sum of all Jacobians within $R$ affected by a penalty function $\varphi_k$, of strength controlled by an index $k$, giving:

\begin{displaymath}
E_R = \sum_{j \in R} \varphi_k( J_j )
\end{displaymath}

where $\varphi_k(t) = 1-exp(-kt)$. As the parameter $k$ is increased during the optimization process, the influence of negative values overbalances the positive ones thus favoring a solution where all Jacobians in the sum are positive.

The aim of the quality maximization, on the other hand, is to find a configuration where all the JRs in a region $R$ are above a predefined level $\text{JR}_{min}$. The associated ``quality energy'' $E_Q$ is defined as the sum of the JRs within $R$ weighted by a penalty function $\psi_k$, thus:

\begin{displaymath}
E_Q = \sum_{j \in R,\,e \in R} \psi_k( \text{JR}_j^e )
\end{displaymath}

where $\psi_k(t) = 1-exp(k(\text{JR}_{min}-t))$. Similarly to the regularization process, the penalty parameter $k$ is used to find a solution where all JRs contributing to the sum $E_Q$ are above the quality threshold $\text{JR}_{min}$. The value of $\text{JR}_{min}$ was chosen in accordance with the quality standard requested by the commercial FE analysis software ANSYS Workbench (ANSYS Inc., USA) i.e. $\text{JR}_{min} = 1/30$ \cite{Kelly98}.

The initial value of $k$ for both penalty functions $\varphi_k$ and $\psi_k$ must be carefully chosen. Indeed, given the formulation of the penalty functions, an excessive penalization level may induce strong numerical instabilities in the optimization process. To avoid this issue, the starting value of $k$ is determined by considering the slope of the penalty function at the most penalized energy term (minimal Jacobian during regularization phase; minimal JR during quality optimization phase) and ensuring that it does not exceed a predefined threshold.

Both regularity and quality optimizations are carried out by gradient ascent. Gradients of both $E_R$ and $E_Q$ energies are computed using the centered differences scheme. Assuming unimodality of the local energy function, the maximum search in the direction of ascent is done using the golden section technique \cite{Press92} between the current nodal configuration and the configuration obtained after applying the maximal amplitude correction.
 
The amplitude and number of iterations for each mesh region are limited so as to restrict loss of surface representation accuracy. Our implementation allows a maximum of 50 iterations, and nodal displacements at each step have a maximal amplitude of 0.1mm. The 50 $\times$ 0.1 = 5mm maximal displacement can only be achieved by moving a unique node in a constant direction throughout the repair procedure, which seldom occurs. During the experimental validation discussed in \S \ref{SecExperimentalevaluation}, nodal corrections with mean amplitude less than 1.2mm and applied to less than 1\% of the nodes were sufficient to repair the 60 registered meshes.

The value of maximal nodal correction used here is not universal and must be determined for each field of application according to the dimensions of the modeled domain and maximal tolerance on the representation of its geometry. In our case, the results were consistent with the desired submillimetric mean surface representation accuracy, as shown in tables \ref{TabResultsCompleteFemora1}, \ref{TabResultsPartialFemora1} and \ref{TabResultsFaces}.

\section{Experimental evaluation}
\label{SecExperimentalevaluation}

The MMRep technique was evaluated in three different situations where:
\begin{itemize}
	\item \textbf{complete organ geometry} can be retrieved from the available data, as discussed in \S \ref{SecCompleteFemora};
	\item only \textbf{partial organ geometry} is accessible, as illustrated in \S \ref{SecPartialFemora};
	\item \textbf{distinct organ features} need to be taken into account by the generated model, as shown in \S \ref{SecFaces}.
\end{itemize}

The mesh adaptation technique alone is presented here and the discussion about the biomechanical simulations illustrating the three following ``use cases'' falls out of the scope of this article.

\subsection{Complete pre-operative femora CT scans}
\label{SecCompleteFemora}

In this section, we demonstrate our technique in the context of total knee arthroplasty (TKA). The prosthesis placement can be optimized to avoid unsealing or femur fracture using biomechanical modeling and FE analysis of the stresses within the tissues. These tissues are modeled based on the bone mechanical properties, geometry and prescribed loads inferred from the patient morphology and gait \cite{Zalzal08}.

\subsubsection{Mesh registration procedure}

\begin{figure}[tb]
\begin{center}
	\subfigure[]{\includegraphics[height=0.12\linewidth]{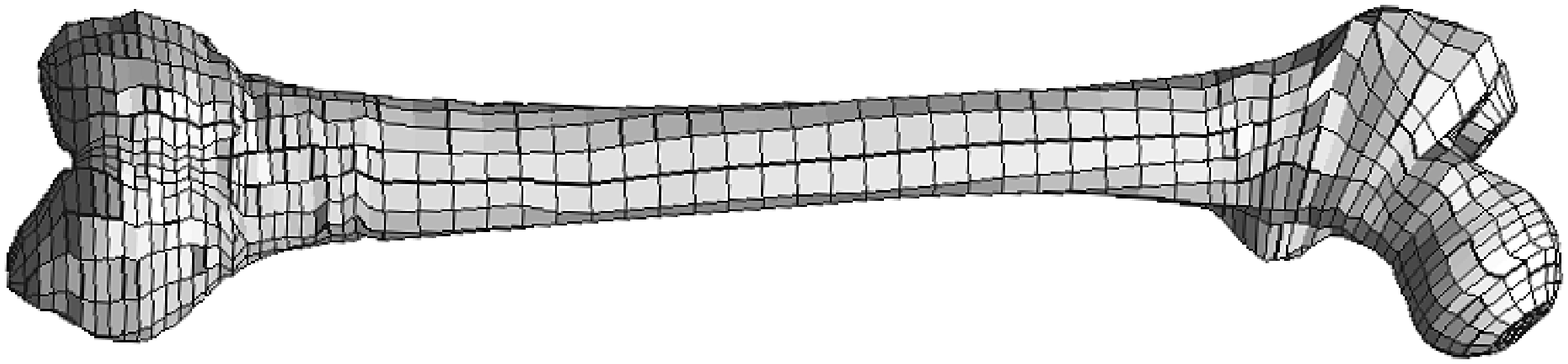}} \\
	\subfigure[]{\includegraphics[height=0.24\linewidth]{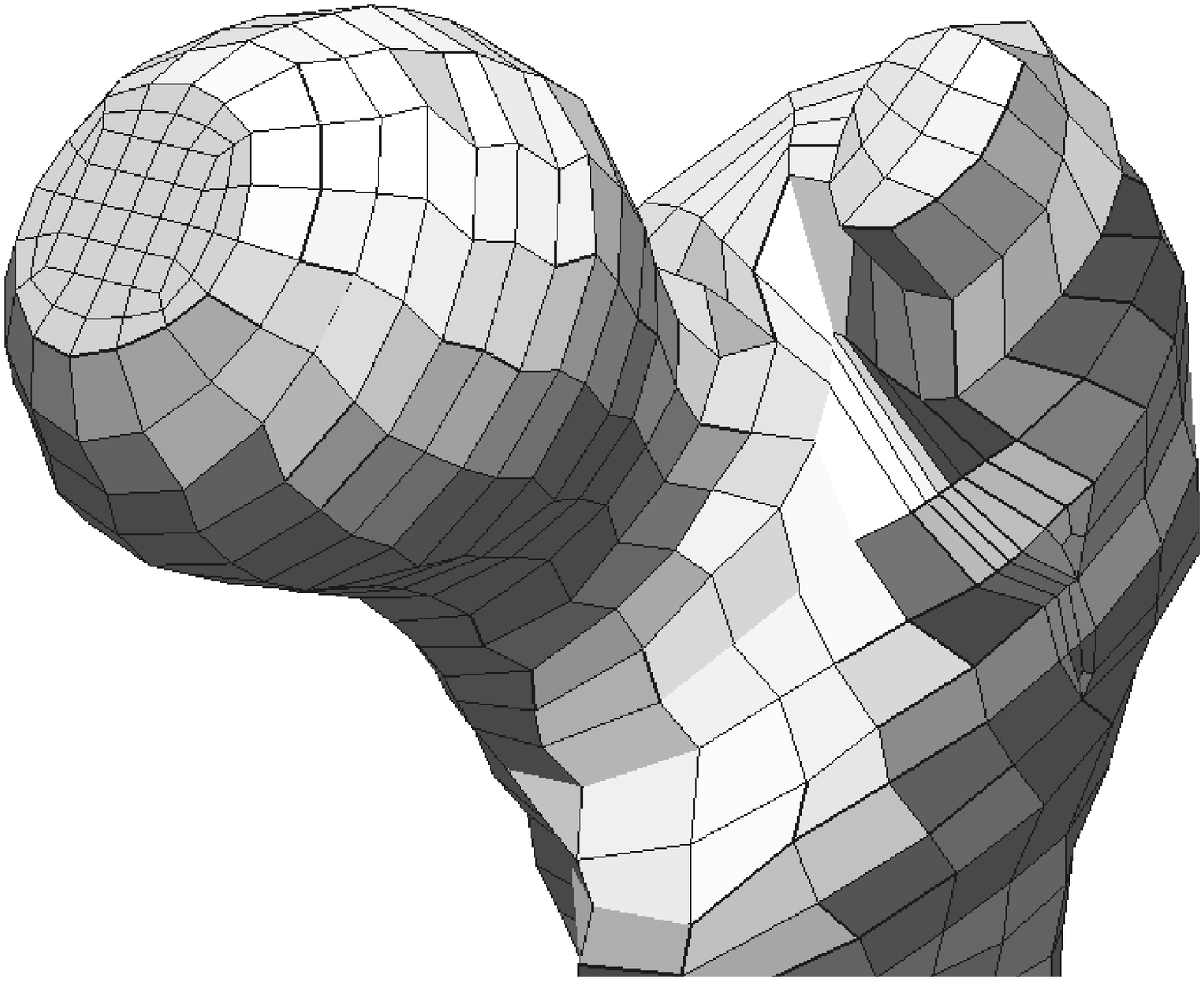}}
	\subfigure[]{\includegraphics[height=0.24\linewidth]{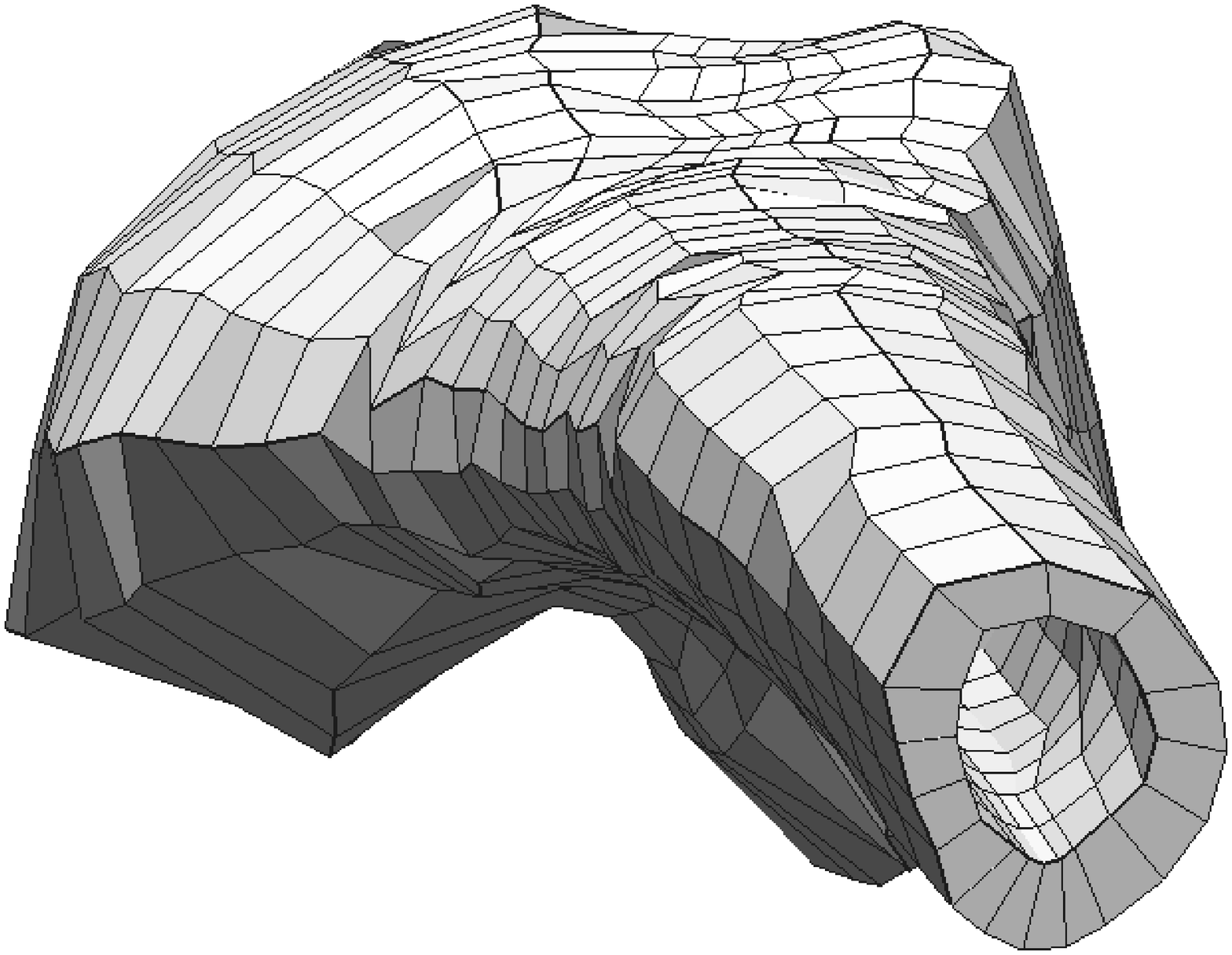}}
	\subfigure[]{\includegraphics[height=0.24\linewidth]{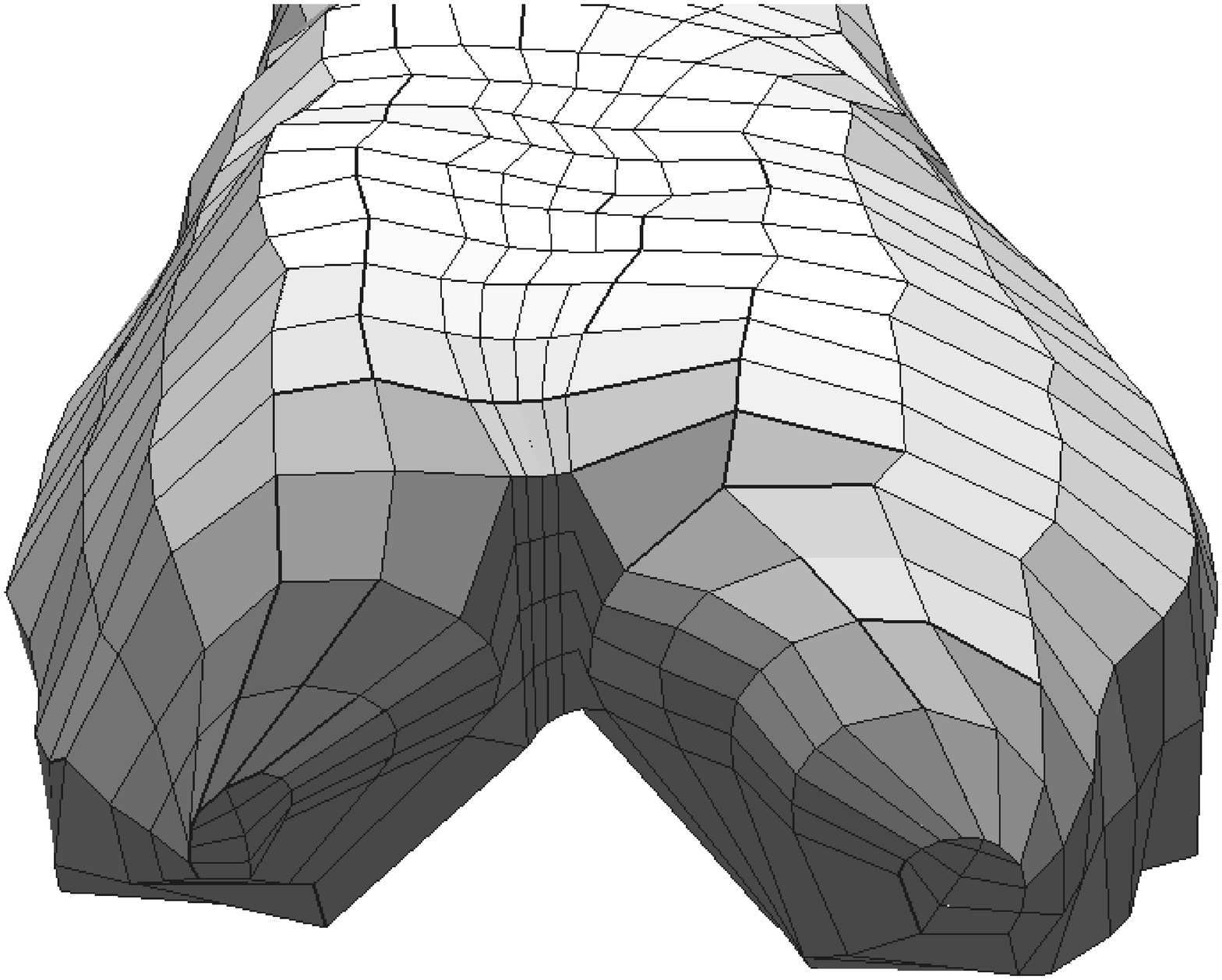}}
	\caption{Femur Atlas mesh from Couteau et al. \cite{Couteau98}. (a) Overview. (b) Femur head and great trochanter. (c) Cut out showing the diaphysis cortex layer of elements. (d) Distal condyles.}
	\label{FigFullFemurMesh}
\end{center}
\end{figure}

The manually assembled right femur Atlas mesh \cite{Couteau98} used in this part of the study is composed of 4052 nodes, forming 3018 elements : 2960 hexahedrons and 58 wedges (6 nodes prisms). The elements are organized so as to reflect the bony structure such as the femoral diaphysis cortex which is discretized by a single layer of elements. The principal mesh features are illustrated in Fig. \ref{FigFullFemurMesh}.

For the 5 considered patients a CT scan of the right leg was pre-operatively acquired and a semi-automatic threshold-based segmentation procedure was performed by a clinician in order to identify the femur cortical surface. Then the resulting points cloud was rigidly registered onto the Atlas bone surface using an Iterative Closest Point algorithm \cite{Besl92}.

Each elastic mesh registration was carried out as described in \S \ref{SecElasticregistration} by taking the source points set $\emph{S}$ as $\emph{S}_i$, the segmented cortical points for patient $i$, and the destination set $\emph{D}$ as the Atlas mesh cortex. Once the segmented points were registered onto the surface of the Atlas, the application of the inverse registration function $\emph{R}^{-1}$ to the generic mesh produced the patient specific FE model. The chosen accuracy level for the inverse computation, as described in \S \ref{SecRegistrationInversion}, was $\epsilon$=0.1mm. Mesh regularity and quality criteria were subsequently analyzed and the mesh repair procedures described in \S \ref{SecMeshreparation} were applied to the model.

\subsubsection{Results}

Table \ref{TabResultsCompleteFemora1} describes the performance of the mesh registration procedure for the 5 data sets. The surface representation error is computed as the distance between the segmented CT points and the generated patient specific FE mesh surface. The registration times include the direct registration $\emph{R}$ computation as well as the application of the inverse registration function $\emph{R}^{-1}$ to the Atlas mesh nodes.

\begin{table}[ht]
\begin{center}
\begin{tabular}{|p{3cm}||p{1cm}|p{1cm}|p{1cm}|p{1cm}|p{1cm}|}
\hline
\textbf{Patient} & \textbf{1} & \textbf{2} & \textbf{3} & \textbf{4} & \textbf{5} \\
\hline
\textbf{Registration (sec)} & 25 & 36 & 42 & 26 & 32 \\
\textbf{CT points} & 10930 & 25980 & 22924 & 20065 & 17886 \\
\hline
\textbf{Mean err. (mm)} & 0.3 & 0.4 & 0.3 & 0.4 & 0.3 \\
\textbf{Max err. (mm)} & 5.4 & 6.6 & 5.2 & 6.0 & 5.5 \\
$\boldsymbol \sigma$ \textbf{(mm)} & 0.5 & 0.5 & 0.4 & 0.5 & 0.4 \\
\hline
\end{tabular}
\caption{Registration (sec): elastic registration times, in seconds; CT points: quantity of segmented points in CT volumes; Mean, Max, $\sigma$ (mm): surface representation mean and maximal error, standard deviation, in millimeters.}
\label{TabResultsCompleteFemora1}
\end{center}
\end{table}

Table \ref{TabResultsCompleteFemora2} gives the performance of the mesh repair procedure carried out on the 5 deformed meshes. Repair computation times and numbers of corrected nodes are given along with nodal displacements statistics.

\begin{table}[ht]
\begin{center}
\begin{tabular}{|p{3cm}||p{1cm}|p{1cm}|p{1cm}|p{1cm}|p{1cm}|}
\hline
\textbf{Patient} & \textbf{1} & \textbf{2} & \textbf{3} & \textbf{4} & \textbf{5} \\
\hline
\textbf{Regularity (sec)} & 2.2 & 0.6 & 0.4 & 0.9 & 0.5 \\
\textbf{Quality (sec)}    & 3.6 & 2.6 & 6.8 & 0.9 & 0.7 \\
\hline
\textbf{\% nodes }& 1.0 & 0.3 & 0.6 & 0.5 & 0.4 \\
\textbf{nodes/4052} & 39 & 27 & 23 & 22 & 17 \\
\hline
\textbf{Mean disp. (mm)} & 1.0 & 0.3 & 0.2 & 0.5 & 0.2 \\
\textbf{Max disp. (mm)} & 3.4 & 2.5 & 1.1 & 1.9 & 0.6 \\
$\boldsymbol \sigma$ \textbf{(mm)} & 1.0 & 0.5 & 0.2 & 0.5 & 0.2 \\
\hline
\end{tabular}
\caption{Regularity, Quality (sec): mesh repair times for both phases, in seconds; \% nodes, nodes/4052: fraction and number of nodes moved by the repair procedure; Mean, Max, $\sigma$ (mm): mean and maximal nodal displacements, standard deviation, in millimeters.}
\label{TabResultsCompleteFemora2}
\end{center}
\end{table}

The MMRep procedure succeeded in generating a quality femur mesh for all 5 patients. All computations were carried out in less than one minute, which is acceptable even in an intraoperative context. The surface representation figures remained unchanged after the application of the mesh repair procedure (see Table \ref{TabResultsCompleteFemora1}) as the proportion of displaced nodes did not exceed 1\% and most of them were inner nodes which could be freely moved without affecting the mesh surface shape. The reported maximal errors are mainly due to manual segmentation irregularities and lack of local refinement of the Atlas mesh, making it difficult to capture some local shape variations. If necessary, this issue could be solved by further refining the generic mesh.

\begin{figure}[tb]
\begin{center}
	\subfigure[]{\includegraphics[height=0.40\linewidth]{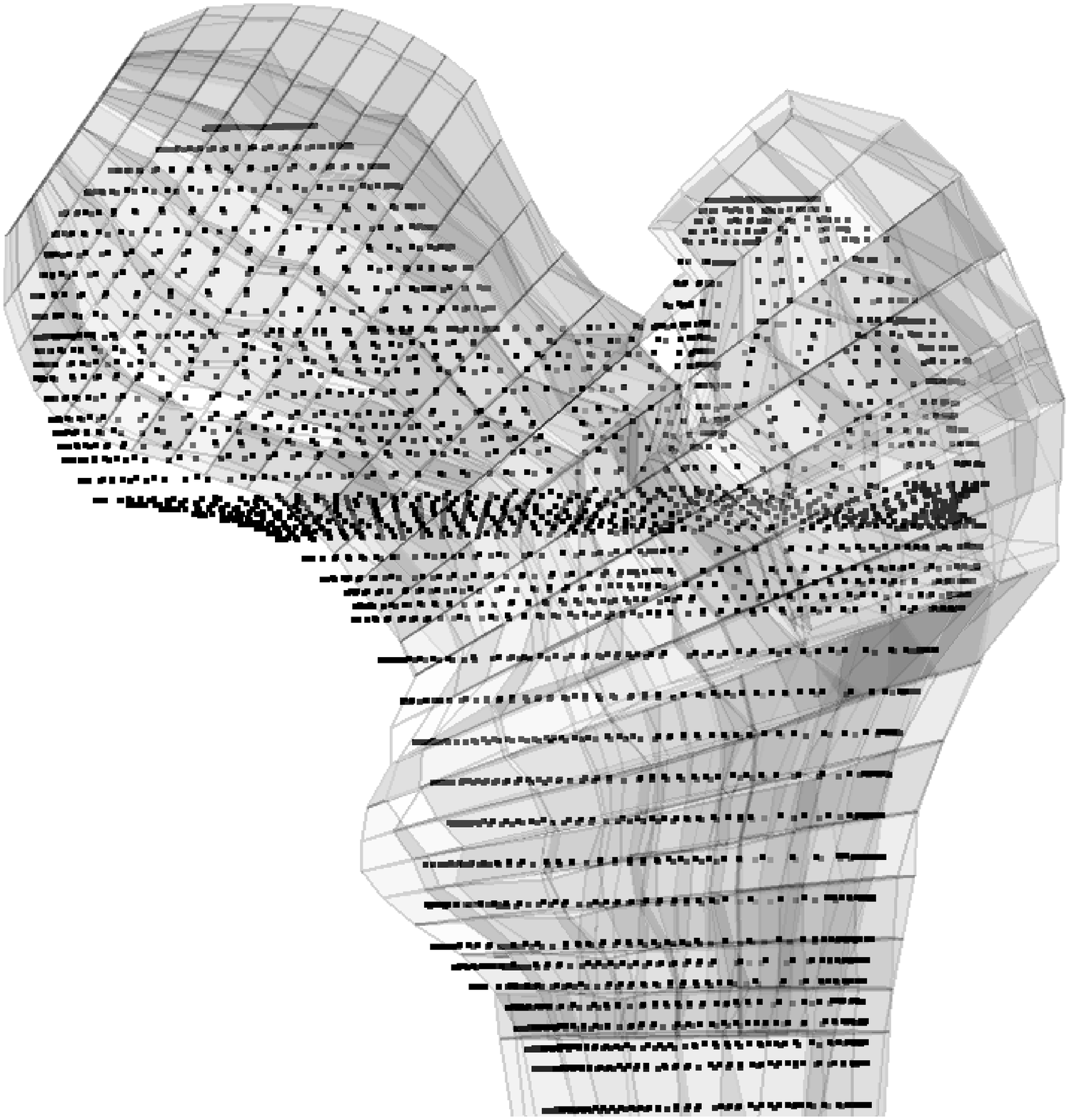}}
	\hspace{1cm}
	\subfigure[]{\includegraphics[height=0.40\linewidth]{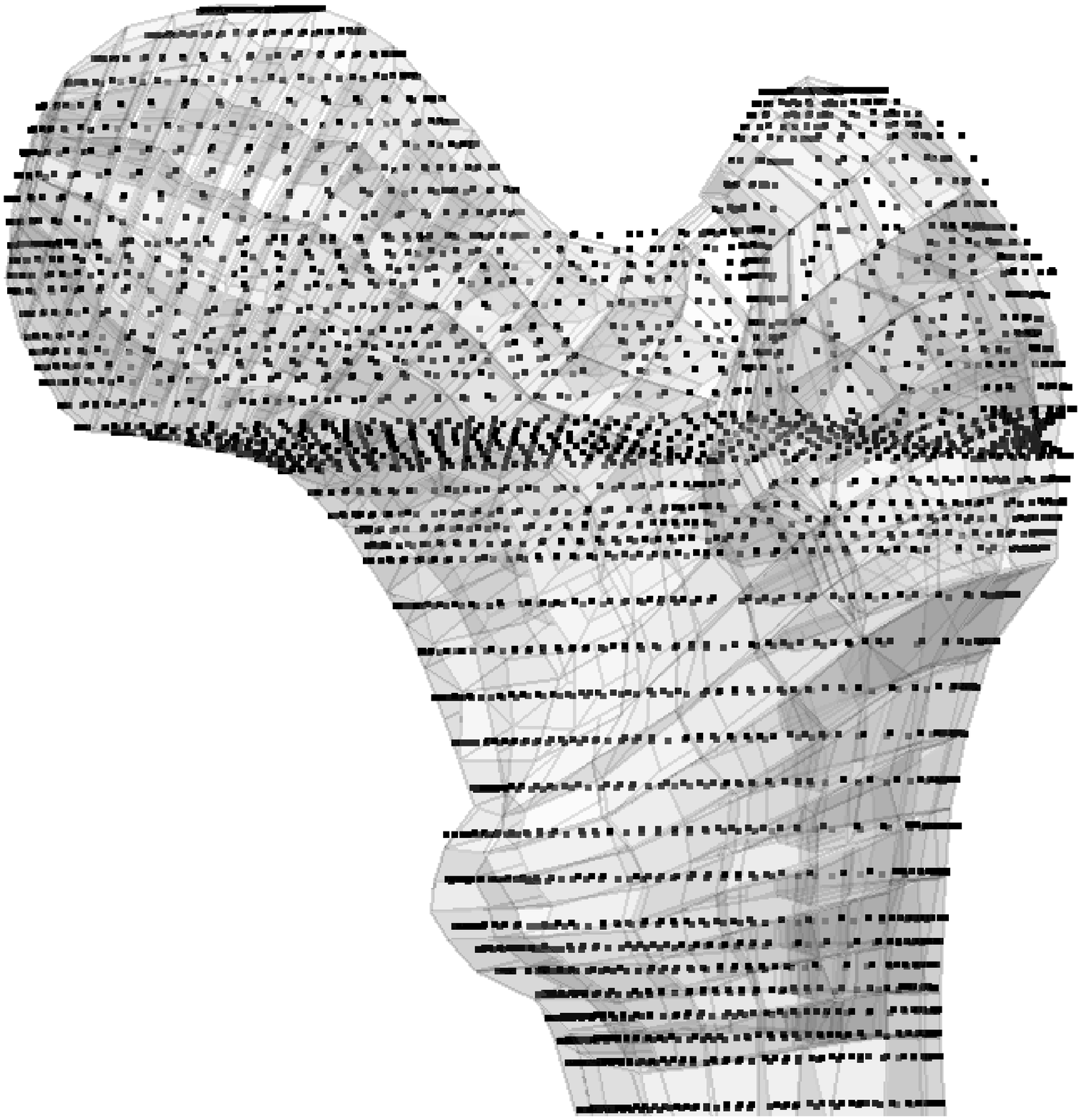}} \\
	\subfigure[]{\includegraphics[height=0.40\linewidth]{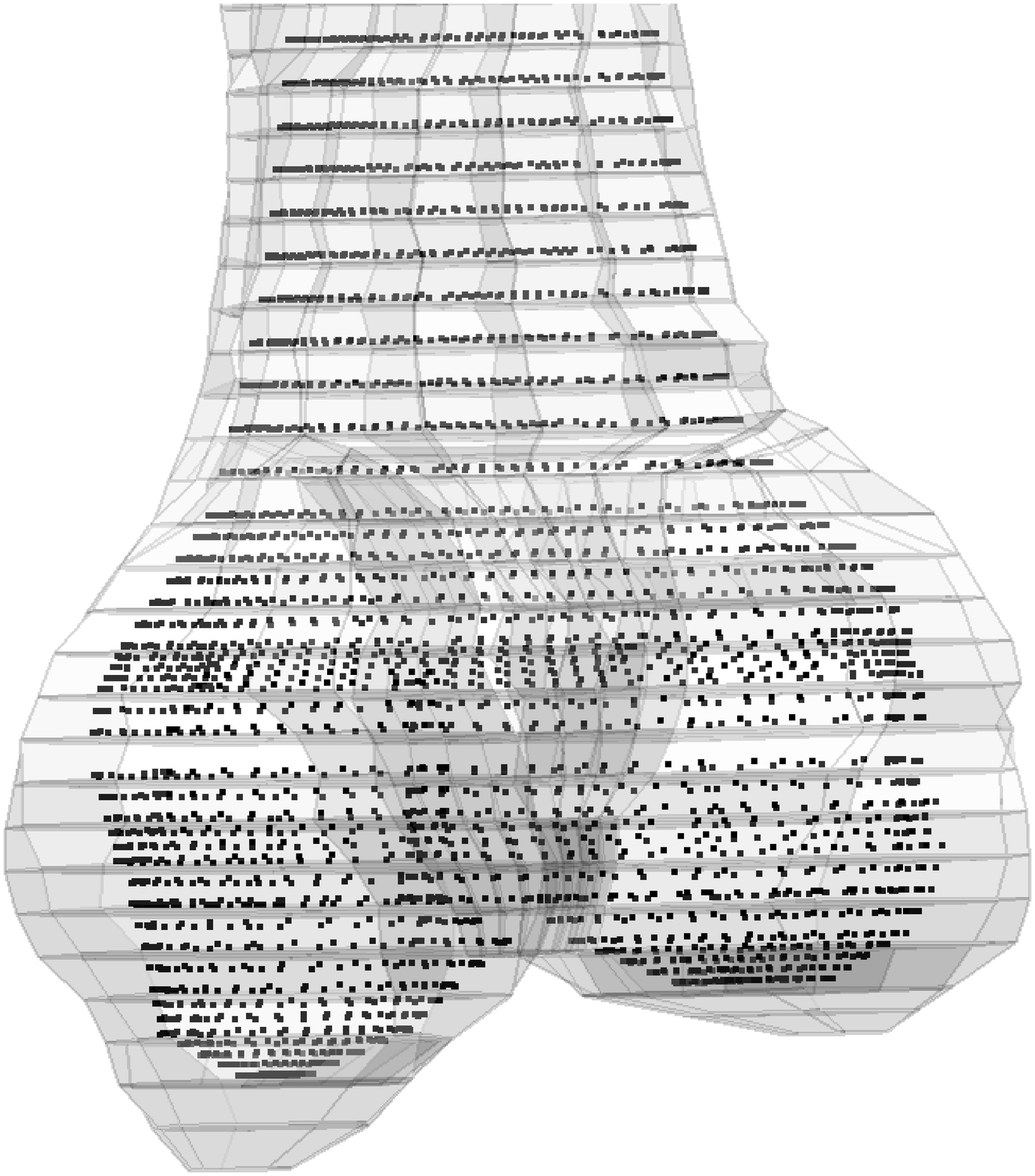}}
	\hspace{1cm}
	\subfigure[]{\includegraphics[height=0.40\linewidth]{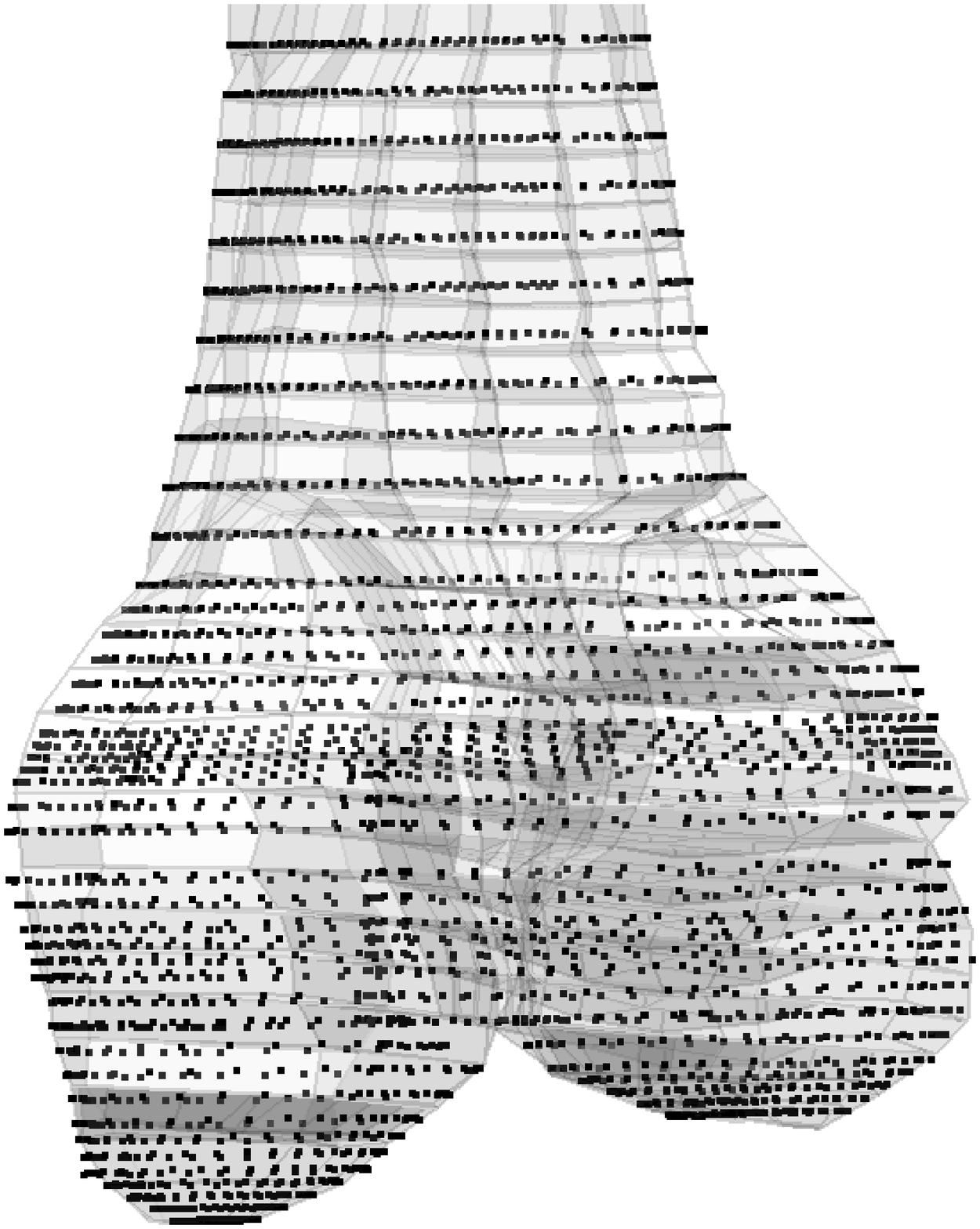}}
	\caption{Sample patient specific mesh. Proximal epiphysis: (a) segmented patient data (black dots) and Atlas mesh; (b) patient specific mesh fitting the femur head surface. Distal epiphysis: (c) segmented patient data (black dots) and Atlas mesh; (d) patient specific mesh fitting the condyles.}
	\label{FigFullFemurMatch}
\end{center}
\end{figure}

A sample result of the procedure is shown in Fig. \ref{FigFullFemurMatch} with focus on proximal and distal parts of the automatically generated femur mesh.

\begin{figure}[tb]
\begin{center}
	\subfigure[]{\includegraphics[height=0.30\linewidth]{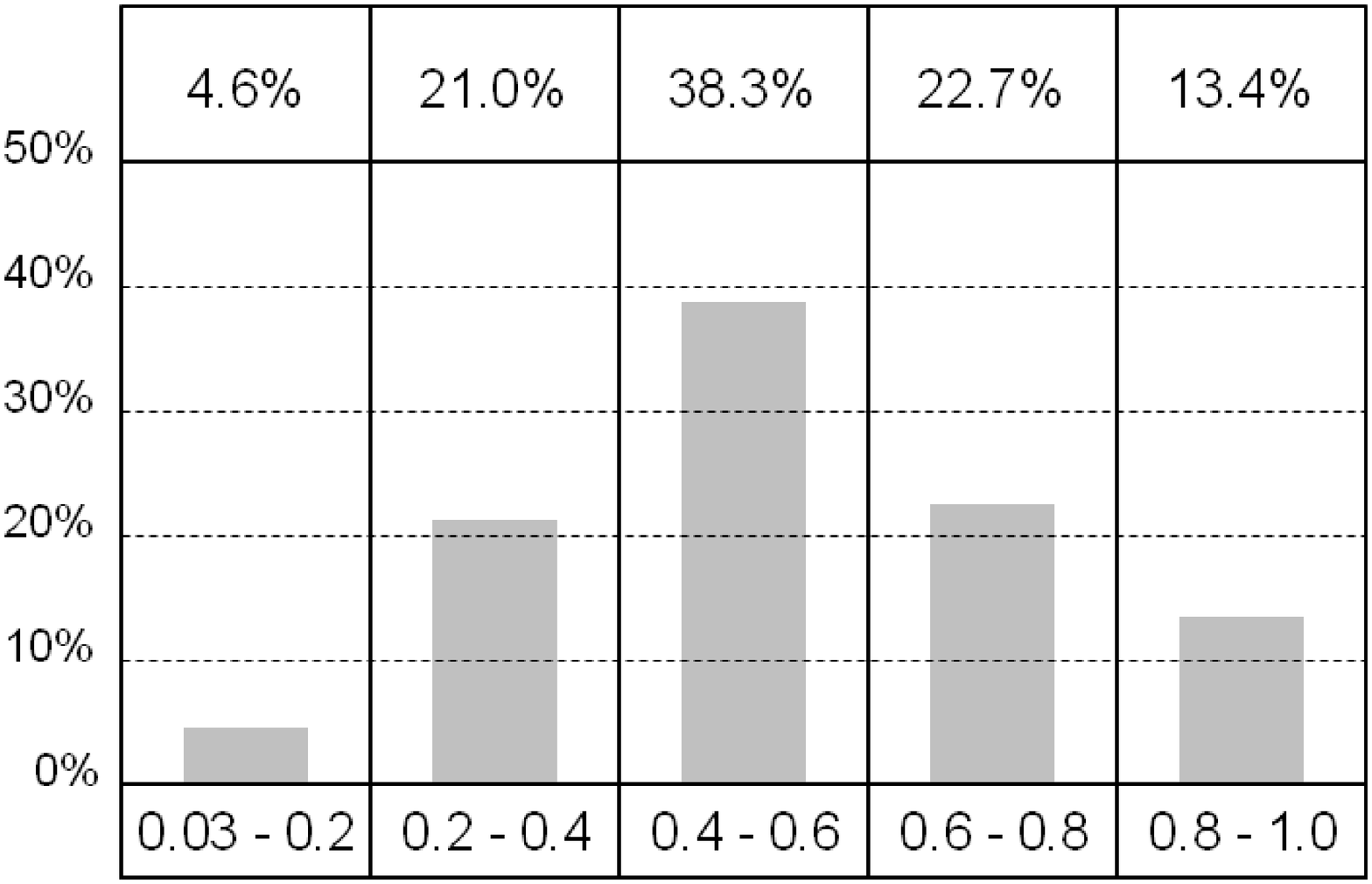}}
	\hspace{0.5cm}
	\subfigure[]{\includegraphics[height=0.30\linewidth]{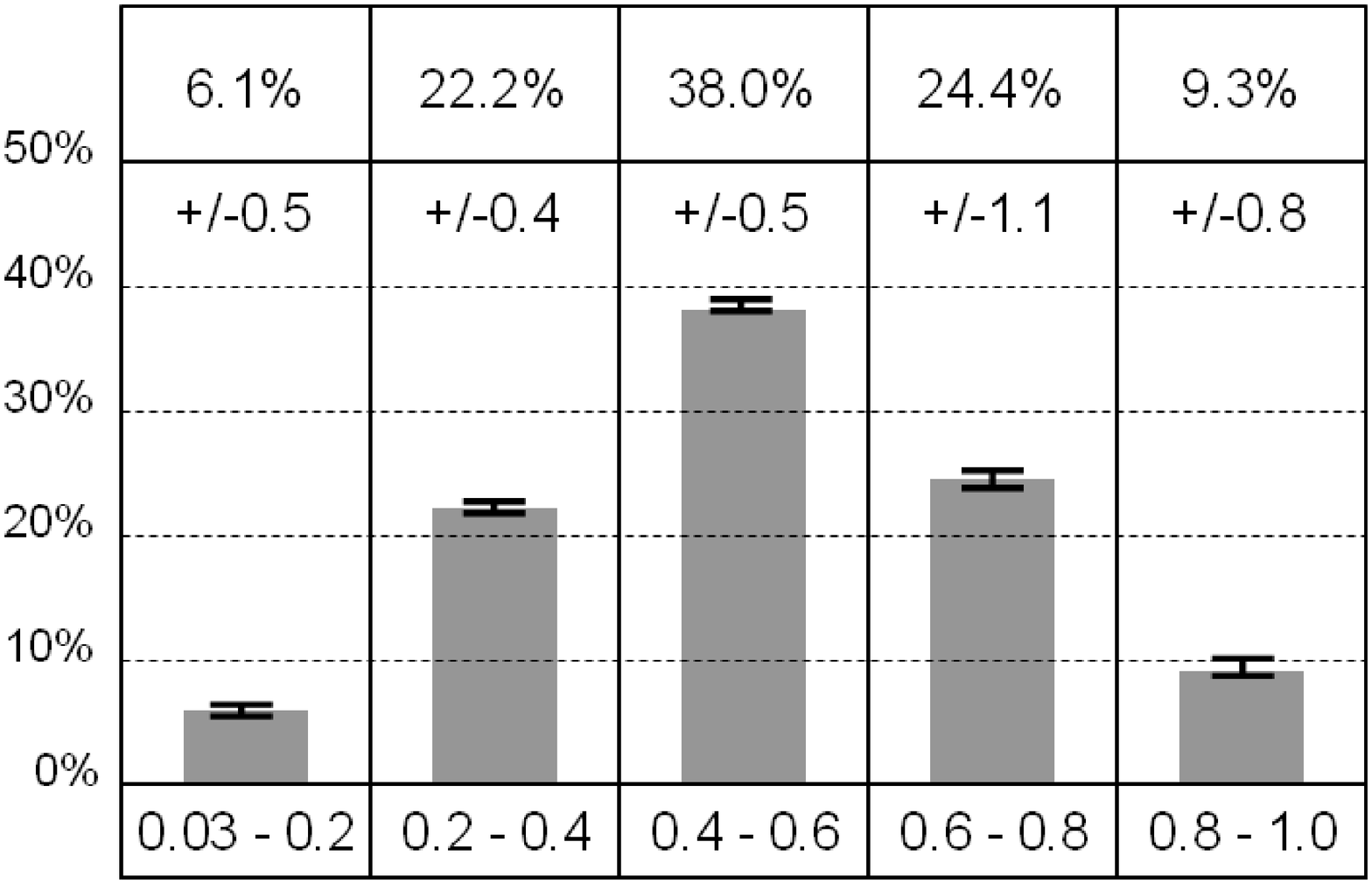}}
	\caption{Mesh quality statistics for Atlas (a) and patient-specific (b) full femur models.}
	\label{FigHistTKA}
\end{center}
\end{figure}

Mesh quality distribution measured on the 3018-element full femur Atlas mesh is shown in Fig. \ref{FigHistTKA}-a, and Fig. \ref{FigHistTKA}-b gives the mean quality distribution in the 5 generated patient-specific meshes, along with standard deviations.

The histograms presented in this article, in Figs. \ref{FigHistTKA}, \ref{FigHistTHA} and \ref{FigHistFACES}, classify the elements into 5 Jacobian ratio categories. The first interval $[0.03,0.2]$ lists the elements with a quality level that is acceptable from the point of view of our target application ANSYS, but is usually considered to be ``questionable''. Due to the manual assembly process, a small number of questionable elements is present in the Atlas meshes used in this study. As shown by the figures, this proportion only slightly increases after the application of the elastic deformation and mesh repair procedures.

\subsection{Partial intraoperative femora digitizations}
\label{SecPartialFemora}

In this part of the study, we demonstrate the possibility of intraoperative FE mesh generation during a total hip arthroplasty (THA) procedure. As in TKA, FE analysis can be used here to optimize femoral stem placement so as to minimize internal stresses and maximize the implant lifetime \cite{Lengsfeld05}.

\subsubsection{Mesh generation procedure}

Due to the surgical procedure limitations, the complete shapes of the patients' femoral heads were not available pre-operatively and each bone geometry was acquired intraoperatively by sliding a calibrated pointer on the cortical surface of the partially exposed hip. The pointer position was tracked in space by means of an optical localization system (Polaris, NDI, Canada) and the position of its tip was recorded continuously.

\begin{figure}[tb]
\begin{center}
	\subfigure{\includegraphics[height=0.21\linewidth]{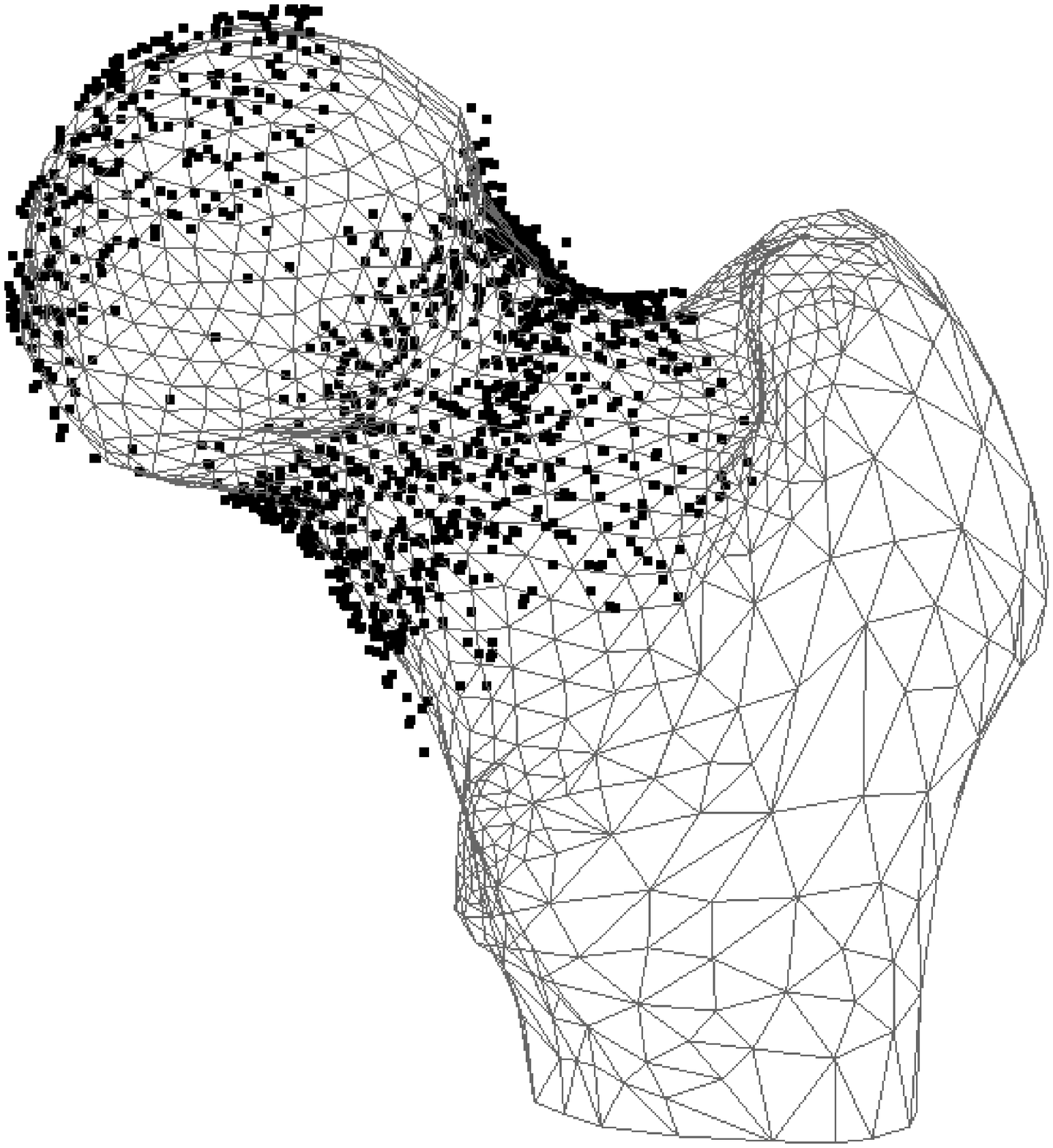}}
	\subfigure{\includegraphics[height=0.21\linewidth]{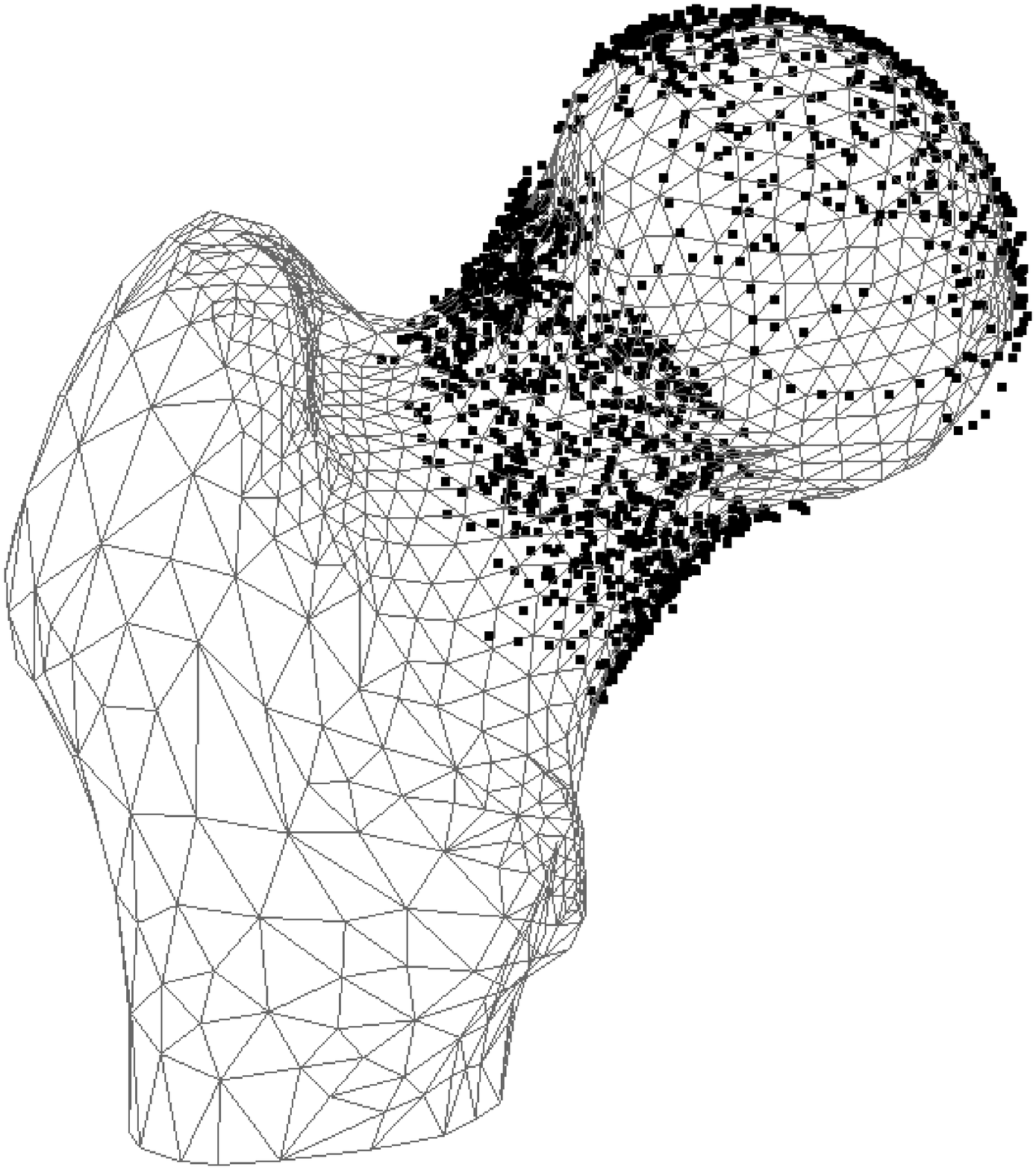}}
	\subfigure{\includegraphics[height=0.21\linewidth]{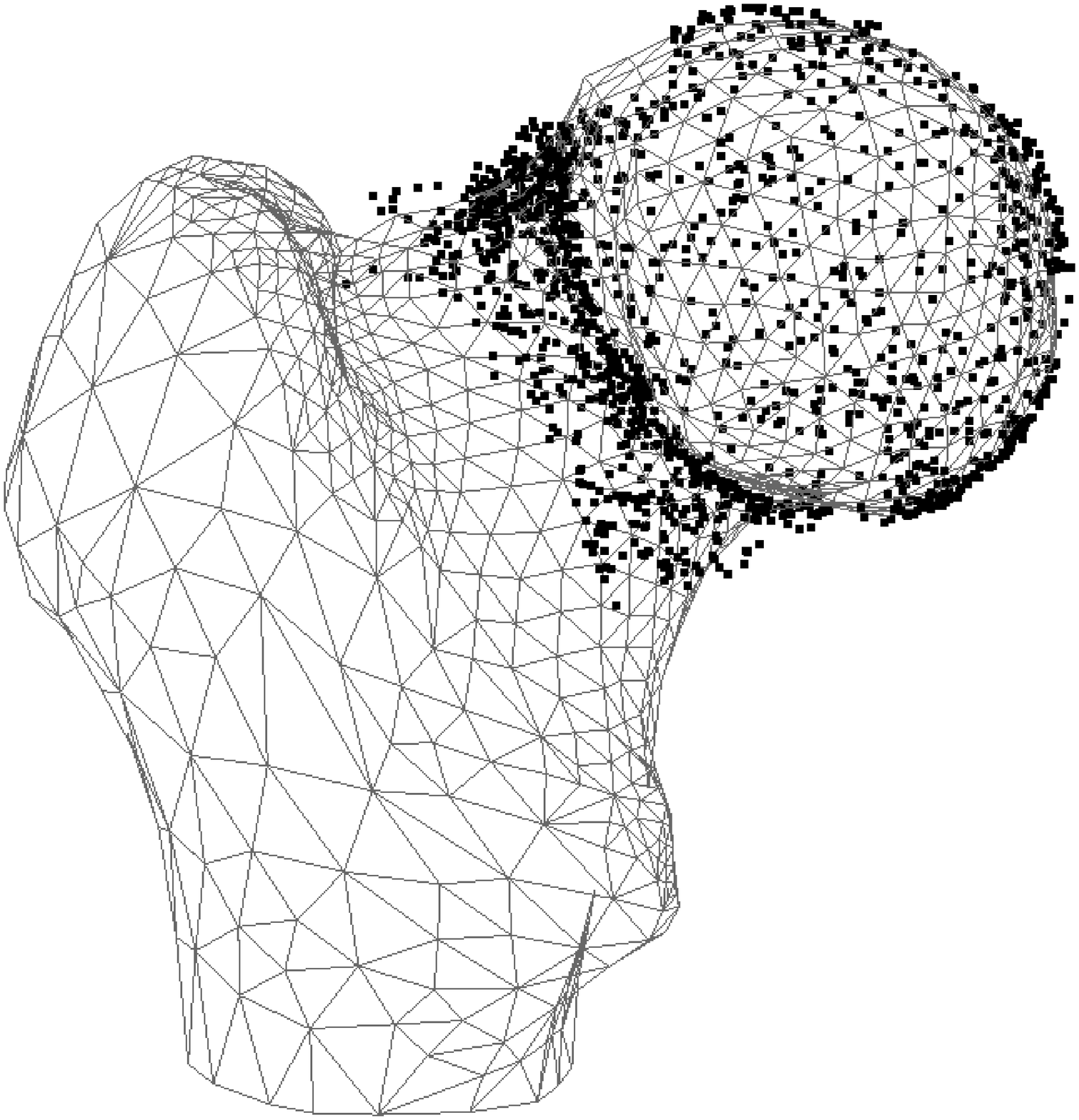}}
	\subfigure{\includegraphics[height=0.21\linewidth]{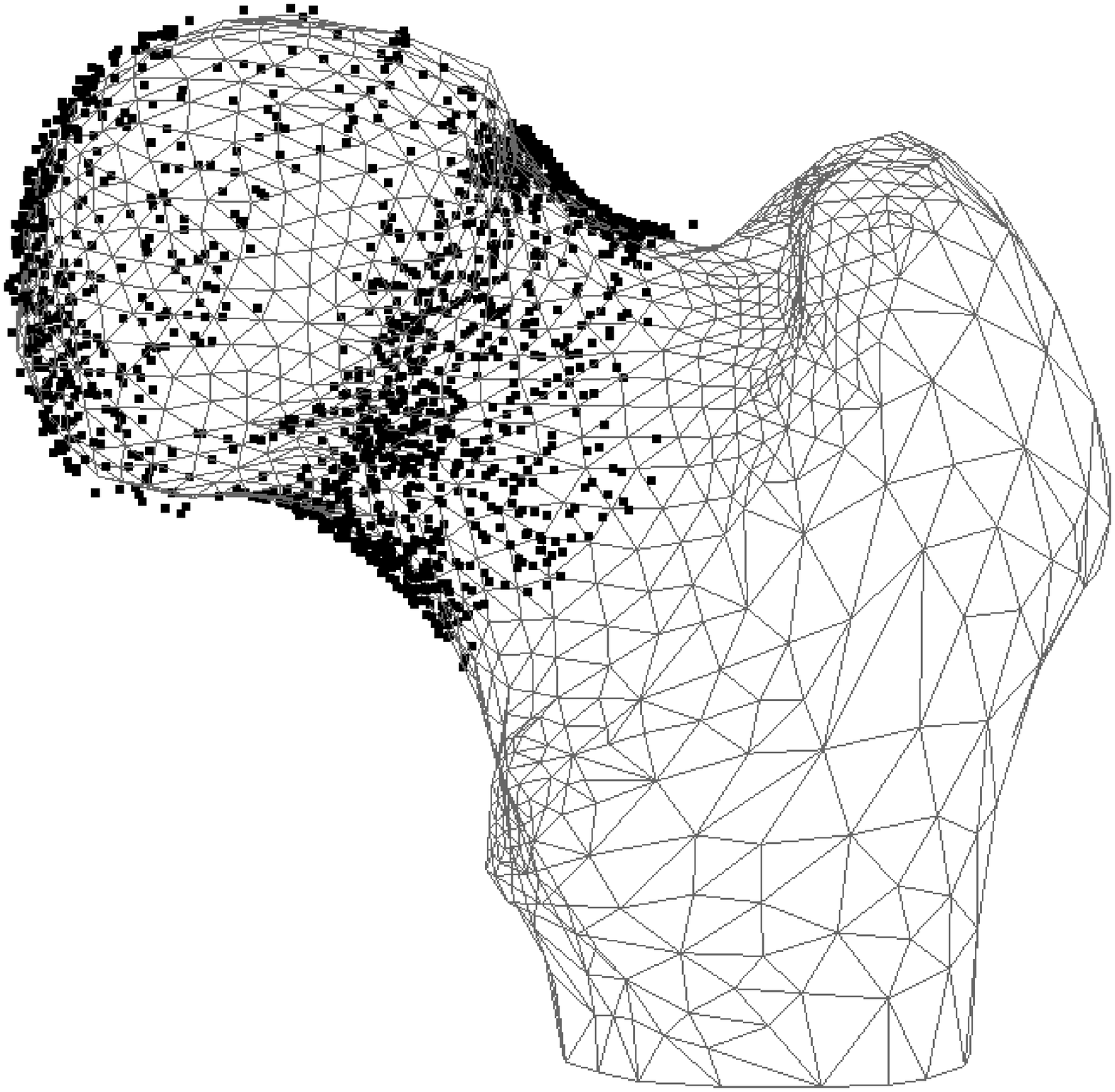}}
	\subfigure{\includegraphics[height=0.21\linewidth]{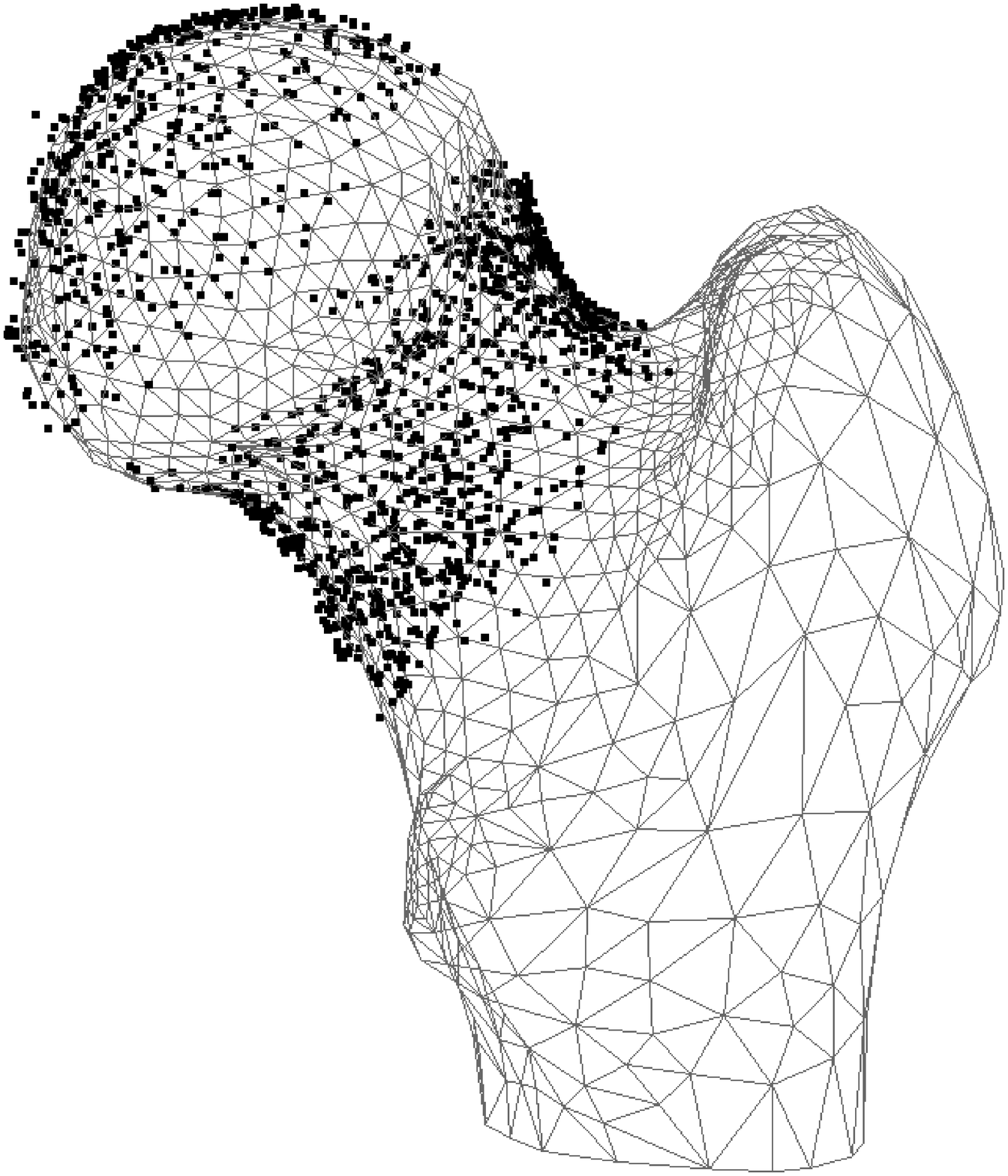}}
	\caption{Intraoperatively acquired point clouds for the 5 considered patients (black dots). To localize the intraoperatively accessible bone region, a surface mesh of the femoral head is showed with each data set (wireframe surface).}
	\label{FigPartialFemoraPointClouds}
\end{center}
\end{figure}

The initial positions of the recorded points with respect to the Atlas model were computed using the correspondence between anatomical landmarks such as the knee center and the piriformis fossa, localized in patient space using the pointer, and defined in the Atlas space by an expert operator. Fig. \ref{FigPartialFemoraPointClouds} shows the similar and very localized distributions of the digitized points for all 5 patients with respect to an approximative surface model of the proximal femur (3 right and 2 left hips were included in this study).

The hip Atlas mesh used here is the upper part of the complete right femur Atlas presented in \S \ref{SecCompleteFemora}. After truncation, the right hip model was mirrored with respect to the sagittal plane to produce the left hip generic model. It is composed of 2105 nodes, forming 1738 hexahedrons and 16 wedges organized so as to reflect the hip principal mechanical structures such as the femoral head and neck.

Mesh registration and repair was carried out as described in \S \ref{SecCompleteFemora}, taking $\emph{S}$ as $\emph{S}_i$, the digitized points cloud for patient $i$, and $\emph{D}$ as the Atlas hip bony surface. Each patient specific mesh was created by applying the inverse $\emph{R}^{-1}$ of the registration function computed with an accuracy level $\epsilon$=0.1mm, as described in \S \ref{SecRegistrationInversion}.

\subsubsection{Results}

The two result tables below are similar to those included in \S \ref{SecCompleteFemora}. Table \ref{TabResultsPartialFemora1} gives the performance of the mesh registration procedures carried out on the 5 data sets, the surface representation error being, this time, computed by considering the distance between all digitized points and the generated patient specific hip surface. 

\begin{table}[ht]
\begin{center}
\begin{tabular}{|p{3cm}||p{1cm}|p{1cm}|p{1cm}|p{1cm}|p{1cm}|}
\hline
\textbf{Patient} & \textbf{1} & \textbf{2} & \textbf{3} & \textbf{4} & \textbf{5} \\
\hline
\textbf{Registration (sec)} & 15 & 17 & 20 & 11 & 23 \\
\textbf{Points} & 1204 & 1433 & 1421 & 1450 & 1437 \\
\hline
\textbf{Mean err. (mm)} & 0.3 & 0.4 & 0.4 & 0.3 & 0.3 \\
\textbf{Max err. (mm)} & 2.1 & 3.4 & 4.2 & 2.2 & 2.7 \\
$\boldsymbol \sigma$ \textbf{(mm)} & 0.3 & 0.4 & 0.4 & 0.3 & 0.3 \\
\hline
\end{tabular}
\caption{Registration (sec): elastic registration times, in seconds; Points: number of intraoperatively digitized points; Mean, Max, $\sigma$ (mm): surface representation mean and maximal error, standard deviation, in millimeters.}
\label{TabResultsPartialFemora1}
\end{center}
\end{table}

Table \ref{TabResultsPartialFemora2} presents the performance of the mesh repair procedures for the 5 meshes. The repair times are given in seconds. In the case of patient 4, the patient mesh produced by the elastic registration was already regular.

\begin{table}[ht]
\begin{center}
\begin{tabular}{|p{3cm}||p{1cm}|p{1cm}|p{1cm}|p{1cm}|p{1cm}|}
\hline
\textbf{Patient} & \textbf{1} & \textbf{2} & \textbf{3} & \textbf{4} & \textbf{5} \\
\hline
\textbf{Regularity (sec)} & 0.4 & 0.1 & 0.5 & 0   & 0.4 \\
\textbf{Quality (sec)}    & 0.5 & 0.4 & 0.8 & 0.4 & 0.4 \\
\hline
\textbf{\% nodes }& 0.3 & 0.05 & 0.7 & 0.05 & 0.4 \\
\textbf{nodes/2105} & 6 & 1 & 14 & 1 & 8 \\
\hline
\textbf{Mean disp. (mm)} & 1.2 & 0.06 & 0.6 & 0.04 & 0.7 \\
\textbf{Max disp. (mm)} & 3.0 & 0.06 & 2.8 & 0.04 & 2.6 \\
$\boldsymbol \sigma$ \textbf{(mm)} & 1.0 & 0.0 & 0.9 & 0.0 & 0.9 \\
\hline
\end{tabular}
\caption{Regularity, Quality (sec): mesh repair times for both phases, in seconds; \% nodes, nodes/2105: fraction and number of nodes moved by the repair procedure; Mean, Max, $\sigma$ (mm): mean and maximal nodal displacements, standard deviation, in millimeters.}
\label{TabResultsPartialFemora2}
\end{center}
\end{table}

\begin{figure}[tb]
\begin{center}
	\subfigure[]{\includegraphics[height=0.30\linewidth]{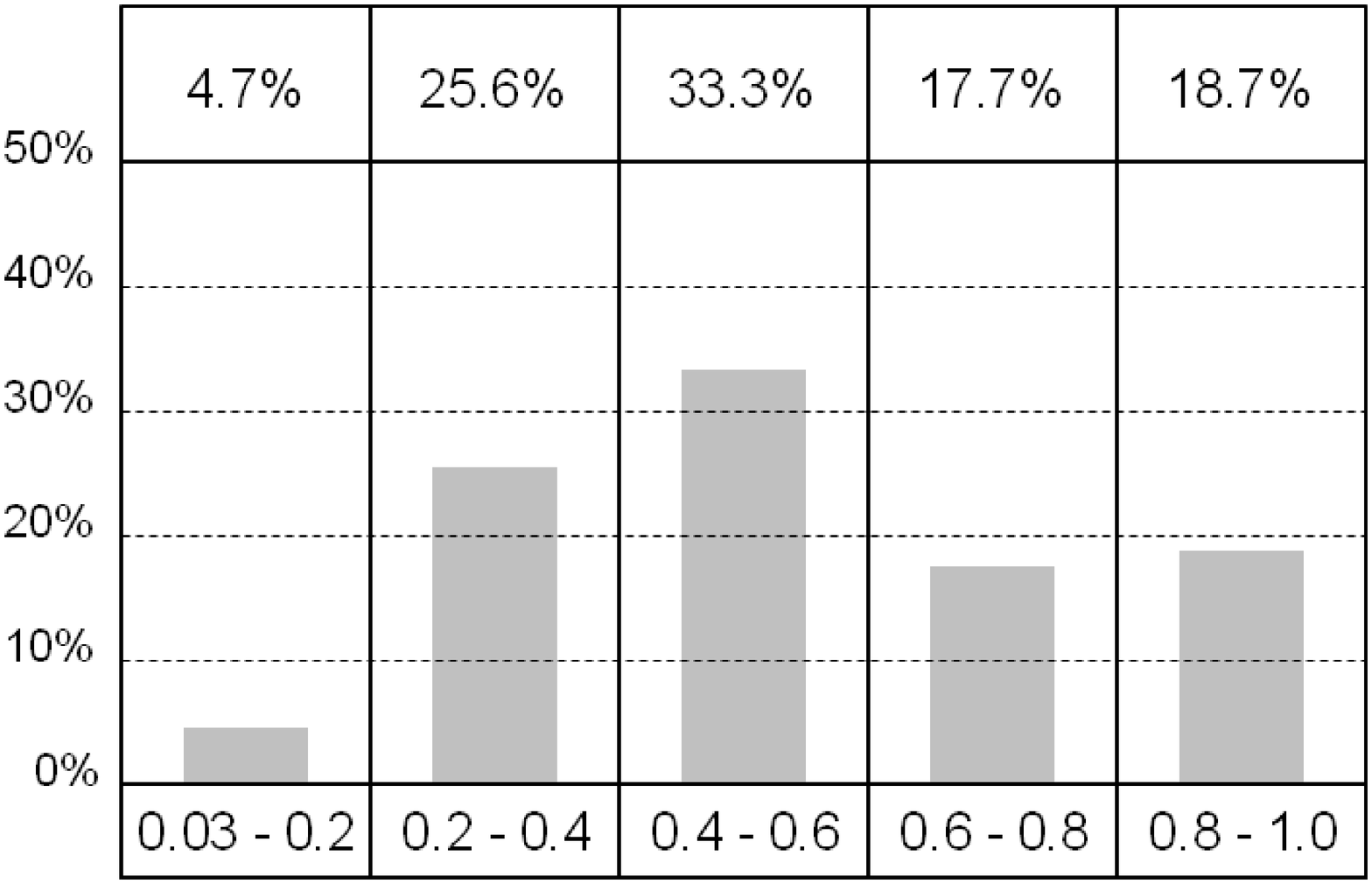}}
	\hspace{0.5cm}
	\subfigure[]{\includegraphics[height=0.30\linewidth]{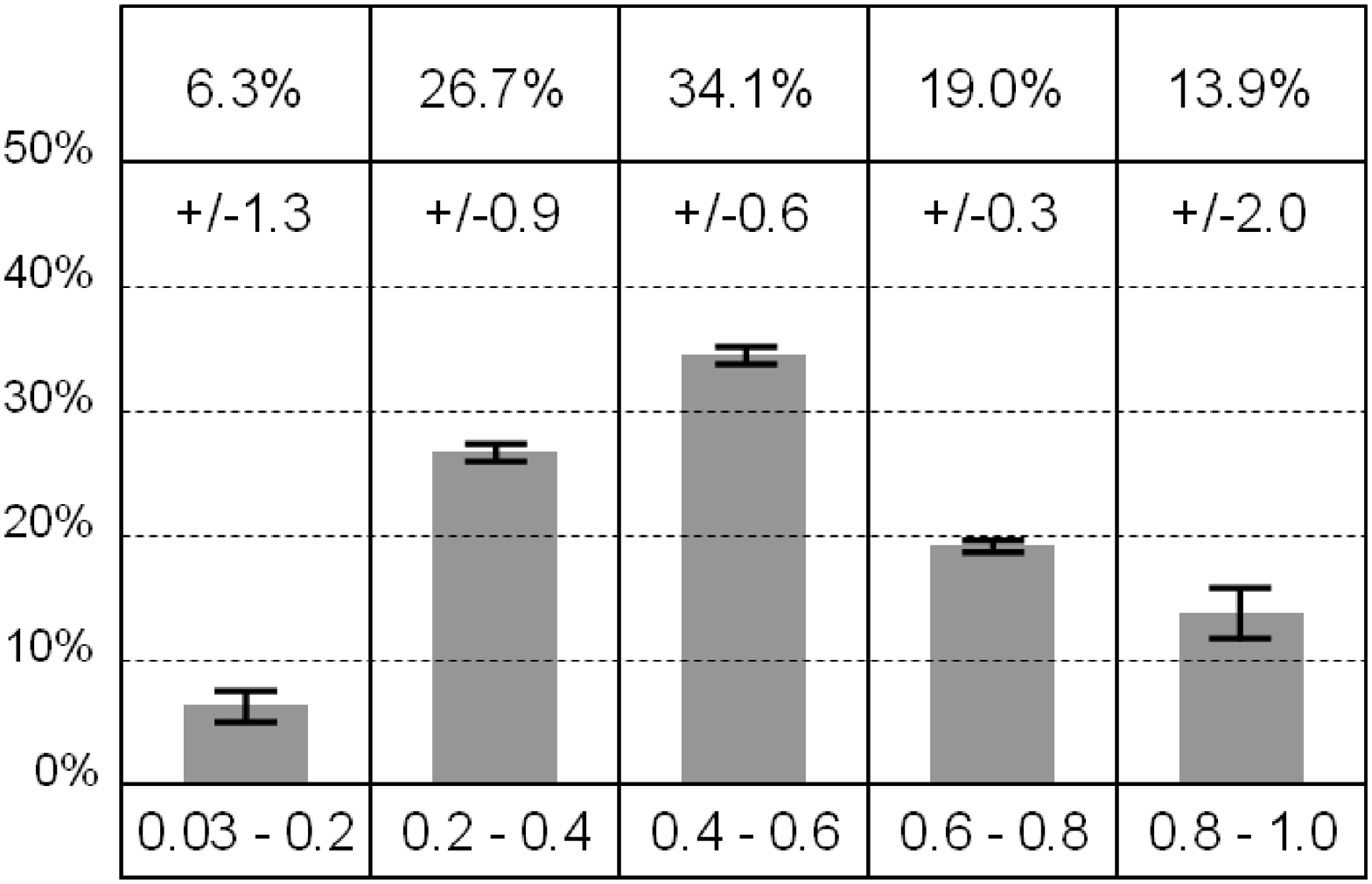}}
	\caption{Mesh quality statistics for Atlas (a) and patient-specific (b) hip models.}
	\label{FigHistTHA}
\end{center}
\end{figure}

Mesh quality distribution measured on the 1754 elements hip Atlas mesh is shown in Fig. \ref{FigHistTHA}-a, and Fig. \ref{FigHistTHA}-b gives the mean quality distribution in the 5 generated patient-specific meshes, along with standard deviations.

As in the previous illustrative case, the MMRep algorithm successfully worked for all 5 patients. All patient specific meshes were generated in less than 25 seconds and had submillimetric surface representation accuracy which remained unchanged by the repair procedures affecting less than 0.7\% of the nodes. The maximal surface representation errors reported here are due to lack of refinement in the template mesh but also to the presence of noise in the digitized point clouds. Indeed, if the hand-held digitization pointer is lifted from the bony surface during the hip surface acquisition process, erroneous points can be recorded. These outliers are eventually averaged out during the elastic registration process, but their presence is revealed by the surface representation error measures. 

This second example demonstrates the adequacy of the MMRep technique in situations where only a fraction of the organ anatomy is known prior to modeling. The elastic registration process, thanks to an a priori knowledge about the organ of interest carried by the Atlas mesh, makes it possible to generate a FE model with high surface representation accuracy in the digitized zones and an approximate yet realistic organ shape in regions where no data is available. The repair phase ensures that the produced model meets the required quality standard and is suitable for FE analysis.

\subsection{Skin and bone face modeling}
\label{SecFaces}

This last use case demonstrates the application of the MMRep procedure to patient specific FE mesh generation in the context of orthognathic surgery, where FE analysis helps predict the consequences of the intervention on the patient's features and facial expressions by simulating the effects of the repositioning of the jaw, maxillary or malar bones \cite{Chabanas02,Chabanas03,Luboz05}.

\subsubsection{Mesh registration procedure}

\begin{figure}[tb]
\begin{center}
	\subfigure[]{\includegraphics[height=0.30\linewidth]{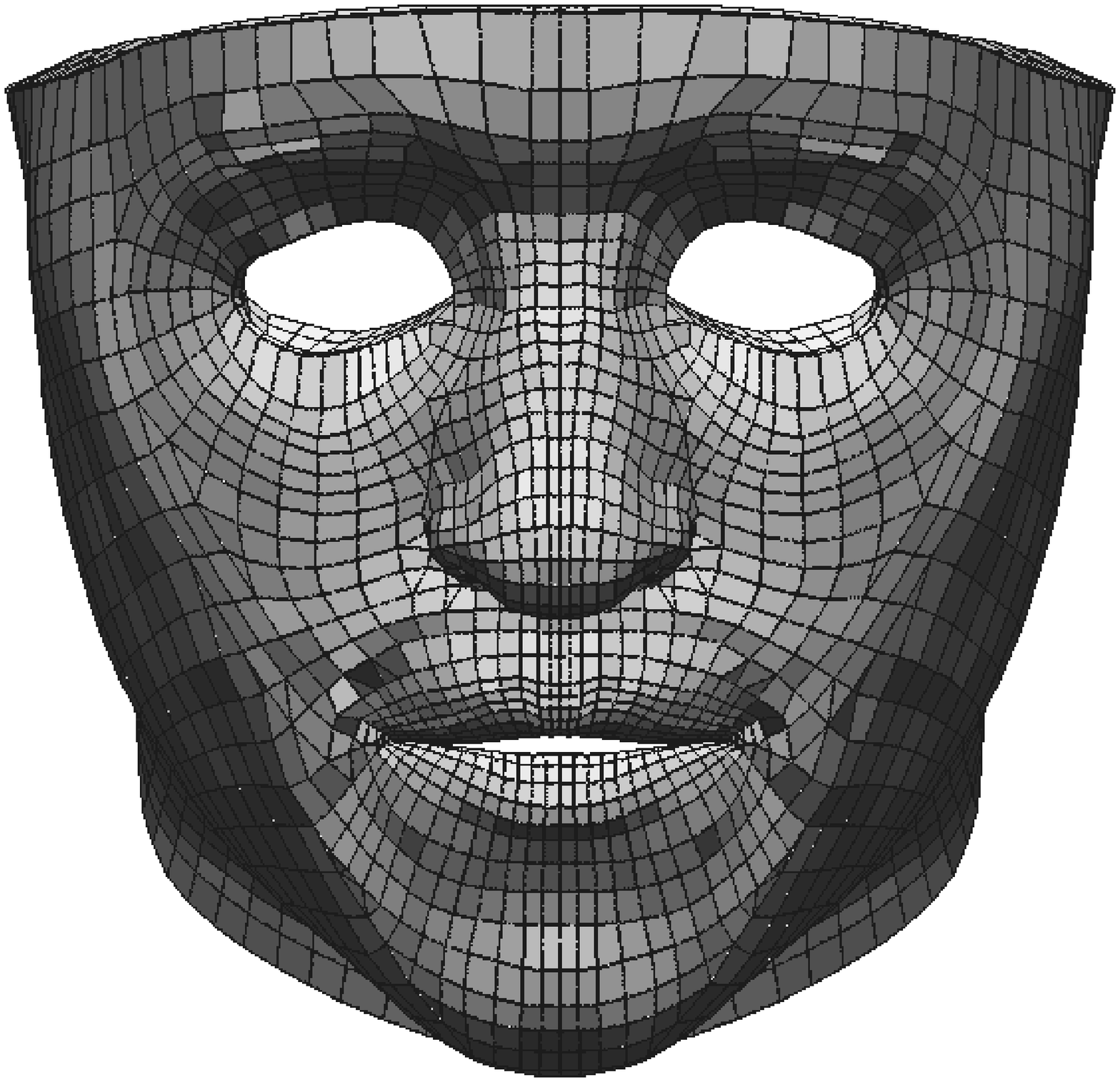}}
	\hspace{0.4cm}
	\subfigure[]{\includegraphics[height=0.30\linewidth]{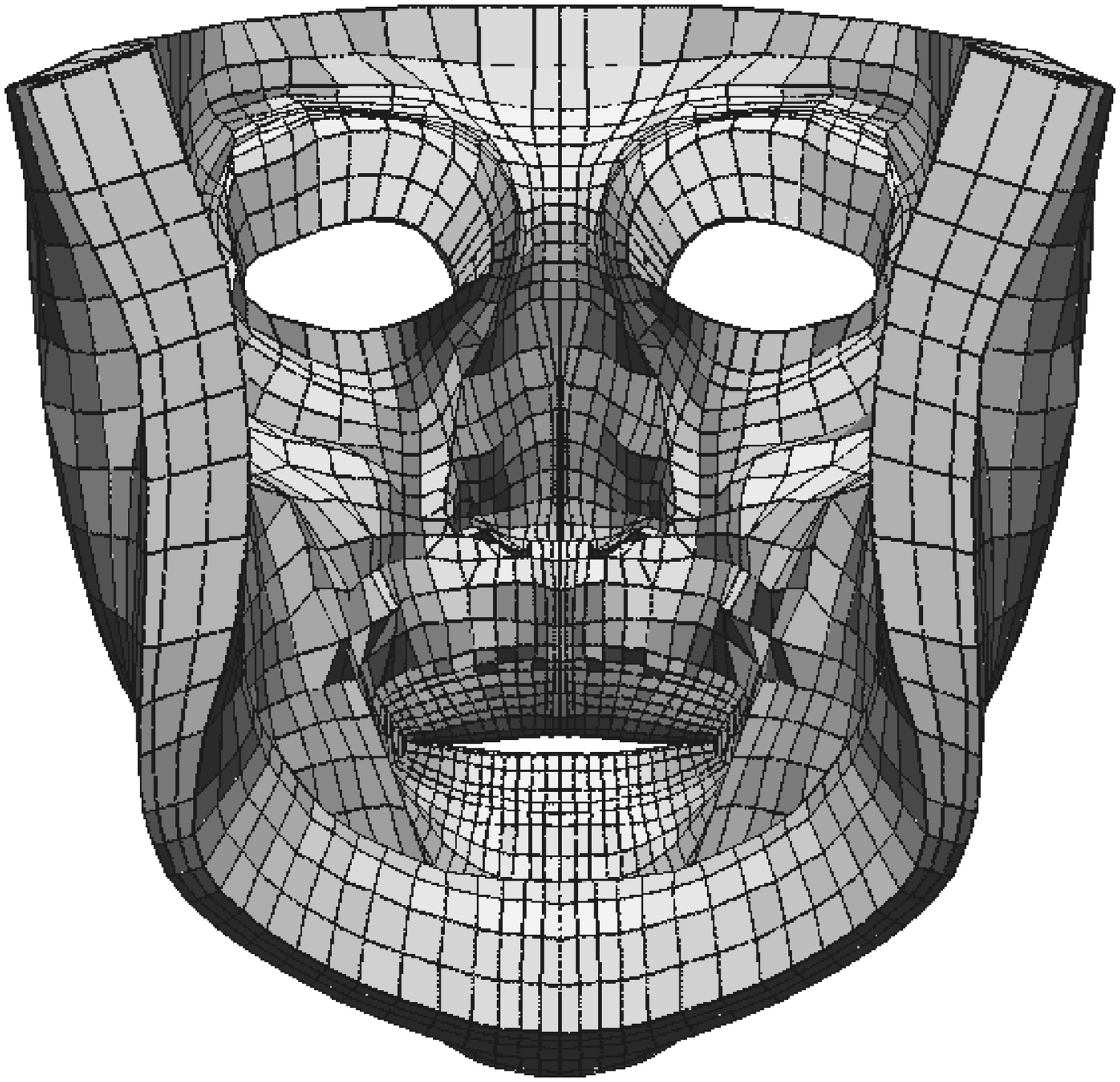}}
	\hspace{0.4cm}
	\subfigure[]{\includegraphics[height=0.30\linewidth]{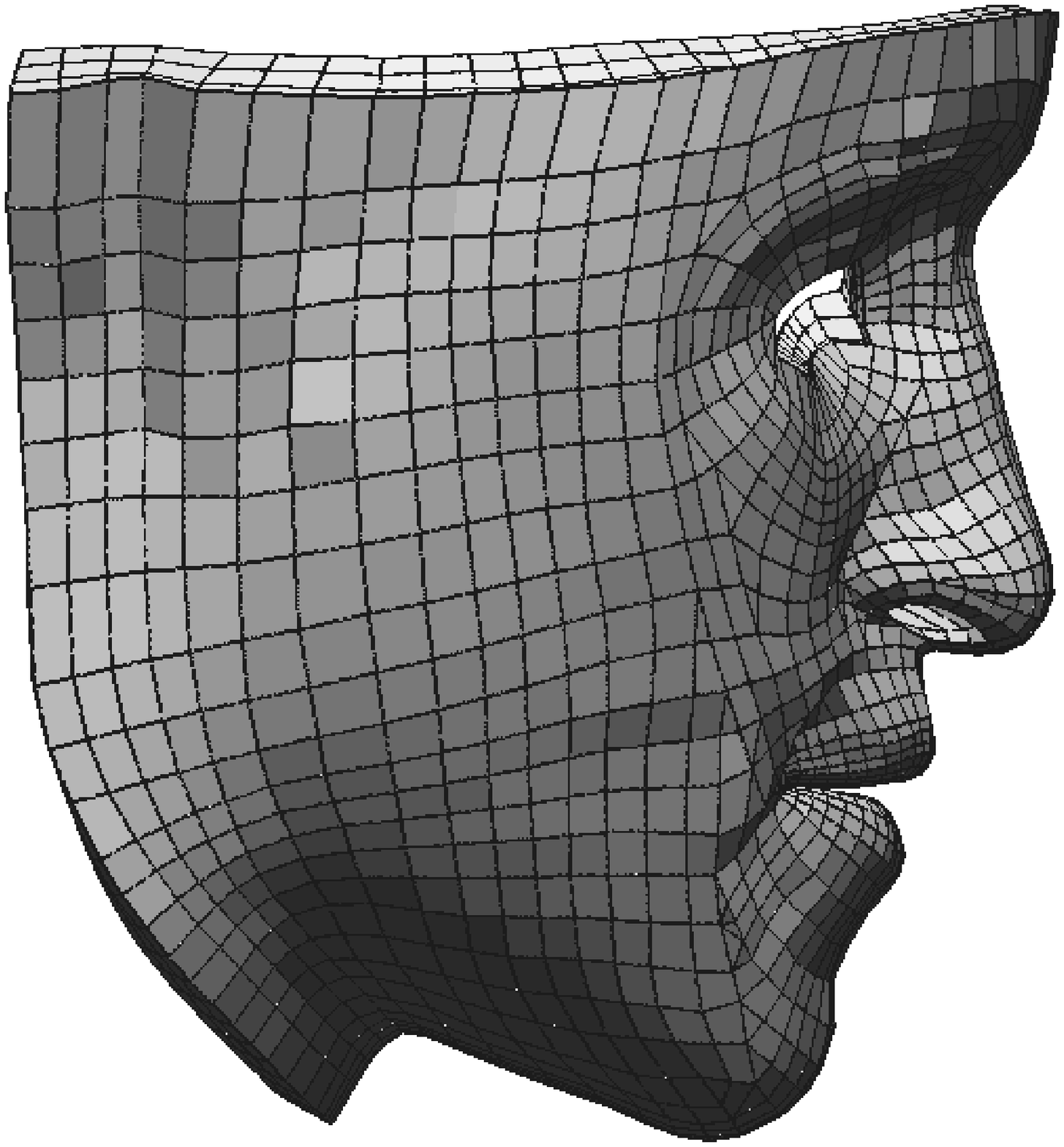}}
	\caption{Face Atlas mesh from Nazari et al. \cite{Nazari08}. (a) Anterior view. (b) Posterior view. (c) Lateral view.}
	\label{FigFaceAtlas}
\end{center}
\end{figure}

In this application a manually assembled 3-layers mesh developed by Nazari et al. \cite{Nazari08}, shown in Fig. \ref{FigFaceAtlas}, is used to model the face muscles and fat. It is made of 8746 nodes forming 6030 hexahedrons and 314 wedges. As we wish to model the interaction between face bones, muscles and features, the patient specific FE model must fit both the skin and skull reconstructed from each CT volume. To this end the Atlas outer and inner layer nodes are labeled ``skin'' and ``bone'' respectively and the multi-labels formulation of the elastic registration discussed in \S \ref{SecMultiplestructuresregistration} is used. The inside Atlas mesh nodes, defining the inner layers of the face tissues, are unlabeled and follow the overall elastic deformation driven by the registration of the nodes labeled ``skin'' and ``bone''.

\begin{figure}[tb]
\begin{center}
	\subfigure[]{\includegraphics[height=0.35\linewidth]{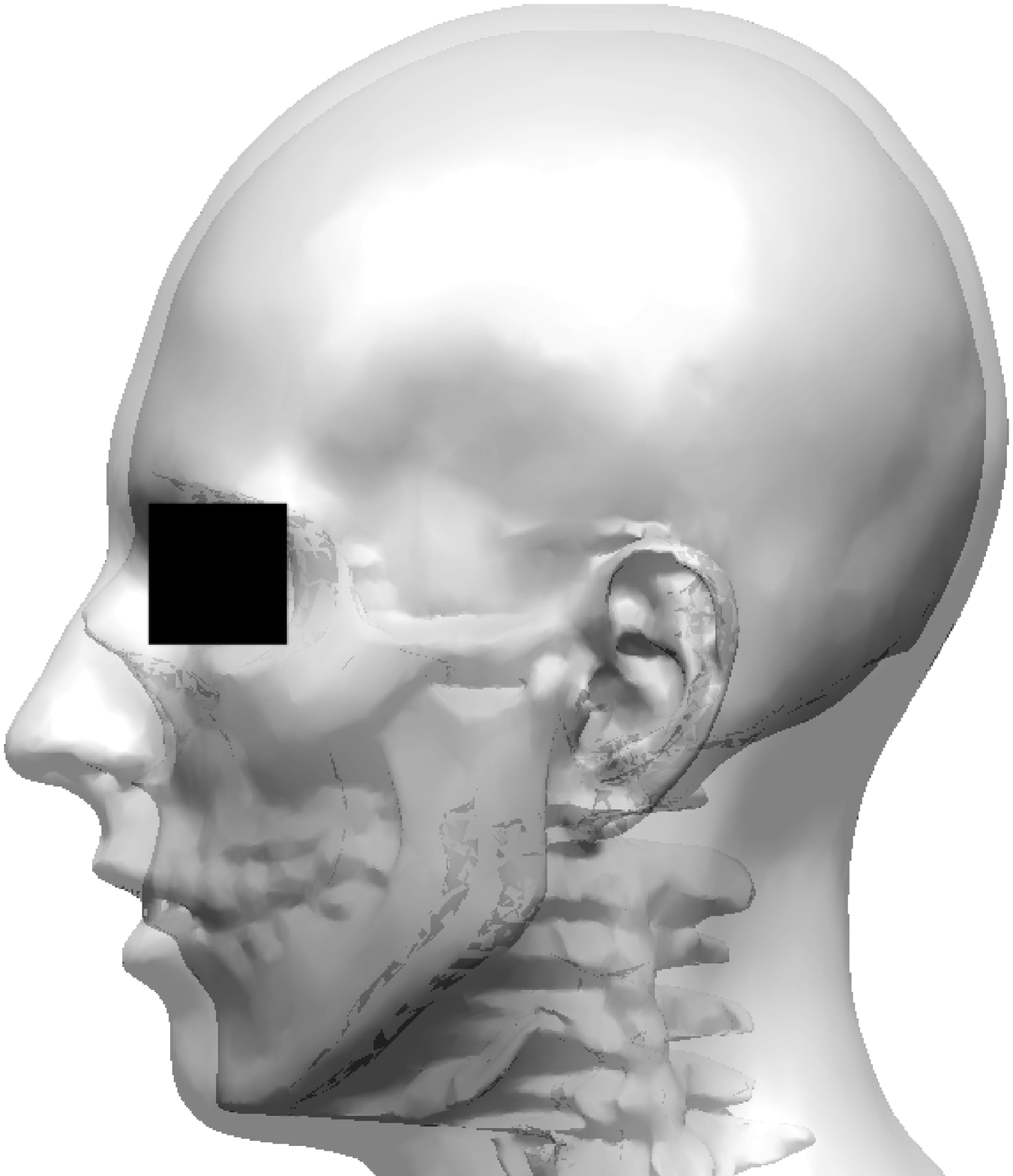}}
	\hspace{1.5cm}
	\subfigure[]{\includegraphics[height=0.35\linewidth]{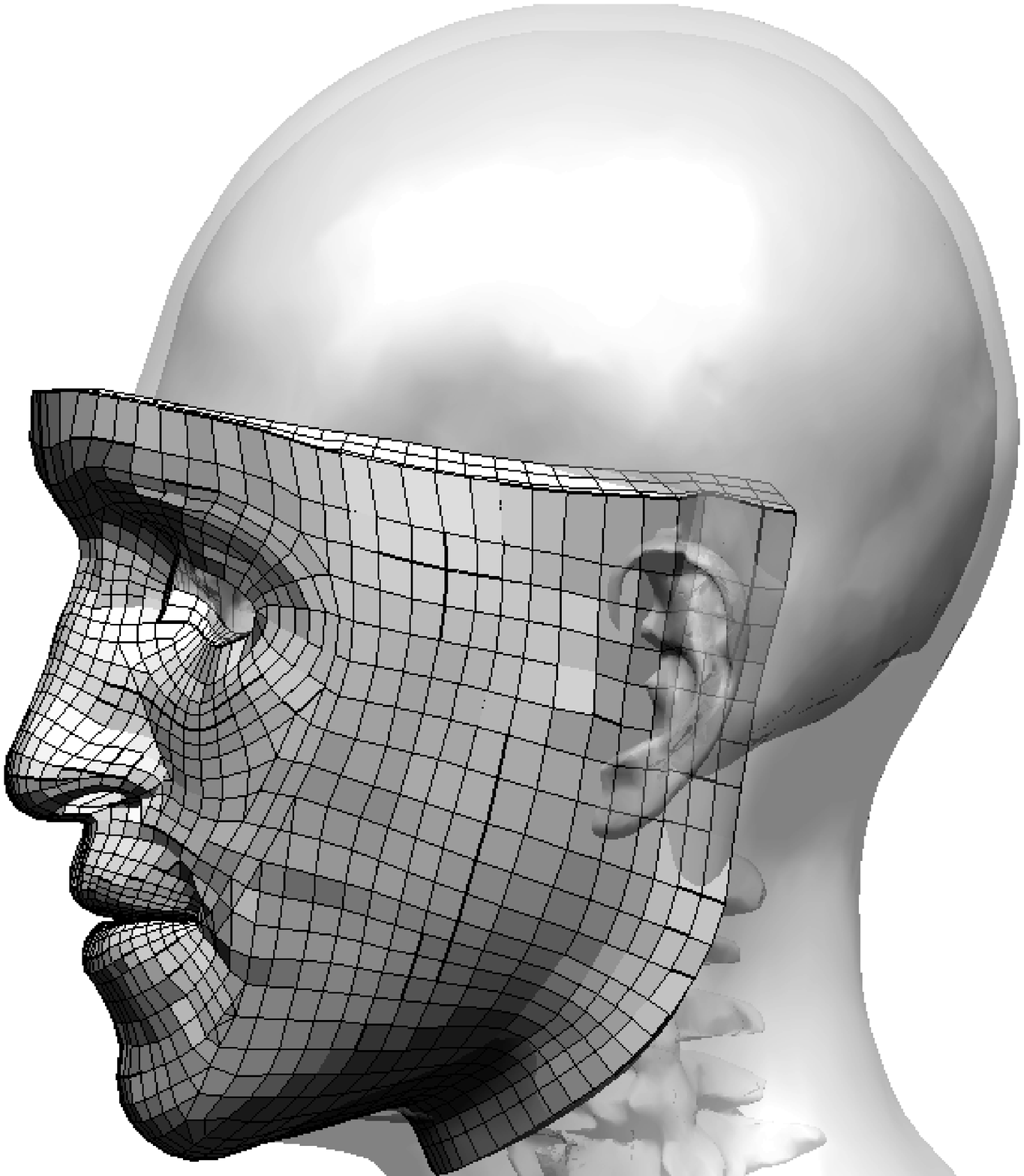}}
	\caption{(a) Skin and skull surfaces reconstruction from a sample patient CT volume. (b) Generic face mesh rigidly registered with the patient data.}
	\label{FigLECMAR}
\end{center}
\end{figure}

For the 50 patients included in this study (data provided by the MAP5 Laboratory, University of Paris V), the bone and skin layers were segmented in the CT volumes using the Hounsfield scale and the resulting surfaces were reconstructed \cite{Tilotta08} and oriented along the anatomical axes in accordance with the generic face model. The Atlas mesh was aligned on the patient data using the nose tip position. Fig. \ref{FigLECMAR} shows sample patient skin and bone surfaces along with the translated Atlas model before elastic registration.

The generic face mesh represents a subset of the complete patient head data, therefore, as discussed in \S \ref{SecAsymmetricregistration}, the computed registration \emph{R} is the one that fits the labeled Atlas mesh nodes to their corresponding destination skin or bone surfaces, producing the transformation that has to be applied to the generic mesh to specialize it for the specific patient. From the computational point of view, as a distinct set of distance maps has to be computed for each patient's skin and skull destination surfaces, the mesh generation process requires more time than in the previously described cases. Yet, in a pre-operative simulation scenario, computational delays are less critical than in an intraoperative FE analysis context.

\subsubsection{Results}

Unlike the previous use cases, this section does not give the detail of MMRep performance figures for each of the 50 patients. Instead, Table \ref{TabResultsFaces} summarizes the overall performance of our technique by presenting the mean registration speed, mesh repair cost and surface reconstruction accuracy. The surface representation errors shown here are the final measures performed after the repair procedure has been applied to the mesh. The nodal displacements amplitudes were evaluated separately for bone and skin layers.

\begin{table}[ht]
\begin{center}
\begin{tabular}{|p{4cm}||p{1.5cm}|p{1.5cm}|p{1.5cm}|}
\hline
 & \textbf{Mean} & \textbf{Max} & $\boldsymbol \sigma$ \\ 
\hline
\textbf{Elastic registration (sec)} & 32 & 96 & 19 \\
\hline
\textbf{Regularity (sec)} & 28 & 50 & 12 \\
\textbf{Quality (sec)} & 4 & 21 & 2 \\
\hline
\textbf{Bone} &&&\\
\hspace{0.5cm}\textbf{Surface err. (mm)} & 0.6 & 13.8 & 0.5 \\
\hspace{0.5cm}\textbf{Moved nodes / 1715} & 39.8 & 105 & 23.9 \\
\hspace{0.5cm}\textbf{Displacements (mm)} & 0.8 & 3.8 & 0.5 \\
\hline
\textbf{Skin} &&&\\
\hspace{0.5cm}\textbf{Surface err. (mm)} & 0.4 & 23.5 & 0.4 \\
\hspace{0.5cm}\textbf{Moved nodes / 2180} & 7.6 & 45 & 10.3 \\
\hspace{0.5cm}\textbf{Displacements (mm)} & 0.3 & 3.1 & 0.2 \\
\hline
\end{tabular}
\caption{Elastic mesh registration, regularization and quality optimization times, in seconds. For both bone and skin layers: final surface representation mean errors, number of nodes corrected by the repair procedure in each layer and nodal displacements amplitudes, in millimeters.}
\label{TabResultsFaces}
\end{center}
\end{table}

The impact of the mesh repair procedure is very limited as it only affects an average of 2\% of the 1715 bone nodes and 0.3\% of the 2180 skin nodes in the mesh. As for the surface representation, submillimetric mean accuracy is achieved for both layers.

\begin{figure}[tb]
\begin{center}
	\subfigure[]{\includegraphics[height=0.06\linewidth]{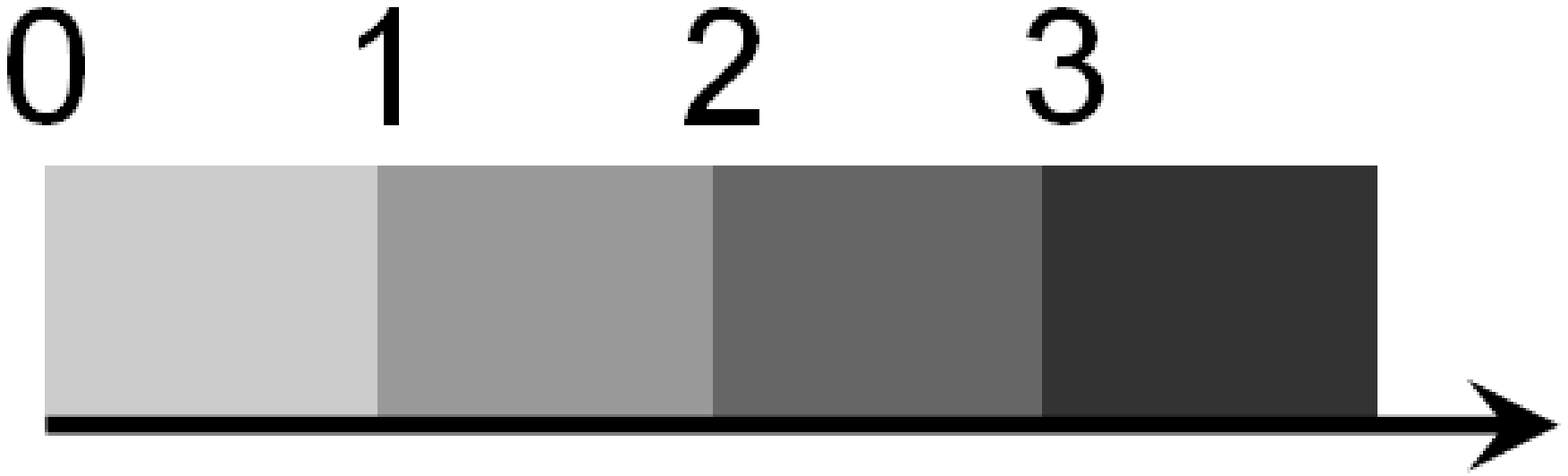}}\\
	\subfigure[]{\includegraphics[height=0.30\linewidth]{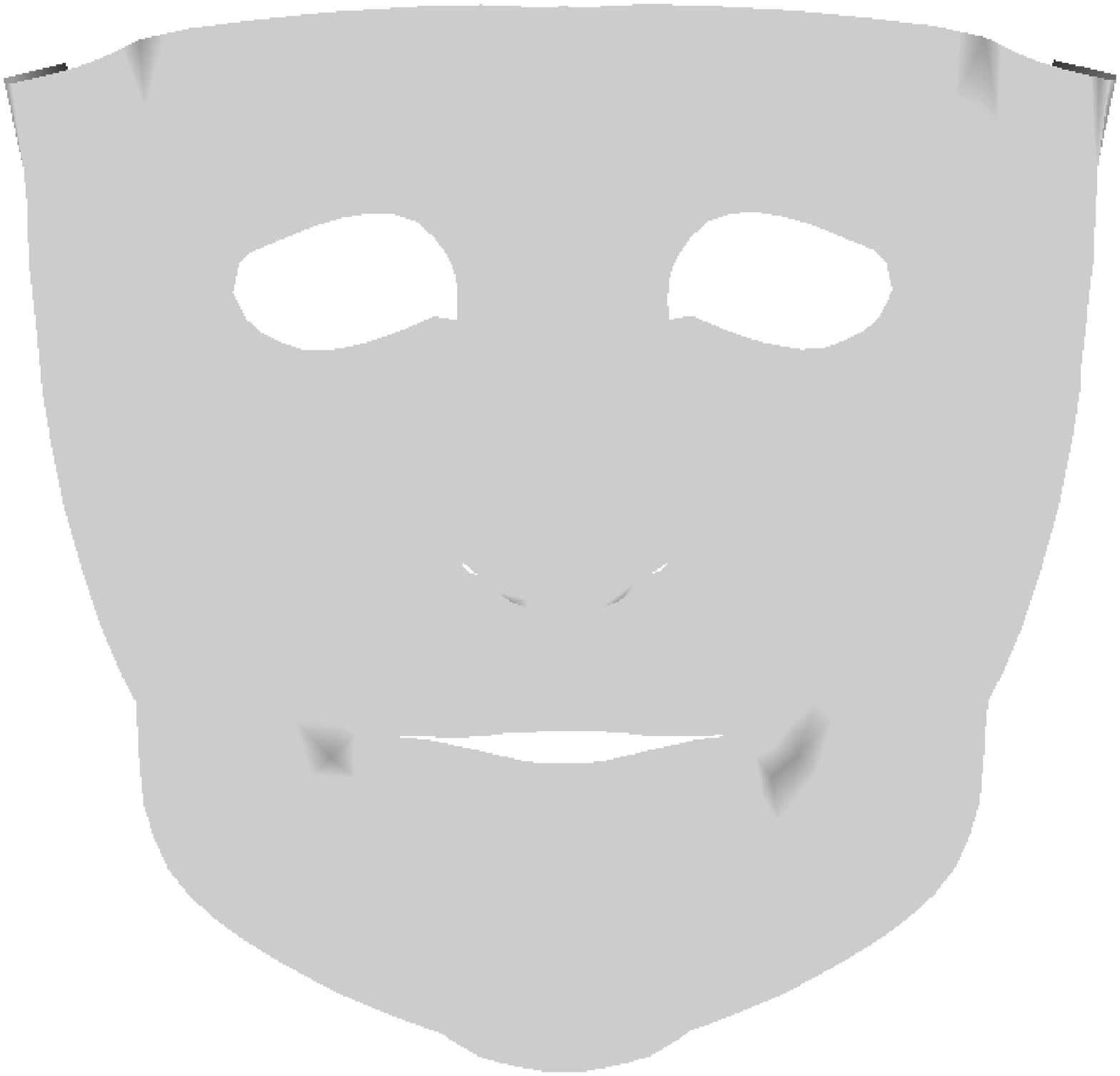}}
	\hspace{1cm}
	\subfigure[]{\includegraphics[height=0.30\linewidth]{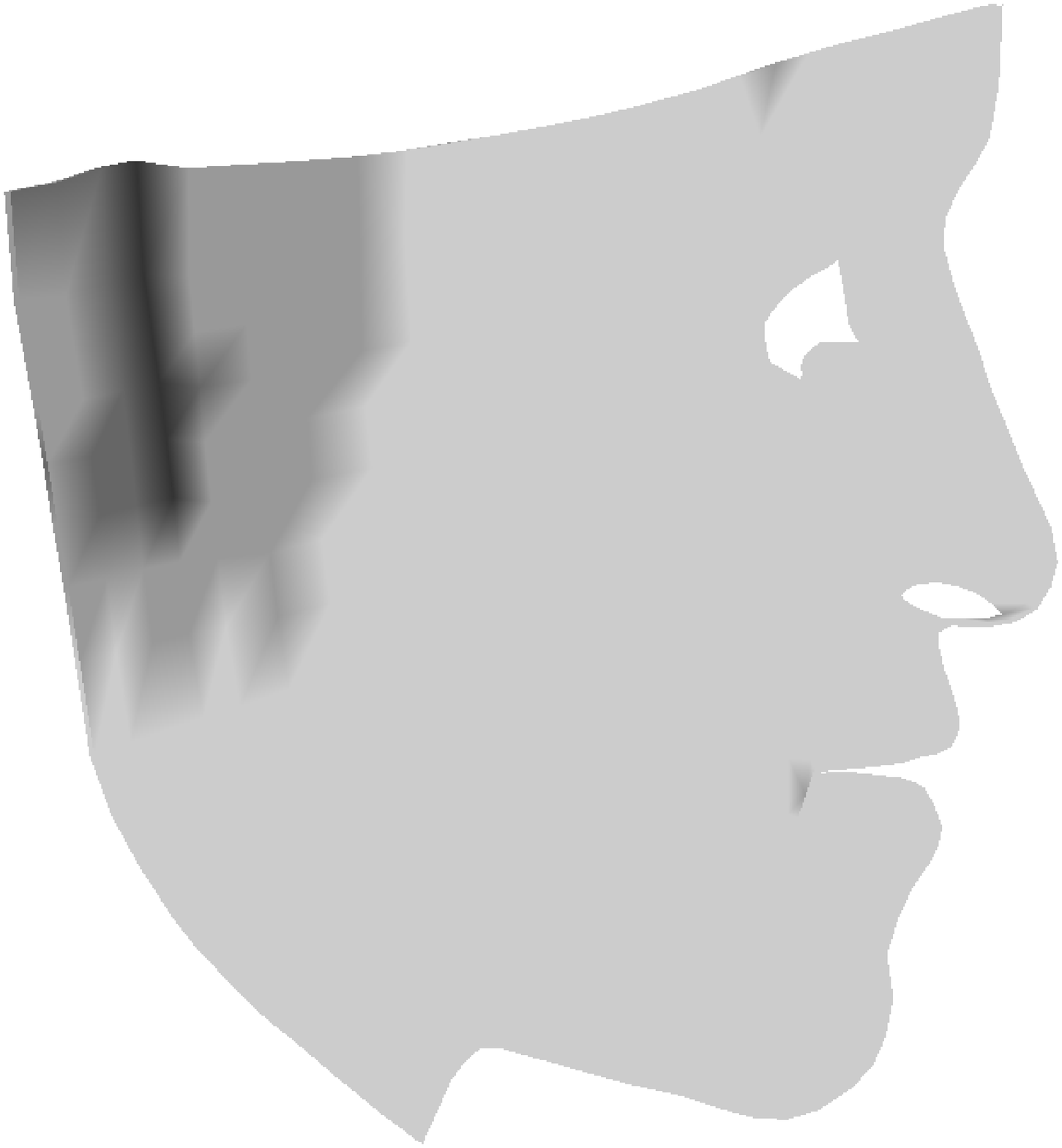}}\\
	\subfigure[]{\includegraphics[height=0.30\linewidth]{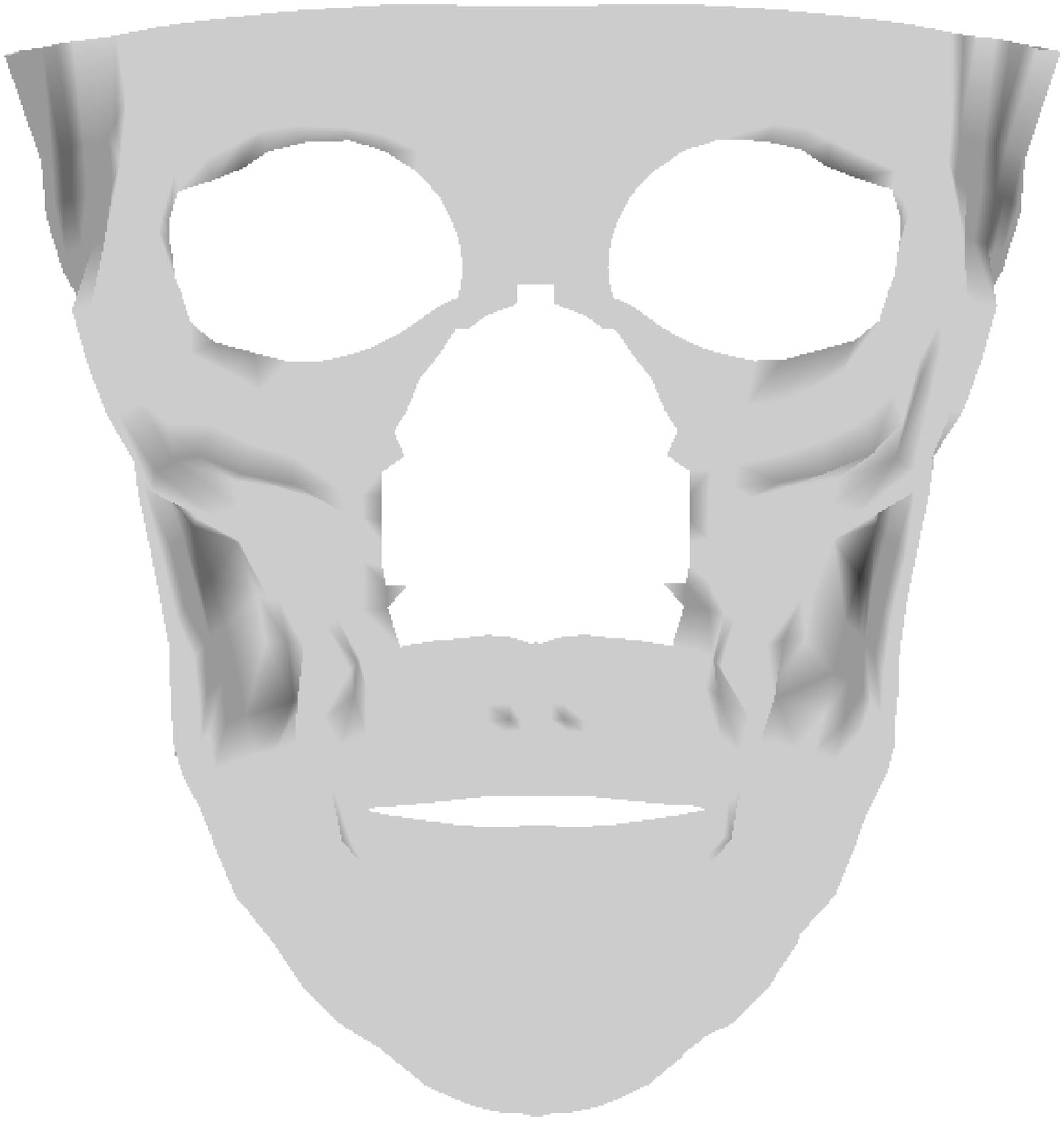}}
	\hspace{1cm}
	\subfigure[]{\includegraphics[height=0.30\linewidth]{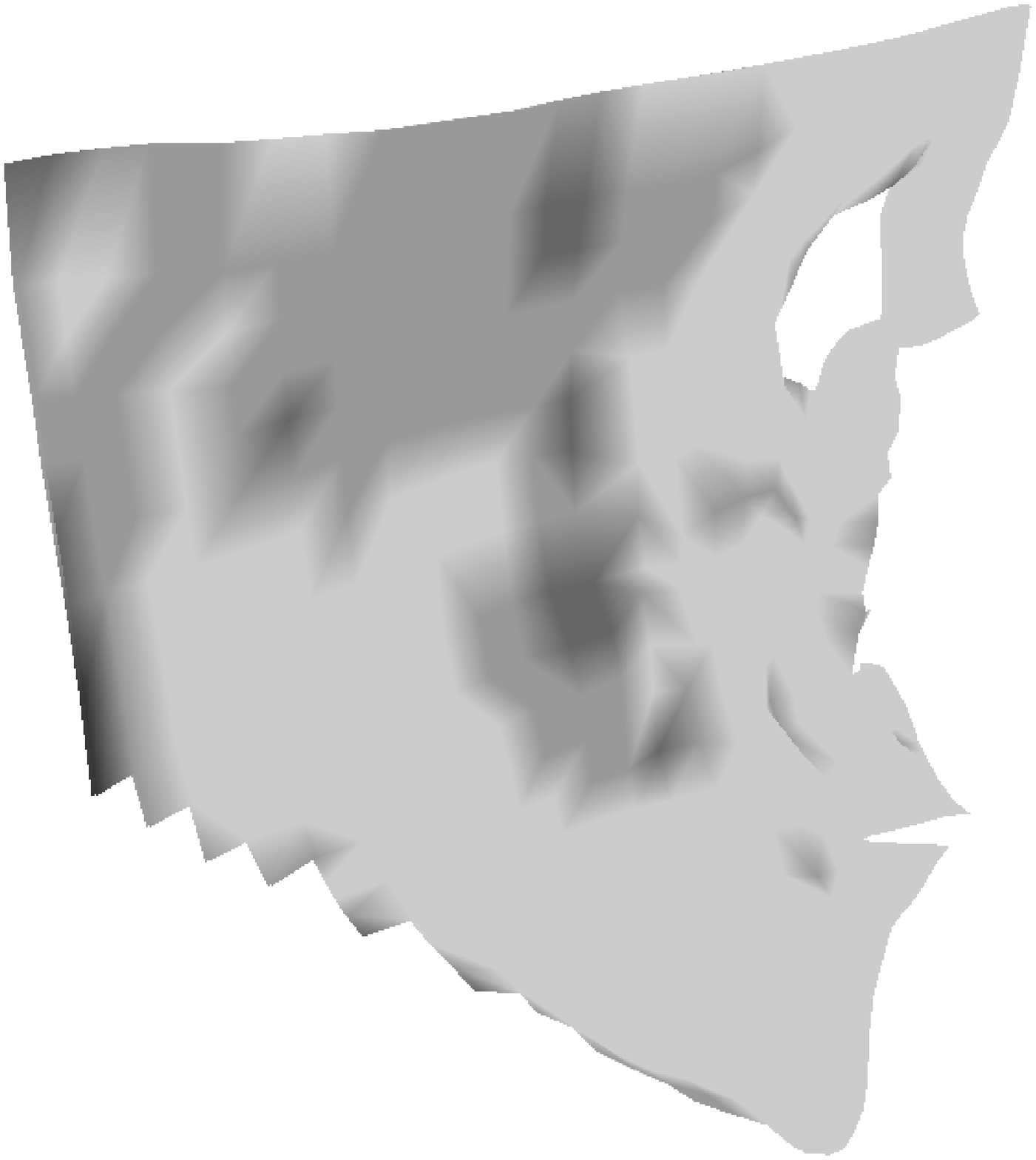}}
	\caption{Face mesh registration mean errors represented as color maps. (a) Error maps color code in millimeters. Skin layer: (b) anterior view; (c) lateral view. Bone layer: (d) anterior view; (e) lateral view.}
	\label{FigFaceErrMaps}
\end{center}
\end{figure}

The reported maximal errors are located in areas where the Atlas mesh lacks refinement. These regions clearly appear on Fig. \ref{FigFaceErrMaps} where the surface representation mean errors computed over the 50 cases are displayed as error maps on the initial Atlas bone and skin layers. The lightest areas represent mean errors between 0 and 1 mm and the darkest areas errors above 3 mm.

Fig. \ref{FigFaceErrMaps} shows that the maximal skin layer errors are located around the ears which are clearly absent from the generic mesh, as can be seen in Fig. \ref{FigFaceAtlas}-c. As for the bone layer, the maximal errors appear near the sphenoid bone and the zygomatic process as both regions have a very approximative representation in the Atlas mesh. 

\begin{figure}[tb]
\begin{center}
	\subfigure[]{\includegraphics[height=0.30\linewidth]{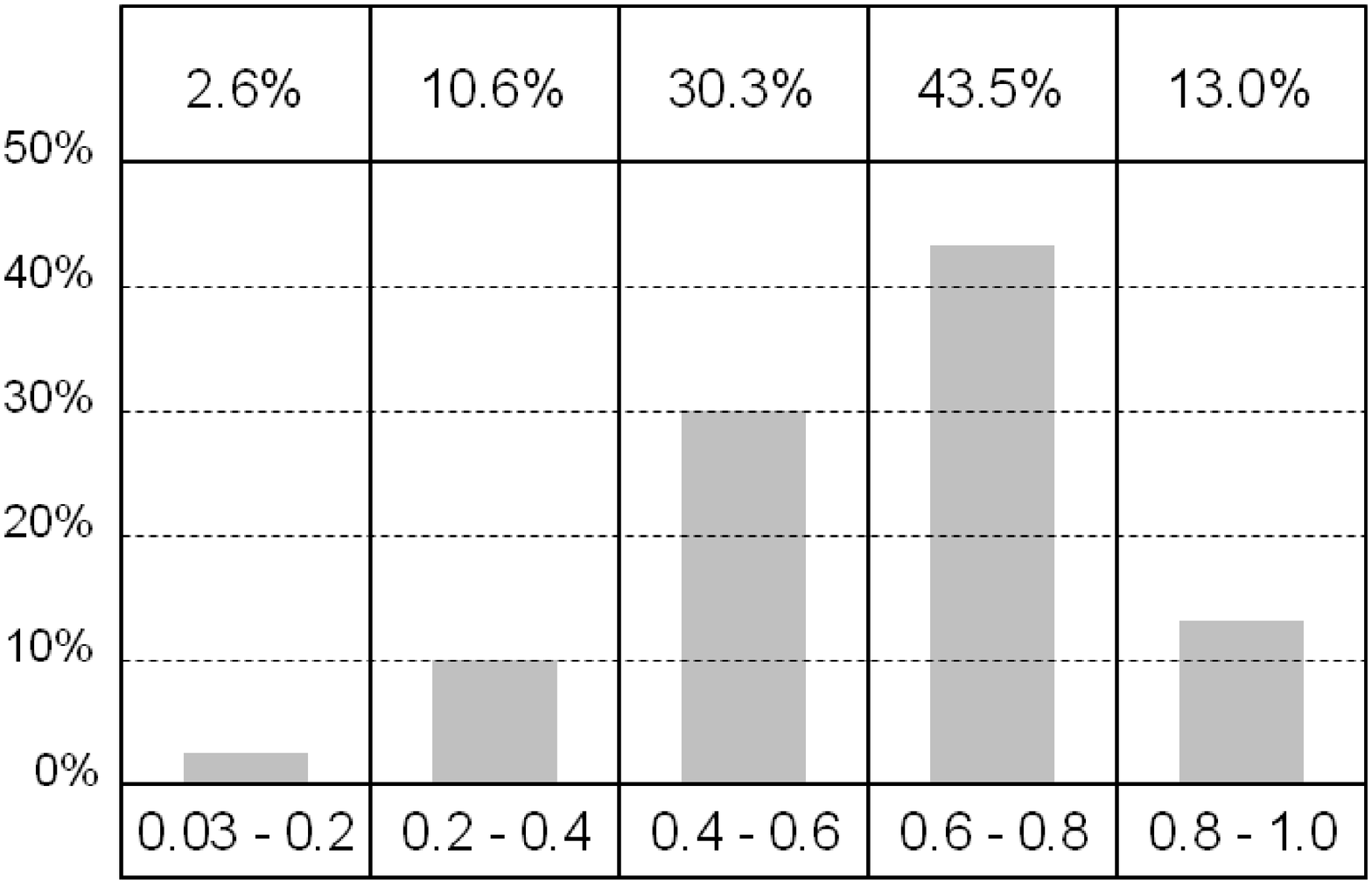}}
	\hspace{0.5cm}
	\subfigure[]{\includegraphics[height=0.30\linewidth]{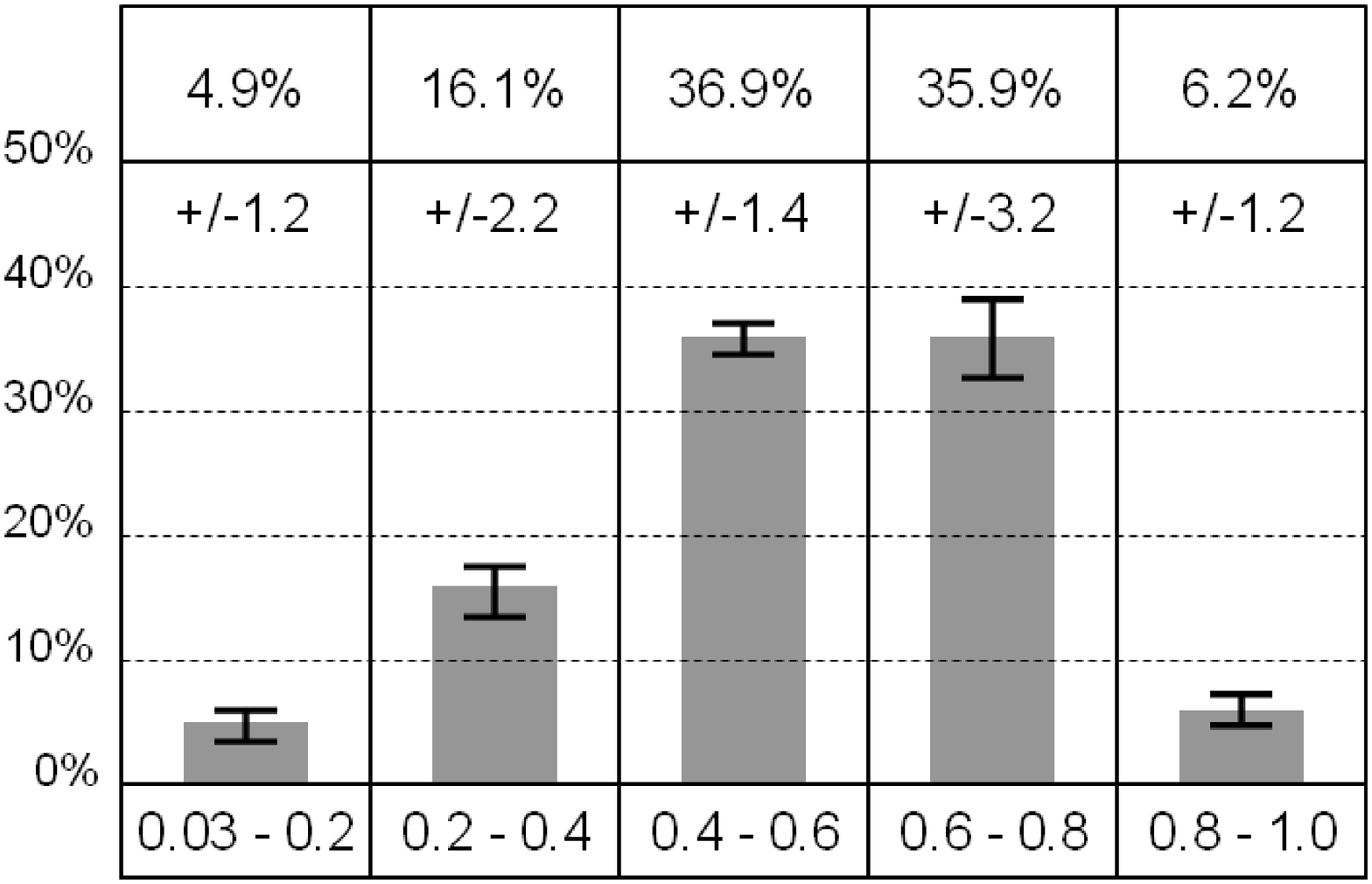}}
	\caption{Mesh quality statistics for Atlas (a) and patient-specific (b) face models.}
	\label{FigHistFACES}
\end{center}
\end{figure}

Mesh quality distribution measured on the 6344 elements face Atlas mesh is shown in Fig. \ref{FigHistFACES}-a, and Fig. \ref{FigHistFACES}-b gives the mean quality distribution in the 50 generated patient-specific meshes, along with standard deviations.

In all cases, the MMRep algorithm was able to generate a patient specific bi-layer mesh suitable for FE analysis within a couple of minutes. The produced meshes exhibited submillimetric mean surface representation accuracy on both skin and bone layers, with larger errors localized around features absent or ill-defined in the generic mesh.

\begin{figure}[tb]
\begin{center}
	\subfigure{\includegraphics[height=0.35\linewidth]{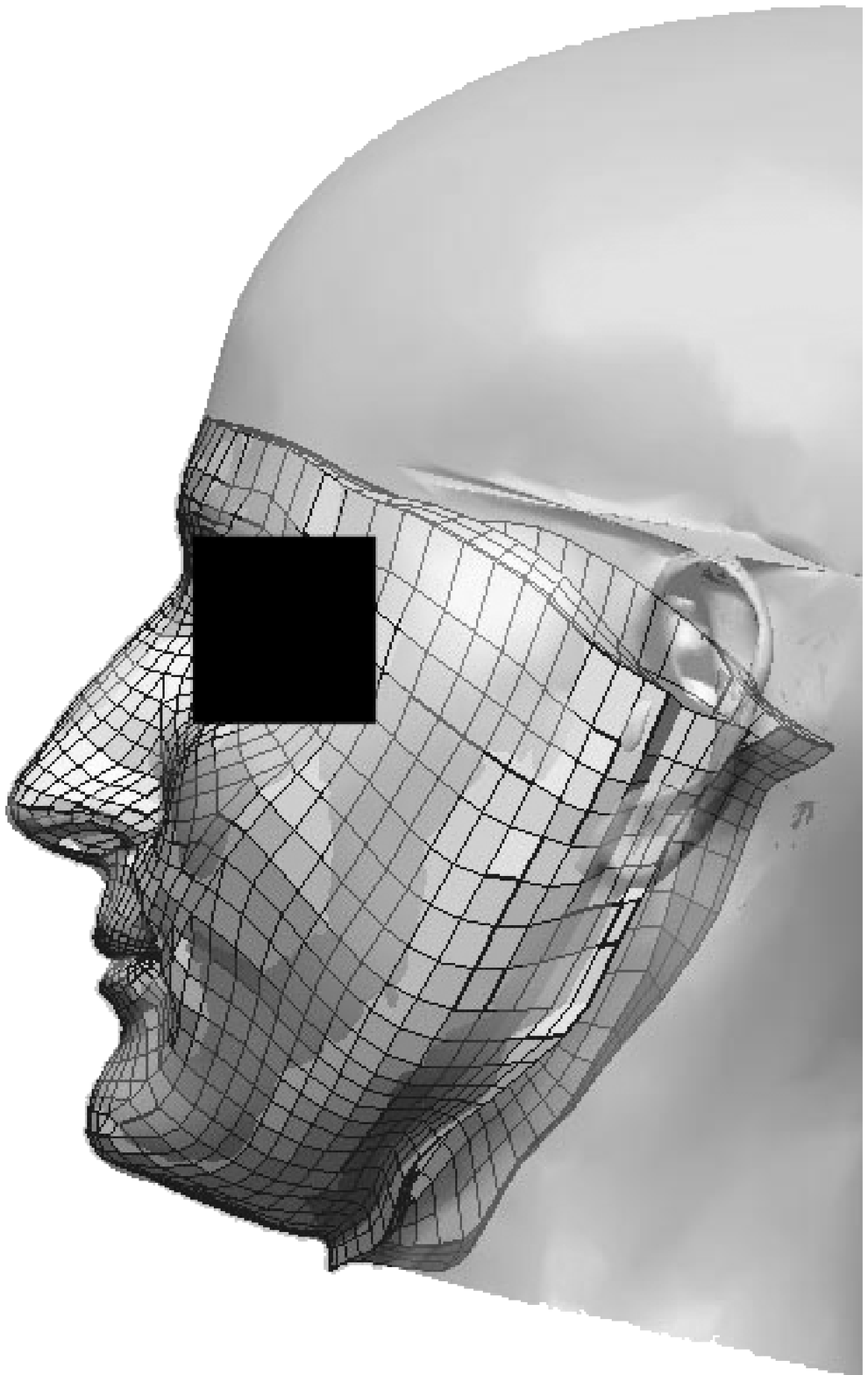}}
	\subfigure{\includegraphics[height=0.35\linewidth]{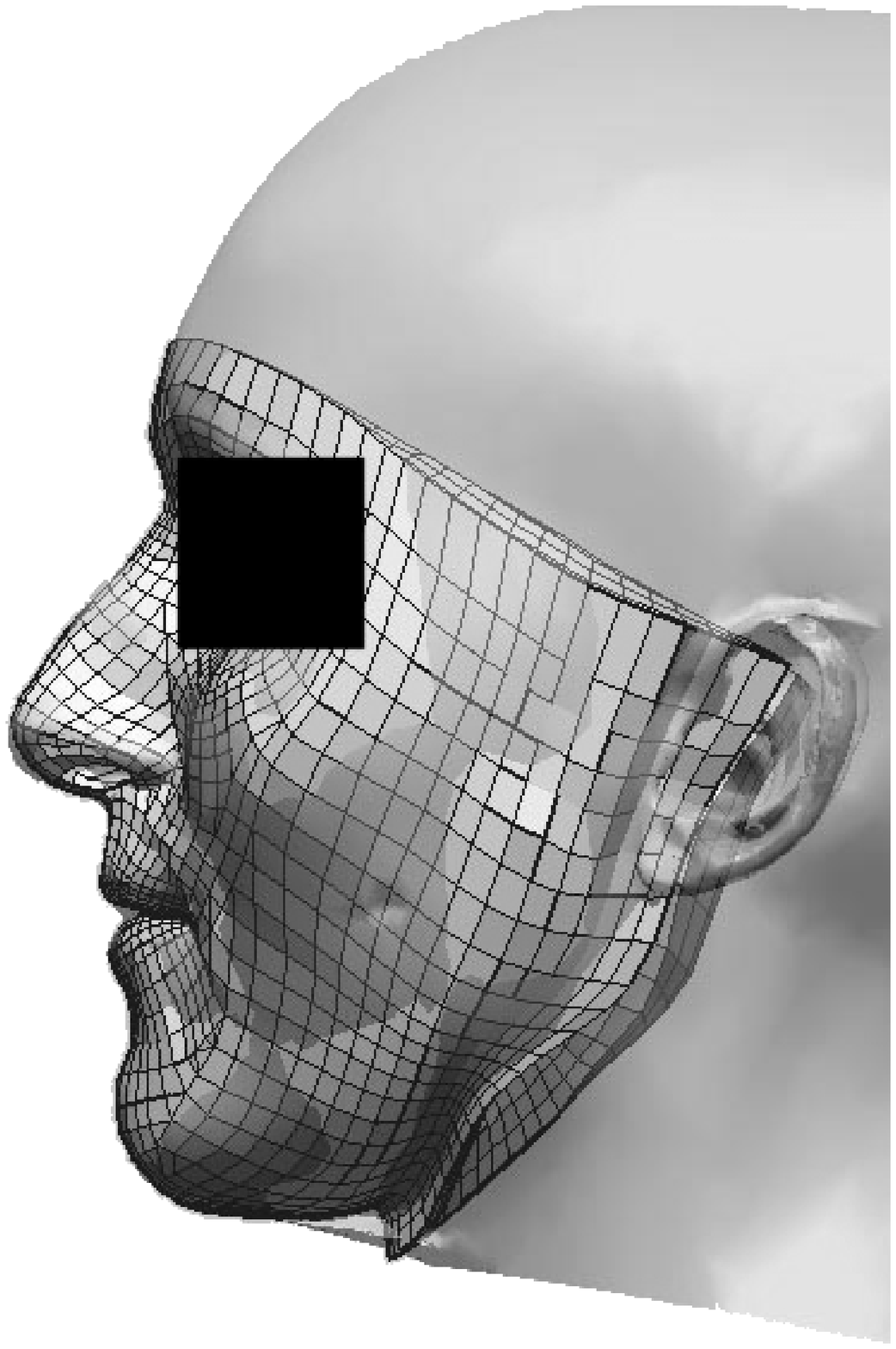}}
	\subfigure{\includegraphics[height=0.35\linewidth]{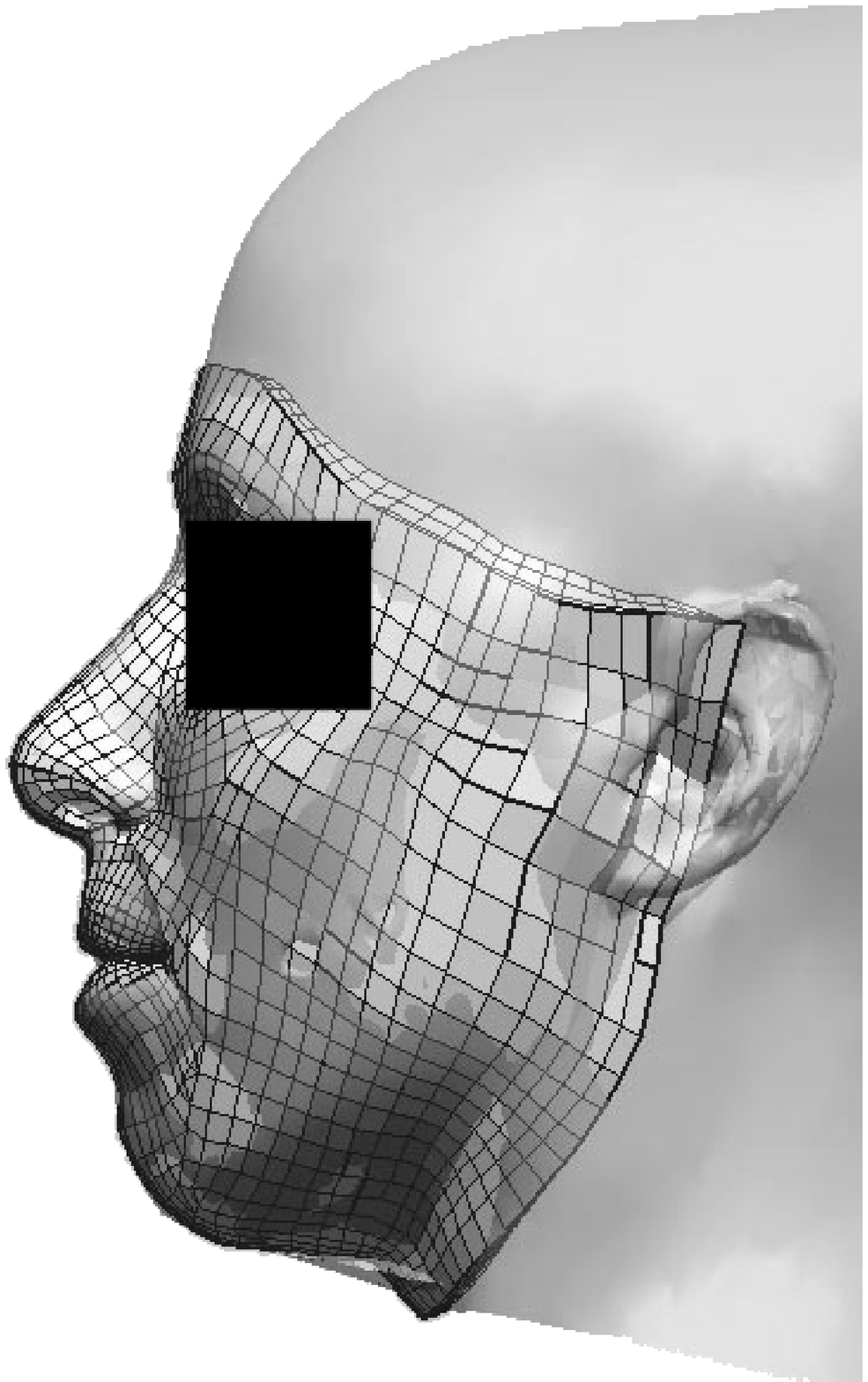}}
	\subfigure{\includegraphics[height=0.35\linewidth]{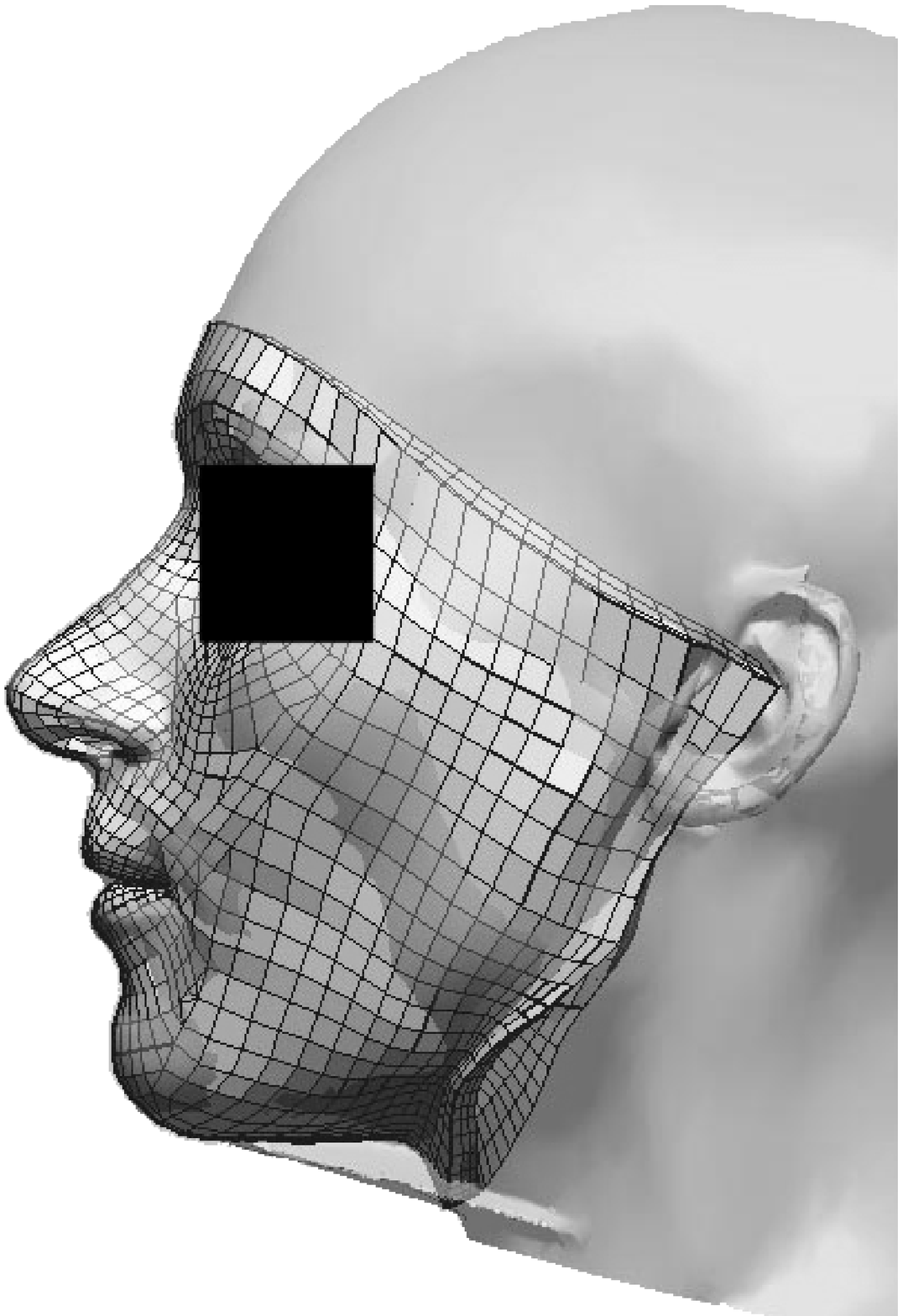}}
	\caption{Four examples of the generated face models. For clarity only the skin surface segmented from the CT volume is shown here although the produced FE models also fit to the underlying skull surface.}
	\label{FigFaces}
\end{center}
\end{figure}

Fig. \ref{FigFaces} shows four examples of meshes fitted onto distinct patient morphologies. The Atlas face mesh used here was constructed based on a unique prognathic\footnote{Having the jaws projecting beyond the upper part of the face.} patient, yet it was successfully used to model both prognathic and retrognathic\footnote{Having a mandible located posterior to its normal position.} patients, which demonstrates the versatility of the registration procedure.

\begin{figure}[tbh!]
\begin{center}
	\includegraphics[width=1.0\linewidth]{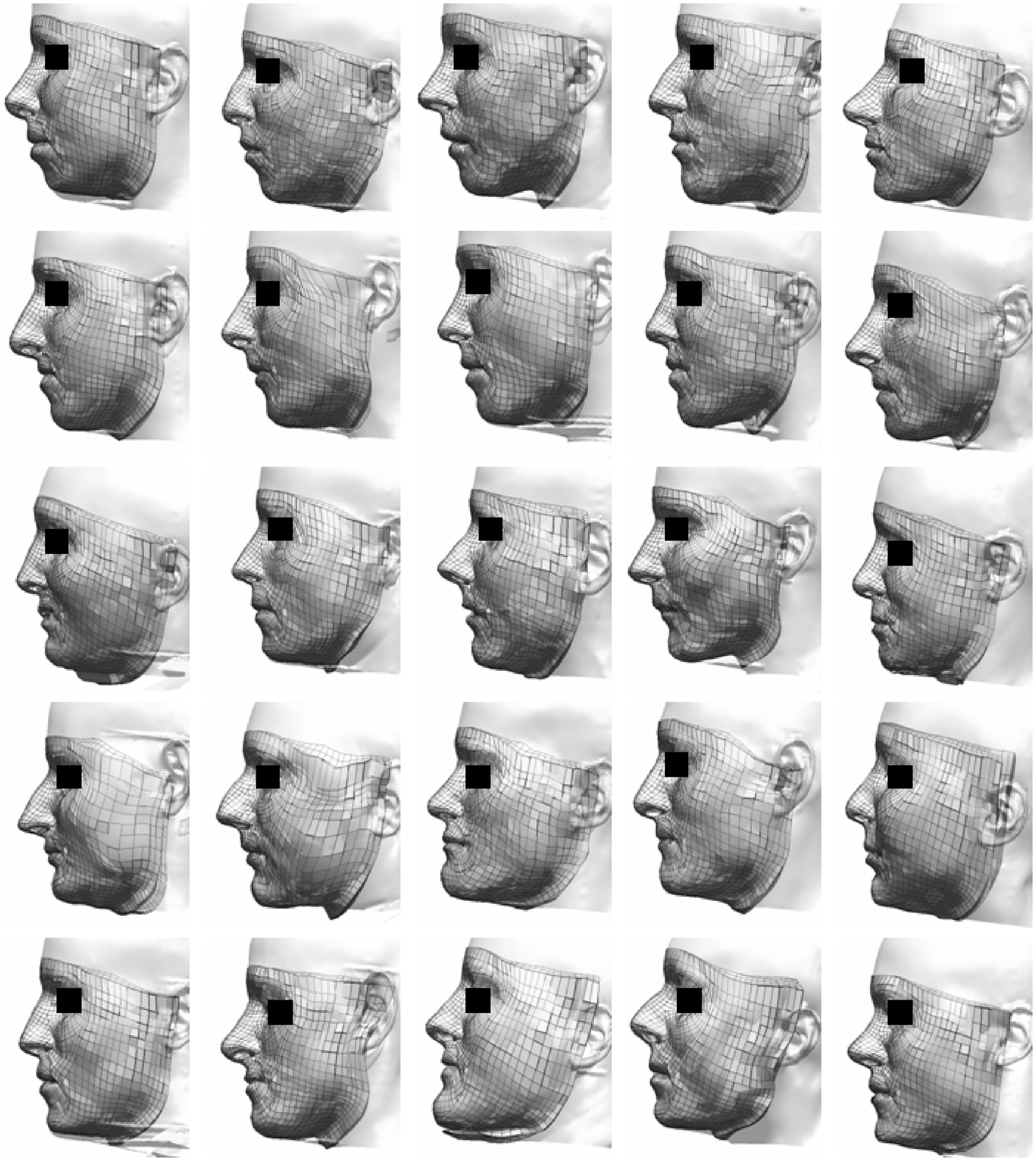}
	\caption{Sample of 25 registered face meshes. For each patient, the transparent skin surface mesh is shown along with the registered FE mesh. For clarity the skulls were omitted.}
	\label{Fig25Faces}
\end{center}
\end{figure}

Finally, Fig. \ref{Fig25Faces} shows 25 thumbnails of registered face meshes demonstrating the variety of clinical cases embraced by the study. The two upper rows show retrognathic cases, the middle row average patients and the two lower rows prognathic morphologies.

\section{Discussion and conclusions}

A fast and automatic mesh generation procedure - the MMRep algorithm - has been presented, based upon the elastic registration of a generic, or ``Atlas'', mesh towards patient specific structures, coupled with a robust mesh repair technique which ensures that element quality standards are met and FE analysis can safely be carried out on the resulting domain discretization.

To our knowledge this is the first evaluation of a FE mesh registration technique carried out on a wide range of 60 clinical data sets, illustrating 3 distinct use cases, raising both pre- and intraoperative biomechanical modeling issues and relying on fully or partially available patient data. In all situations the MMRep technique automatically generated a patient specific quality mesh within minutes, which strongly contrasts with time-consuming manual mesh assembly procedures involving a human expert operator. 

In all studied cases the regularity of the elastic deformation preserved the mesh elements from excessive distortions and the repair algorithm was able to find a suitable nodal configuration while maintaining a satisfactory surface representation accuracy. Only a small fraction, less than 1\%, of the mesh nodes positions needed to be corrected by submillimetric displacements.

Furthermore, the elastic registration formulation made it possible to compute a mesh deformation driven by multiple anatomical structures or sub-structures, as shown in \S \ref{SecFaces}. This feature could easily be used in the femur model generation scenarii, \S \ref{SecCompleteFemora} and \ref{SecPartialFemora}, for example for distinction between cortical and spongious bone layers, which have distinct mechanical properties.

This study shows that the proposed elastic registration technique is well-suited to the addressed meshing problem: computational times are short and submillimetric surface representation accuracy is achieved. However the MMRep procedure could be supported by any elastic registration algorithm \cite{Beg05,Vercauteren07,Rueckert06}, provided that the following properties are ensured:
\begin{itemize}
	\item A smooth deformation field, ideally $C^1$, can be estimated by the registration algorithm within reasonable computational times.
	\item Non-folding and bijection of the registration are ensured.
	\item Accurate registration inverse can be computed if the patient data is registered onto the Atlas and the inverse deformation is applied to the generic mesh in order to make it specific, as described in \S \ref{SecCompleteFemora} and \ref{SecPartialFemora}.
	\item The simultaneous registration of different structures, if required by the biomechanical modeling, can be computed as shown in \S \ref{SecFaces}.
\end{itemize}

The Atlas based approach presented here relies on the definition of a generic model of the target organ with the desired elements layout and possibly some identified sub-structures of interest. This modeling effort only needs to be done once but the resulting Atlas mesh must be carefully designed so that the anatomical features of interest can be properly registered to their patient specific counterparts.

The tests carried out on our database suggest algorithm robustness although no formal proof of convergence could be given. In extreme cases where strongly distorted pathological organs diverge from the average Atlas shape, mesh regularity can be lost and the repair strategies proposed here may fail. This issue can be overcome by working with a pool of Atlas meshes that represent the main deformation classes likely to be encountered. Such Atlas variations can be easily generated by successfully applying the MMRep procedure to a representative case and using the resulting quality mesh as a starting point for subsequent mesh registrations of similar configurations.

A number of enhancements to the presented MMRep technique can be foreseen. Breaking the sequential registration-repair scheme, the algorithm could benefit from the integration of the repair process within the elastic registration computation itself. This could be implemented either as a new energy term replacing the ad-hoc mechanical formulation controlling the deformation regularity, or as a per-iteration post-processing callback which, although more straightforward, would unfortunately raise the issue of the convergence of the elastic registration algorithm.

Future works also include the broadening of our mesh repair approach by taking into account other element types such as pyramids and tetrahedrons, as well as other quality criteria, such as the face warping factor \cite{Robinson82} that measures each element's face nodes coplanarity. Stronger constraints on the evenness of the generated meshes should enhance the numerical stability of the FE analysis carried out.

Finally, we can imagine an ideal FE mesh generation algorithm that performs mesh registration and repair directly in patient 3D image space. This fully integrated organ modeling tool could be achieved by replacing the distance based elastic registration formulation used here by an appropriate segmentation energy suitable for the considered imaging modality.

\section{Acknowledgements}

This work was carried out within the French national research project ANR/TecSan HipSurf. The authors wish to thank Mr. Yves Rozenholc, from the La\-bo\-ra\-toire de Ma\-th\'e\-ma\-tiques Ap\-pli\-qu\'ees (MAP5, UMR CNRS, University Paris 5) for the head CT scan volumes, Mr. Pascal Swider from the Tou\-louse biomechanics laboratory for the complete femora CT scans and Mr. Lionel Carrat (Praxim, France) for providing us with the intraoperative hip digitization data sets.

\bibliographystyle{splncs}
\bibliography{MyBib}

\end{document}